\documentclass[11pt]{article}   
\usepackage{amssymb,amsmath,amsfonts,eurosym,geometry,ulem,graphicx,caption,color,setspace,sectsty,comment,footmisc,caption,pdflscape,subfigure,array,hyperref}
\newtheorem{lemme}{Lemma}

\newtheorem{theor}{Theorem}
\newtheorem{prop}{Proposition}

\newtheorem{defi}{Definition}
\newtheorem{example}{Example}

\newcommand{\pimunu}{\Pi (\mu ,\nu)}
\newcommand{\var}{\hbox{\rm var}}
\newcommand{\cov}{\hbox{\rm cov}}
\newcommand{\erfc}{\hbox{\rm erfc}}
\newcommand{\sxx}{\sigma_X^2}
\newcommand{\syy}{\sigma_Y^2}
\newcommand{\A}{\mu_X-\mu_Y}
\newcommand{\AAA}{\mu_Y-\mu_X}
\newcommand{\mxy}{|\mu_X-\mu_Y|}
\newcommand{\B}{\sigma_X^2+\sigma_Y^2}

\newcommand{\xandy}{X \hbox{ and } Y}
\newcommand{\muandnu}{\mu \hbox{ and } \nu}
\newcommand{\exy}{E|X-Y|}

\newcommand{\probamodel}{(\Omega,{\cal A},P)}

\newcommand{\sumjn}{\sum_{j=1}^n}

\newcommand{\sumin}{\sum_{i=1}^n}

\newcommand{\idotn}{1,\ldots ,n}
\def\real{I\!\!R}

\newcolumntype{L}[1]{>{\raggedright\let\newline\\arraybackslash\hspace{0pt}}m{#1}}
\newcolumntype{C}[1]{>{\centering\let\newline\\arraybackslash\hspace{0pt}}m{#1}}
\newcolumntype{R}[1]{>{\raggedleft\let\newline\\arraybackslash\hspace{0pt}}m{#1}}

\begin{document}

\begin{titlepage}
\title{Various issues around the $L_1$-norm distance}
\author{Jean-Daniel Rolle\thanks{University of Applied Arts and Sciences Western Switzerland \newline
Haute Ecole de Gestion, Chemin du Mus\'ee 4, 1700 Fribourg, Switzerland}}
\date{\today}
\maketitle
\begin{abstract}
Beyond the new results mentioned hereafter, this article aims at familiarizing researchers working in applied fields -- such as physics or economics -- with notions or formulas that they use daily without always identifying all their theoretical features or potentialities. Various situations where the $L_1$-norm distance $\exy$ between real-valued random variables intervene are closely examined. The axiomatic surrounding this distance is also explored.
We constantly try to build bridges between the concrete uses of $\exy$ and the underlying probabilistic model. An alternative interpretation of this distance is also examined, as well as its relation to the Gini index (economics) and the Lukaszyk-Karmovsky distance (physics). The main contributions are the following:
(a) We show that under independence, triangle inequality holds for the normalized form $\exy /(E|X| + E|Y|)$.
(b) In order to present a concrete advance, we determine the analytic form of $\exy$ and of its normalized expression when $X$ and $Y$ are independent with Gaussian or uniform distribution. The resulting formulas generalize relevant tools already in use in areas such as physics and economics. (c) We propose with all the required rigor a brief one-dimensional introduction to the optimal transport problem, essentially for a $L_1$ cost function. The chosen illustrations and examples should be of great help for newcomers to the field. New proofs and new results are proposed.
\\
\vspace{0in}
\vspace{0in}
\bigskip
\end{abstract}
\setcounter{page}{0}
\thispagestyle{empty}
\end{titlepage}
\pagebreak \newpage
\section{Introduction} \label{introduction}
The notion of distance is fundamental in human experience; human beings constantly need to represent some degree of closeness between objects, whether the latter are physical or symbolic, concrete or abstract. Quantifying the closeness between random objects has become a task of vital interest to virtually all researchers working in applied sciences. This text is designated for a broad readership and we tried to make it as self-contained as we reasonably can, with as few ``it can be shown that'' as possible. As a result, the presentation contains a relatively larger amount of background material than it is usually found in articles dealing with comparable subjects. Our hope is that readers mainly interested in applications will find a theory that is accessible to them. 

Conceptual metric spaces are usually meant to have properties similar to those  of the ``natural'' metric $|x-y|$ of the real line. We can ask ourselves the following question: which distance should we use when the real numbers $x$ and $y$ are replaced by real-valued random variables $\xandy$? The answer is not unique, but it immediately comes to mind to look for it in the family of distances resulting from the $L_p$ norm. Let us first recall the context in which this norm is used. In what follows a probability space will be denoted by $\probamodel$, where the sample space $\Omega$ is endowed with a $\sigma$-field ${\cal A}$ and a probability function $P$, and ${\cal B}_1$ will designate the $\sigma$-field of Borel subsets of $\real$. If $p \in [1, \infty )$, which is the range of interest of most applications and the range we will consider in this article, the space $L_p (\Omega, {\cal A}, P)$ -- also denoted by $L_p (\real)$ or even more simply by $L_p$ -- consists of all $p$-(Lebesgue) integrable random variables $(\Omega, {\cal A},P) \rightarrow (\real, {\cal B}_1)$, i.e. random variables that satisfy $E|X|^p < \infty$. Then, if $X \in L_p$, we define the $L_p$ norm of $X$ by $||X||_p = (E|X|^p )^{1/p}$. Note that we are facing the following technical point: $||X||_p = 0$ does not imply that $X=0$, but only that $X=0$ almost surely. To be quite precise, the set of $p$-integrable real-valued random variables together with the function $||\cdot ||_p$ is a seminormed vector space denoted by ${\cal L}_p (\Omega, {\cal A}, P)$, ${\cal L}_p(\real)$ or simply ${\cal L}_p$ . Then the quotient space $L_p (\Omega, {\cal A}, P)$ is defined as the normed vector space of the equivalence classes for the equivalence relation: $X \sim Y $ if and only if $X = Y$ almost surely, where $X,Y \in {\cal L}_p(\Omega, {\cal A}, P)$. In other words  the random variables which agree almost surely are identified. The passage of ${\cal L}_p$ to $L_p$ is convenient, but also a bit confusing. However, many authors use the notation $L_p$ to refer to either space. In practice, we often forget that we are in the presence of equivalence classes rather than random variables. In short, ${\cal L}_p$ refers to a set of random variables while $L_p$ is a set of classes for the a.s. equality relation. We will make a clear distinction when other equivalence relations are involved, e.g. in Subsection~\ref{inconsistency}.

The spaces $L_p$ are complete normed vector spaces, that is Banach spaces. ``Complete'' means that the limit of any Cauchy sequence is within the space itself. Among all $p \in [1, \infty )$, the case $p=2\ $ is special: the norm $||X||_2$ follows from the inner product of $L_2$, and $(L_2,||X||_2)$ is the only Hilbert space\footnote{A Hilbert space is a Banach space whose norm is determined by an inner product.   } of the family. Finally $1 \le p < q < \infty$ implies that $||X||_p = (E|X|^p )^{1/p} \le (E|X|^q )^{1/q} = ||X||_q$ and hence $L_q \subset L_p$ (in particular $L_2 \subset L_1$).

We still haven't answered the question of replacing the distance $|x-y|$ between two real numbers $x$ and $y$ when instead of them we have to deal with real-valued random variables $\xandy$. Let $\delta_x$ (resp. $\delta_y$) denote the Dirac delta measure supported on the singleton $\{ x \}$ (resp. $\{ y \}$). If $X \sim \delta_x$ and $Y \sim \delta_y$, then the distance $||X-Y||_p$ equalizes to $|x-y|$. Indeed, 
$\exy^p = \int_{\real} \ \{ \int_{\real} |s-t|^p \delta(t-y) dt\}\ \delta(s-x) ds = |x-y|^p$, and thus 
$||X-Y||_p = (\exy^p )^{1/p} = |x-y|$. So all distances $||X-Y||_p$ generalize $|x-y|$ in the sense described above. Which $p$ should we choose? A serious candidate is $p = 2$, by virtue of the Hilbertian character of the space $(L_2,||\cdot ||_2)$. But if in a given study we are not interested in the mathematical facilities that the existence of an inner product allows (ability to treat random variables as vectors, projection on a closed convex set, orthogonality, etc.), then the choice of $(L_1,||\cdot ||_1)$ seems more suitable: let us look in this respect at the $L_2$-norm distance $||X-Y||_2 = (E|X-Y|^2 )^{1/2}$. Its use gives rise to a distortion in that it squares the differences: this distance tends to underestimate small differences (smaller than 1) and overestimate large differences (larger than 1), although the problem is mitigated by taking the square root at the end of the calculation. As the $L_1$ distance is not subject to this type of deformation, it comes first in terms of simplicity and interpretation. And all the more so since it is less than or equal to all other $L_p$ distances: $||X-Y||_1 \le ||X-Y||_p$ $\forall p > 1$. For all these reasons we will mainly focus on the distance $||X-Y||_1 = \exy$ between real-valued integrable random variables $\xandy$ defined on the same probability space $\probamodel$, namely
\begin{equation}
E|X-Y| =\int_{\omega \in \Omega} |X(\omega)- Y(\omega)|dP(\omega)\ \ \ \ \ (= \int_{\real^2}|x-y|dP_{(X,Y)} (x,y)\ ),         \label{L1}
\end{equation}
where $P_{(X,Y)}$ is the pushforward probability of $P$ induced on $\real^2$ by the random pair $(X,Y)$. Equation~(\ref{L1}) can be given the following interpretation: a pair of random values $(x,y)$ taken by the jointly distributed $\xandy$ is observed. The absolute difference between $x$ and $y$ is recorded. The sampling procedure is repeated independently an infinite number of times and the observed absolute differences $|x-y|$ are averaged. This endless process will yield $\exy$.

The fields concerned with the expected absolute difference are numerous and various. They include in particular data analysis, clustering, optimal transport, physics, biology, economics, finance, engineering, image analysis. 
Interestingly, an identical distance measurement is subject to sporadic reappearances in areas that have \emph{a priori} nothing in common. For example, the \emph{Gini mean difference} (GMD) used in inequality economics -- and sometimes also considered as an $L_1$ alternative to the standard deviation  --, or the so-called \emph{Lukaszyk-Karmowski metric} used in mechanical physics or in quantum physics, have been proposed independently, according to the specific needs of their domain. Both are expressions of the statistical distance $\exy$ between two random variables $X$ and $Y$. In the case of GMD, $\xandy$ are independent and identically distributed (i.i.d), whilst in the case of the Lukaszyk-Karmowski metric they are usually assumed to be independent.
Not surprisingly, $\exy$ appears under different names in the literature, including expected (mean, average) absolute difference (deviation) between variables, ${\cal L}_1$- distance, $L_1$-distance, $L_1$-norm distance, $L_1$-metric, 1-average compound metric. In this document, $\exy$ will almost always be referred to as the \emph{expected absolute difference} between $\xandy$.

As we will frequently encounter the notion of ``distance'', ``semimetric'', ``metric'' and ``metametric'', it is certainly not useless to specify the mathematical properties that these words cover. Let $E$ be a set and consider a function $d:E\times E\longrightarrow\real_+ := [0,\infty )$. This non-negative function may have various properties that must hold for all $x$, $y$, $z\in E$:  
\begin{enumerate}
\item $d(x,x)=0$ \hskip .3cm(reflexivity)
\item $d(x,y)=0 \Rightarrow x=y$ \hskip .3cm(reverse reflexivity)
\item $d(x,y)=d(y,x)$ \hskip .3cm(symmetry)
\item $d(x,y) \leq d(x,z) + d(z,y)$ \hskip .3cm(triangle inequality)
\end{enumerate}
If $d$ satisfies reflexivity and symmetry, it is called a \emph{distance} and the ordered pair $(E,d)$ is called a \emph{distance space}. 
If $d$ satisfies reflexivity, symmetry and triangle inequality, it is a \emph{semimetric} and $(E,d)$ is a \emph{semimetric space}. 
If $d$ satisfies reflexivity, reverse reflexivity, symmetry and triangle inequality, it is a \emph{metric} and $(E,d)$ is a \emph{metric space}. Reflexivity and reverse reflexivity together constitute the so-called \emph{identity of indiscernibles}. We will exercise some latitude when using the word ``distance'' even if we are actually talking about metrics or semimetrics.

This work organizes the reflexion around $\exy$ in the following way: \textbf{Section~\ref{absdiff}} shows how the $L_1$-distance between distribution functions\footnote{which is also the 1-Wasserstein distance between the corresponding distributions.} and the expected absolute difference are related and how two separate experiments can be consistently unified. \textbf{Section~\ref{axioms}} deals with the axiomatic of probability metrics, $\exy$ being what is called a \emph{compound metric}. On this occasion, a well-hidden logical inconsistency tainting published work is brought to light. \textbf{Section~\ref{indep}} focuses on the behavior of $\exy$ in relation to independence, almost sure equality, or equality of distribution of the variables $\xandy$. The \emph{normalized} expected absolute difference is discussed in \textbf{Section~\ref{normalized}}, where the prominent role of independence is emphasized. A primary metric defined on the distributions of pairs of random variables is specified in \textbf{Section~\ref{interpretation}}, in order to provide a new interpretation of the expected absolute difference. This leads to a very general expression for the \emph{Gini mean difference} and the \emph{Gini index}. In \textbf{Section~\ref{appl}}, we give in analytic form the expected absolute difference between two independent normally distributed random variables. We end up with a result generalizing formulas used in applied physics an in economics. In the process, we also give the analytic form of the average distance between coordinates of points falling at random into a proper rectangle of $\real^2$. We envision that \textbf{Section~\ref{formulation}} can provide the basic background material to understand the main concepts of the optimal transport theory. The latter is consciously presented in a restricted framework, as a first step in the access to a complex field in full expansion. It is precisely these restrictions that allow to bring to light very telling results, sometimes even spectacular, in any case of a indeniable mathematical beauty. In this context, the presence of closed-form solutions to the optimal transport problem allows - or at least greatly facilitates - a good understanding of the subject through important special cases.

Here are some of the notations used throughout this article: we write $\real_+$ for the set of non-negative real numbers, and ${\cal B}_d$ refers to the $\sigma$-field of Borel subsets of $\real^d$, $d\ge 1$. Moreover, 
${\cal P}(\real^d)$ denotes the set of probability measures on $(\real^d,{\cal B}_d)$ and ${\cal P}_p(\real^d)$ the set of probability measures on $(\real^d,{\cal B}_d)$ with finite $p$-th moment. We will be mainly interested in random variables $X,Y,Z,\ldots$ defined on $\probamodel$ taking their values in $(\real,{\cal B}_1)$. Let $V:\probamodel \rightarrow (\real^d,{\cal B}_d)$ be a random variable or a random vector defined on a given probability space. 
We denote by $P_V$ the pushforward probability of $P$ induced by $V$ on $(\real^d,{\cal B}_d)$. For example, $P_X$, $P_{(X,Y)}$, $P_{(X,Y,Z)}$ refer to pushforward probability measures of $P$ on $(\real,{\cal B}_1)$ (resp. $(\real^2,{\cal B}_2)$, $(\real^3,{\cal B}_3)$) induced by $X$ (resp. $(X,Y)$, $(X,Y,Z)$). The notation ${\cal L}_1(\real^d)$, $d\ge 1$, will refer to the space of integrable random variables or vectors defined on a probability space $\probamodel$ which take their values in $(\real^d,{\cal B}_d)$.
 Almost sure equality (resp. equality of distribution) of random variables $\xandy$ are denoted by $X\stackrel{a.s.}{=} Y$ (resp. $X\stackrel{d}{=} Y$).
The respective abbreviations \emph{cdf} and \emph{pdf} stand for \emph{cumulative distribution function} and \emph{probability density function}. 
\section{On distances based on absolute difference} \label{absdiff}
Subsection~\ref{linklit} focuses on the relationship between the \emph{Gini-Kantorovich distance }(a $L_1$-distance between cdf's)  and the expected absolute difference (a $L_1$-norm distance between random variables). We discuss properties of these distances and examine the historical premises of the optimization problem at the origin of the link that unites them. Subsection~\ref{unif} discusses how two separate experiments can be consistently unified.
\subsection{How $L_1$-distance between cumulative distribution functions and ${\cal L}_1$-distance between random variables are related} \label{linklit}
A statement such as ``random variables $X$ and $Y$ are defined on the same probability space'' implies that $X$ and $Y$ have a joint distribution, in which case we say that they are \emph{coupled}. First, consider two integrable real-valued random variables $X$ and $Y$ that \emph{may not} be defined on the same probability space. If one knows their (individual) distributions only, one can define a distance between them by using their respective cumulative distribution functions $F_X$ and $F_Y$. A rather intuitive way of measuring this distance is to calculate the Gini-Kantorovich distance
\begin{equation}
GK(X,Y) = \int_{\real} |F_X(x) - F_Y(x)|dx ,    \label{Kant}      
\end{equation}
which can be easily visualized as a surface between two curves. One can show that
\begin{equation}
GK(X,Y) = \int_{\real} |F_X(x) - F_Y(x)|dx = \int_0^1 |F_X^{-}(t) - F_Y^{-}(t)|dt ,    \label{coincint}      
\end{equation}
where $F_X^-$ and $F_Y^-$ are the \emph{quantile functions} or \emph{generalized inverses} of $F_X$ and $F_Y$, respectively.
A proof of this remarkable coincidence is given in Thorpe (2018)\nocite{Thorpe2018}, see also Rachev and Rueschendorf (1998)\nocite{Rachev1998}. The generalized inverse is defined in Subsection~\ref{onedim} (Definition~\ref{geninv}). For more details about quantile functions, see Karr (1993)\nocite{Karr1993} p. 63, or Embrechts and Hofer (2014)\nocite{Embrechts2014}.  

$GK$, also known as the \emph{Gini index of dissimilarity}\footnote{$GK$ is also called the $L_1$-metric between distribution functions or Monge-Kantorovich metric or Hutchinson metric, 1-Wasserstein metric, Fortet-Mourier metric, see Deza and Deza (2014)\nocite{Deza2014}.}, is a special case of a more general Gini-Kantorovich metric $GK_p$ when $p=1$ (Ortobelli et al. (2006)\nocite{Ortobelli2006}). The Gini index of dissimilarity should not be confused with the\emph{ Gini mean difference} (GMD) or the \emph{Gini index} discussed later in this article (Subsections~\ref{gindex}, \ref{probGMD} and \ref{formulas}).
Rachev et al. (2013)\nocite{Rachev2013}, note that (\ref{Kant}) is the explicit solution of a minimization problem studied by Gini (1914)\nocite{Gini1914a} and solved by Salvemini (1943)\nocite{Salvemini1943} for discrete cdf's and Dall'Aglio (1956)\nocite{Dallaglio1956} in the general case. More precisely, let ${\cal F}(F_1,F_2)$ denote the set of all bivariate cdf's $F$ with marginal cdf's $F_1$ and $F_2$. Then the analytic solution of the minimization problem
\begin{equation}
\hbox {inf} \{ \int_{\real^2} |x - y| dF(x,y):  F \in {\cal F}(F_1,F_2) \}         \label{inf}
\end{equation}
is ``simply''
\begin{equation}
\int_{\real} |F_1(x) - F_2(x)|dx.         \label{sol}
\end{equation}
The optimization problem (\ref{inf}) and its solution are often expressed as
\begin{equation}
GK(X,Y) = \hbox {inf} \{ E|\tilde{X} - \tilde{Y}|:\tilde{X}\stackrel{d}{=} X, \tilde{Y}\stackrel{d}{=} Y \},    \label{inf2}
\end{equation}
where $\xandy$ are given random variables and where ``$\stackrel{d}{=}$'' refers to equality of distribution (Rachev et al. (2007), Eq. 3.23\nocite{Rachev2007})\footnote{Rachev et al. (2013)\nocite{Rachev2013} formulate concisely the more general problem of mass transportation studied by Kantorovich, of which the classic transportation problem in linear programming and the minimization problem (\ref{inf}) are special cases.}. Actually, as the map $c(x,y) := |x - y|$ is continuous\footnote{or even lower semi-continuous, a weaker condition. In optimal transport, $c(x,y)$ denotes the transportation cost function.}, a minimimizer does exist (Gangbo (2004), Th. 2.4)\nocite{Gangbo2004} and we can replace ``inf'' by ``min'' in (\ref{inf}) and (\ref{inf2}). Typically, (\ref{inf2}) shows that the infimum runs over all \emph{couplings}\footnote{Couplings are defined in Subsection~\ref{coupldef}. }
$(\tilde{X},\tilde{Y})$ of $\xandy$, where $\xandy$ may or may not be defined on the same probability space. Now suppose that $\xandy$ are both defined on a probability space $\probamodel$, i.e. are jointly distributed. Then a look at (\ref{Kant}) -- where $GK$ depends only on the individual cdf's of $\xandy$ --
confirms that $GK$ ignores any structure of dependence or independence inside the pair $(X,Y)$. In the process of minimization described in (\ref{inf2}), of which $GK(X,Y)$ is the solution, any dependence or independence structure is swept away: we are left only with a probability metric measuring the $L_1$-distance between the cdf's $F_X$ and $F_Y$ of $\xandy$, respectively. Such a situation is unsatisfactory because it ignores valuable information that can be available in practice, for example when one can assume that $\xandy$ are independent. This can be illustrated with a simple example: Table~\ref{gk} shows two joint distributions of binary $\{0,1\}$-valued random variables.
\begin{table}[ht]
	\centering
   \begin{tabular}{cc}
   \begin{tabular}{|r|c|cc|c|} \hline
          \textbf{(a) }    &   & $Y$  &  & \\ \hline
                  &   & 0  & 1 & $P_X$\\ \hline
     $X$          & 0 & 0.1& 0.6& \textbf{0.7} \\
                  & 1 & 0.1  & 0.2 & \textbf{0.3}\\ \hline
									& $P_Y$ & \textbf{0.2}  & \textbf{0.8} & \textbf{1} \\ \hline
   \end{tabular}  &  \begin{tabular}{|r|c|cc|c|} \hline
          \textbf{(b) }    &   & $Y$  &  & \\ \hline
                  &   & 0  & 1 & $P_X$\\ \hline
     $X$          & 0 & 0.14& 0.56& \textbf{0.7} \\
                  & 1 & 0.06  & 0.24 & \textbf{0.3}\\ \hline
									& $P_Y$ & \textbf{0.2}  & \textbf{0.8} & \textbf{1} \\ \hline
   \end{tabular}

   \end{tabular} 
	\caption{\small Distributions of two pairs $(X,Y)$ of binary random variable $\xandy$. The two distributions have the same marginals. In \textbf{(a)} $\xandy$ are dependent while in \textbf{(b)} they are independent.} 
	\label{gk}
\end{table}
Distribution \textbf{(a)} in Table \ref{gk} reflects a dependence between $X$ and $Y$, while distribution \textbf{(b)} corresponds to independence. In both cases, $GK(X,Y) = 0.5$, ignoring the dependence structure between the variables. Note that $\exy = 0.7$ (case \textbf{(a)}) and 0.62 (case \textbf{(b)}).
A probability metric such as $GK(X,Y)$ makes sense if $X$ and $Y$ are uncoupled, i.e. if we only know their one-dimensional cdf's. When $\xandy$ are coupled, there are more informative ways of determining how far apart they are from each other. As $GK$ cannot take full account of the information of the model, it may be replaced by the expected absolute difference. As a matter of fact, $E|X-Y|$ uses all the information contained in the probability space $\probamodel$ governing the distribution of the pair $(X,Y)$ to determine how far $X$ is from $Y$. An emblematic case occurs when $\xandy$ can be assumed to be independent. It is well-known that under the independence assumption 
$X: (\Omega_1,{\cal A}_1,P_1) \rightarrow (\real, {\cal B}_1)$ and $Y: (\Omega_2,{\cal A}_2,P_2) \rightarrow (\real, {\cal B}_1)$ can be defined trivially on the same probability space, namely the product space
$(\Omega,{\cal A},P) = (\Omega_1 \times \Omega_2,{\cal A}_1 \otimes {\cal A}_2,P_1 \otimes P_2)$. Consequently, $\xandy$ are jointly distributed and the use of $GK$ (or any other similar metric) would be inappropriate.

\subsection{Consistent unification of two separate experiments} \label{unif}
(The reader familiar with measure or probability theory may skim this subsection).
Fundamental probabilistic concepts, although often trivialized in applied papers, are not always sufficiently understood. We have seen in the previous section how the fact that random variables are jointly distributed or not can affect the choice of an adequate distance function. In connection with the content of Subsection~\ref{linklit}, we recall the rules that must be respected so that two separate experiments can be adequately combined into a joint experiment.

Consider two random experiments ${\cal E}_1$ and ${\cal E}_2$. Suppose that the information on the experiments is captured by real numbers; that is, ${\cal E}_1$ and ${\cal E}_2$ are completely described by the respective probability spaces $(\real, {\cal B}_1,P_1)$ and $(\real, {\cal B}_1,P_2)$\footnote{For simplicity, we suppose that $\real$ is the common sample space of ${\cal E}_1$ and ${\cal E}_2$ and that $\real$ is endowed with 
${\cal B}_1$, to obtain the common measurable space $(\real,{\cal B}_1)$.   }. 
If the unification operation is conducted in a coherent way, then ${\cal E}_1$ and ${\cal E}_2$ can be seen as ``marginal'' experiments of ${\cal E}$. A probability model $\probamodel$ has to be defined to describe the ``joint experiment'' ${\cal E}$.

Random variables $X$, $Y$ -- and the resulting pair $(X,Y)$ taking values in $(\real^2, {\cal B}_2)$ -- can be defined on a common probability space $\probamodel$ so that $P_1$ and $P_2$ are the marginal probability measures of $P$. Indeed, we can define $\Omega = \real^2$, ${\cal A} = {\cal B}_1 \otimes {\cal B}_1 = {\cal B}_2 $ and  $(X,Y)= Id_2 = (q_1,q_2)$, where $Id_2$ is the identity map on $\real^2$ and the $q_i$'s are the corresponding projection functions (i.e. for $(x_1,x_2)\in \real^2$, $q_i(x_1,x_2) = x_i$, $i=1,2$). So $\xandy$ can be interpreted indifferently as random variables or as projections. In order for $P_1$ and $P_2$ to be the marginals of some probability measure $P$, and noting that $P(q_1^{-1} B_1) = P(B_1 \times \real)$ and $P(q_2^{-1} B_2) = P(\real \times B_2)$, the following consistency conditions are imposed:
\begin{equation}
P(B_1 \times \real) = P_1(B_1) \hskip .5cm \hbox{ and }\hskip .5cm 
P(\real \times B_2) = P_2(B_2)         \label{marginals}
\end{equation}
for all $B_1, B_2 \in {\cal B}_1$. Of course the conditions in (\ref{marginals}) are not sufficient to fully determine $P$, a feature that was predictable since the link between the two experiments was not specified. In the particular case where the two experiments (and hence the two random variables $X$ and $Y$) are assumed to be independent, $P$ must satisfy the additional condition 
\begin{equation}
P(B_1 \times B_2) = P_1(B_1) P_2(B_2)       \label{produit}
\end{equation}
for all $B_1,B_2  \in {\cal B}_1$. In other words, $P$ is the product probability $P_1\otimes P_2$, which is uniquely defined on ${\cal B}_2$. Moreover $P_1 = P_X$, $P_2 = P_Y$ and $P = P_{(X,Y)}$ in the above construction. The process just described consists of two steps that are worth distinguishing.
\vskip .2cm \noindent
\emph{First step:} Two random experiments ${\cal E}_1$ and ${\cal E}_2$ are united to obtain a measurable space for the resulting joint experiment ${\cal E}$.
\vskip .2cm \noindent
\emph{Second step:} A probability measure (and the resulting dependence structure between $\xandy$) is enforced on the model set up in the first step.

Figuratively, one could say\footnote{without reference to the Marxist phraseology...} that the first step corresponds to an ``infrastructure" on which a probabilistic ``superstructure" is built in the second step.
\vskip .2cm

\begin{figure}[t]
  \centering
  \includegraphics[width=8cm,height=6.3cm]{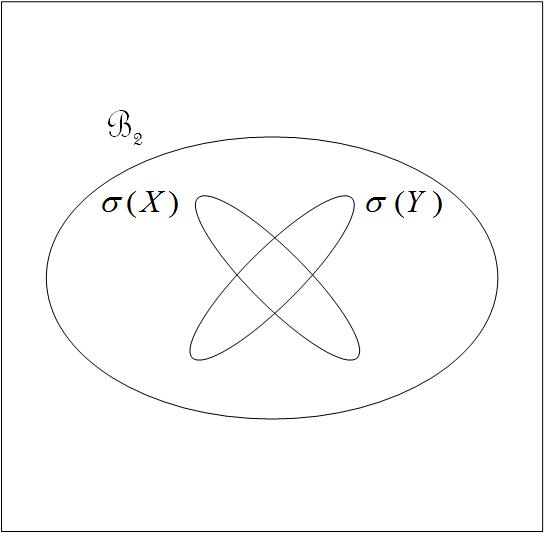}
  \caption{\small Venn diagram representing $\sigma (X) = \{B_1 \times \real : B_1 \in {\cal B}_1  \}$, \newline $\sigma (Y) = \{\real  \times B_2: B_2 \in {\cal B}_1  \}$ and the $\sigma$-field ${\cal B}_2$  generated by the union of $\sigma (X)$ and $\sigma (Y)$.}
  \label{fig:sigxy}
\end{figure}

One can illustrate, in terms of $\sigma$-fields, the qualitative leap following the coupling of two previously separate (stand-alone) random experiments ${\cal E}_1$ and ${\cal E}_2$. Under the above assumptions and the ensuing construction, the information that ${\cal E}_1$ and ${\cal E}_2$ provide is carried by the random variables $X:(\real^2,{\cal B}_2) \longrightarrow (\real,{\cal B}_1 )$  and 
$Y:(\real^2,{\cal B}_2) \longrightarrow (\real,{\cal B}_1 )$. The minimal $\sigma$-fields generated by $X$ and $Y$ are denoted by       $\sigma (X)$ and $\sigma (Y)$, respectively. In turn, $\sigma (X)$ and $\sigma (Y)$ generate the $\sigma$-field
\begin{equation}
\sigma (X) \vee \sigma (Y) := \sigma (\sigma (X)\cup \sigma (Y)) .   \label{sigunion}
\end{equation} 
As by construction $X = q_1$ and $Y = q_2$, we have: $\sigma (X) = \{B_1 \times \real : B_1 \in {\cal B}_1  \}$ and  $\sigma (Y) = \{\real  \times B_2: B_2 \in {\cal B}_1  \}$.
It can be shown without much difficulty that $\sigma (X) \vee \sigma (Y) = \sigma ((X,Y))$, the $\sigma$-field on   $\real^2$ induced by the pair $(X,Y)$, noting that $\sigma ((X,Y)) = {\cal B}_2 = \sigma({\cal C})$, where ${\cal C} = \{B_1 \times B_2 : B_1, B_2 \in {\cal B}_1 \}$\footnote{That is, ${\cal B}_2$ is generated by the $(B_1 \times B_2)$'s, but ${\cal B}_2 = \sigma (X) \vee \sigma (Y)$ implies that ${\cal B}_2$ is also generated by the union of the $(B_1 \times \real)$'s and the $(\real \times B_2)$'s.}. Considering $X$ and $Y$ together as a pair of random variables instead of two stand-alone random variables \emph{allows to prepare a much wider portion} of ${\cal B}_2$ (infrastructure) on which a probability measure (superstructure) can be defined. The representation $\sigma (X) \vee \sigma (Y)$, probably more telling than $\sigma ((X,Y))$, is symbolized in Figure~\ref{fig:sigxy}. 
Incidentaly, note that the intersection $\sigma (X) \cap \sigma (Y)$ is not empty, since it contains $\emptyset$ and $\real^2$.
\section{Axiomatic approach of probability metrics}  \label{axioms}
Errors of interpretation about the notion of distance between random variables occur because important concepts are not formally stated or not sufficiently explained. The purpose of this section is to eliminate ambiguities encountered here and there in applied science works published in journals or on the Internet. The role of a metric is to define a distance between elements of a same set. We mentioned in the introduction that to verify precisely whether a functional is adequate to cover what is commonly understood by the notion of distance, it is unavoidable to specify certain elementary, natural and intuitive rules -- axioms -- that this functional should satisfy. Again, we limit the discussion to real-valued random variables. Rachev (1991)\nocite{Rachev1991} provides a more general treatment.

The idea of distance between two random variables $\xandy$ is linked in a decisive way to what is meant by $\xandy$ "are the same", "are coincident", "are indistinguishable". And the concept of sameness, coincidence or indistinguishably -- treated here as synonyms -- can be mathematically captured by an equivalence relation: $\xandy$ are declared to be the same, coincident or indistinguishable if they are in the same equivalence class. It is therefore natural to define a metric not on an initial set of random variables, but on the quotient space resulting from the adequate equivalence relation.

The theory of probability metrics considers three categories of metrics defined according to the type of equivalence deemed useful in a given context.
\subsection{Primary, simple and compound metrics} \label{prisicom}
We assume throughout this section that real-valued random variables are defined on one and the same probability space $\probamodel$. Denote by ${\cal V}$  the set of all random variables on $\probamodel$ taking values in $(\real, {\cal B}_1)$.\footnote{${\cal V} = {\cal L}_1(\real)$ when the random variables are assumed to be integrable.} So ${\cal V}^2 = {\cal V}\times {\cal V}$ is the set of all random pairs defined on $\probamodel$ taking values in $(\real^2, {\cal B}_2)$.

Let $\psi : {\cal V} \rightarrow \real$ be a functional candidate to be a distance. We would like $({\cal V},\psi )$ to be a metric space from the outset, but this wish is thwarted by the fact that we are working with random variables -- that is relatively complex mathematical objects. It is nevertheless possible to use $\psi$ to build a metric space defined on equivalence classes. For this purpose, let us define an equivalence relation on ${\cal V}$ by
\begin{equation}
 X \stackrel{\psi}{\sim} Y \Leftrightarrow \psi(X,Y) = 0.   \label{eqrel}
\end{equation}
What properties of $\psi$ will ensure that (\ref{eqrel}) is a reflexive, symmetric and transitive relation, i.e. is an equivalence relation? It is easily verified that the following axioms meet this requirement: for all $X,Y,Z \in {\cal V}$,
\newline
(a) $\psi (X,X) = 0$ \hskip .3cm(reflexivity) \newline
(b) $\psi (X,Y) = \psi (Y,X)$ \hskip .3cm(symmetry) \newline
(c) $\psi (X,Z) \leq \psi (X,Y) + \psi (Y,Z)$ \hskip .3cm(triangle inequality), \newline
noting that these axioms imply the non-negativity of $\psi$. So (a), (b) and (c), beyond their natural and intuitive content, are sufficient to ensure that 
$\stackrel{\psi}{\sim}$ is an equivalence relation. We call a functionnal satisfying (a), (b) and (c) a \emph{probability metric}, although it is rigorously a semimetric\footnote{The terminology is still fluctuating: in topology, what we define here as a semimetric is called a \emph{pseudometric}. }.
Now, $X \stackrel{\psi}{\sim} Y$ means different things depending on the choice of $\psi$, because $\psi(X,Y) = 0$ may be true if and only if $\xandy$ \newline
(i) share a given set $c_1, c_2,\ldots$ of characteristics such that
$c_i(X) = c_i(Y)$, $i= 1,2,\ldots$ (e.g. $\xandy$ have the same mean and the same variance, as in (\ref{primary2}) below), or\newline  
(ii) have the same distribution, or \newline 
(iii) are almost surely equal. 

So $\stackrel{\psi}{\sim}$ in (\ref{eqrel}) is a generic notation for specific equivalence relations:
\newline
$X \stackrel{c}{=} Y$ when $\xandy$ share a given set of characteristics, or    \newline
$X \stackrel{d}{=} Y$ when $\xandy$ have the same distribution, or   \newline
$X \stackrel{a.s.}{=} Y$ when $\xandy$ are almost surely equal. 
\begin{defi}\label{prisico} (Rachev et al. (2011) \nocite{Rachev2011})\newline
If $\stackrel{\psi}{\sim}$ means $\stackrel{c}{=}$, then $\psi$ is called a \emph{primary (probability) metric}.\newline
If $\stackrel{\psi}{\sim}$ means $\stackrel{d}{=}$, then $\psi$ is called a \emph{simple metric}.\newline
If $\stackrel{\psi}{\sim}$ means $\stackrel{a.s.}{=}$, then $\psi$ is called a \emph{compound metric}\footnote{The intervention of three random variables (instead of two) in the triangle inequality axiom raises a theoretical issue in the compound metrics case. The pairs $(X,Y)$, $(X,Z)$ and $(Y,Z)$ can be chosen in such a way that there exists a random vector $(X,Y,Z)$ ensuring that the three pairs are its two-dimensional projections. For more information, see the so-called "gluing lemma" (Thorpe (2018, Lemma~5.5)) which allows to "glue" two (or more) bivariate (multivariate) distributions so as to respect the different marginals. We refer to our discussion in paragraph~\ref{trieqgen}, where the \emph{consistency rule} is stated. The triangle inequality does not hold for all random variables $X,Y,Z$, but only for those satisfying this rule.  }.
\end{defi}
Primary metrics correspond to the weakest form of equivalence. Random variables $\xandy$ can be considered equivalent if they have the same mean and the same standard deviation. A plain example of primary metric is
\begin{equation}
\psi (X,Y) = |E(X) - E(Y)| + |\sigma(X) - \sigma(Y)|,                \label{primary2}
\end{equation}
where $\sigma$ refers to the standard deviation\footnote{Note that (\ref{primary2}) makes sense because the standard deviation and the mean are defined in the same unity.  }.

The simple metrics imply a stronger form of sameness: $\xandy$ are considered equivalent if their cdf's are identical (remembering that a random variable is completely described by its cdf). An example of simple metric is the Gini-Kantorovich metric GK given in (\ref{Kant}). GK measures the distance between $\xandy$ -- which are assumed to have finite first moment -- by a distance between their respective cdf's.\footnote{If $\xandy$have    respective probability distributions $\mu$ and $\nu$, it is remarkable that $GK(X,Y)$ coincides with the 1-Wasserstein metric $W_1(\mu,\nu)$. It turns out that the latter is none other than the minimal cost in the Monge-Kantorowich transport problem (Kantorovich (1942)\nocite{Kantorovich1942}, see e.g. Villani (2008)\nocite{Villani2008} or Thorpe (2018)). Rachev (2007)\nocite{Rachev2007} uses the notation $GK(X,Y) = \min E|X - Y|$. This is very telling since (i) due to the fact that $|x - y|$ is a continuous cost function in the Monge-Kantorovich optimal transport problem, the infimum is realized (Gangbo (2004)\nocite{Gangbo2004}) and (ii) it remains us that $GK(X,Y)$ may be seen as the solution of this celebrated minimization problem.   }

The compound metrics represent the strongest form of sameness. The simplest example of compound metric is probably the expected absolute difference $\psi (X,Y) = E|X - Y|$. Importantly, here $\xandy$ are necessarily defined on the same probability space (i.e. are jointly distributed) and have finite first moment.

We will now show that there is a (true) metric $\tilde{\psi}$ derived from the semimetric $\psi$, where $\tilde{\psi}$ is defined on equivalence classes. The classes, denoted by $[\cdot]$, stem from the equivalence relation 
$\stackrel{\psi}{\sim}$, which can mean, as we have seen above, $\stackrel{c}{=}$, $\stackrel{d}{=}$ or $\stackrel{a.s.}{=}$. 
Define in a canonical way
\begin{equation}
\tilde{\psi} ([X],[Y]) = \psi (X,Y). \label{psitilde}
\end{equation}
We must show that $\tilde{\psi}$ is well-defined, i.e. does not depend on the representatives chosen to designate the classes. To show that (\ref{psitilde}) makes sense, we need the following lemma.
\begin{lemme} \label{inequadri} (quadrilateral inequality, proof in the appendix) \newline
Let $\psi : {\cal V} \rightarrow \real$ be a functional satisfying non-negativity, symmetry and triangle inequality. Then for any $X,Y,X_1, Y_1 \in {\cal V}$,
\begin{equation}
|\psi (X,Y) - \psi (X_1,Y_1) | \leq \psi (X,X_1) + \psi (Y,Y_1). \label{quadrineq}
\end{equation}
\end{lemme} 
Assume that $\psi$ satisfies reflexivity, symmetry and triangle inequality (and therefore nonnegativity, so that the conditions of Lemma~\ref{inequadri} are fulfilled) and let the equivalence relation on ${\cal V}$ be given by (\ref{eqrel}). Suppose that $X_1\stackrel{\psi}{\sim}X$ and $Y_1\stackrel{\psi}{\sim}Y$. Then, using (\ref{quadrineq}) and the symmetry of $\psi$, we get $\psi (X_1,Y_1) = \psi (X,Y)$, that is $\tilde{\psi} ([X_1],[Y_1]) = \tilde{\psi} ([X],[Y])$, which proves that $\tilde{\psi}$ in (\ref{psitilde}) is well-defined. 

Let us write $\tilde{{\cal V}}$ for the quotient space ${\cal V}/\stackrel{\psi}{\sim} \ = \{[X]: X \in {\cal V} \}$.
We are now able to state that $(\tilde{{\cal V}},\tilde{\psi})$ is a metric space, i.e. that $\tilde{\psi}$ satisfies the following axioms:
\begin{enumerate}
\item $[X] = [Y] \Rightarrow \tilde{\psi} ([X],[Y])=0 $ \hskip .3cm(reflexivity)
\item $\tilde{\psi} ([X],[Y])=0 \Rightarrow [X] = [Y]$ \hskip .3cm(reverse reflexivity)
\item $\tilde{\psi}([X],[Y])=\tilde{\psi} ([Y],[X])$ \hskip .3cm(symmetry)
\item $\tilde{\psi} ([X],[Z]) \leq \tilde{\psi} ([X],[Y]) + \tilde{\psi} ([Y],[Z])$ \hskip .3cm(triangle inequality).
\end{enumerate}
The first two axioms -- known as the identity of indiscernibles when taken together -- are the consequence of the definitions of $\stackrel{\psi}{\sim}$ and $\tilde{\psi}$, while axioms 3 and 4 stem directly from the symmetry and triangle inequality property of $\psi$.

Note that if ${\cal V} = {\cal L}_1(\real)$ and $\psi (X,Y) = \exy$, then $\stackrel{\psi}{\sim}$ means     $\stackrel{a.s.}{=}$. In this case $\tilde{{\cal V}} = L_1(\real)$ and $\tilde{\psi}$ is the metric induced by the $L_1$-norm.
\subsection{Uncovering a logical inconsistency} \label{inconsistency}
We limit our discussion to $E|\cdot -\cdot |$, which implies that ${\cal V} = {\cal L}_1(\real)$ (${\cal V}$ defined in Section~\ref{prisicom}), but our conclusions can be generalized to other compound metrics. 

Focusing on the two equivalence relations $\stackrel{a.s.}{=}$ and $\stackrel{d}{=}$ which group variables belonging to ${\cal L}_1(\real)$, we denote by $[\cdot]_{a.s.}$ and $[\cdot]_{d}$, the corresponding equivalence classes. 
Why are we so eager to identify certain elements of ${\cal L}_1(\real)$? It is because it allows $\exy$ to switch from a semimetric to a metric. Indeed, if we define $\tilde{\psi} ([X]_{a.s.},[Y]_{a.s.}) = \psi (X,Y) = \exy$, then
$E|\cdot - \cdot|$ represents both $\tilde{\psi}$ (a metric defined on classes) and $\psi$ (a semimetric defined on random variables). We saw in Section~\ref{prisicom} that 
$\tilde{\psi}$ realizes the identity of indiscernibles (reflexivity and reverse reflexivity), whereas $\psi$ satisfies reflexivity ($\psi (X,X) = 0$), but not reverse reflexivity ($\psi (X,Y) = 0$ does not imply $X = Y$). Moreover, $\psi$ and $\tilde{\psi}$ both satisfy symmetry and triangle inequality.

That said, some authors using $\exy$ have fallen into the trap of identifying within ${\cal L}_1(\real)$ the identically distributed random variables rather than the almost surely equal random variables. Unfortunately, this leads to a logical impasse. Seeking a contradiction, suppose that we set 
\begin{equation}
\tilde{\psi} ([X]_d,[Y]_d) = \exy , \label{interdit}
\end{equation}
where $X,Y \in {\cal L}_1(\real)$ are identically distributed without being almost surely equal. We have 
$[X]_d = [Y]_d$ and $\exy > 0$ (since $\exy = 0 \Leftrightarrow X\stackrel{a.s.}{=} Y$), and we end up with the following contradiction: \newline
$\tilde{\psi} ([X]_d,[X]_d) \stackrel{(\ref{interdit})}{=} E|X-X|=0 < \exy \stackrel{(\ref{interdit})}{=}
\tilde{\psi} ([X]_d,[Y]_d) = \tilde{\psi} ([X]_d,[X]_d)$, meaning that (\ref{interdit}) is ill-defined.

To convince ourselves that the above discussion is not in vain, take the case of the so-called Lukaszyk-Karmowski metric (Lukaszyk 2004)\nocite{Lukaszyk2004} which is actually the functional 
$\tilde{\psi} ([\cdot]_d,[\cdot]_d) = E|\cdot - \cdot|$ set in (\ref{interdit}). It is only when this author asserts that the identity of indiscernibles property is not realized by the metric $E|\cdot - \cdot|$ he uses that we end up understanding he implicitly identifies the identically distributed random variables of ${\cal L}_1(\real)$. In other words, he reasons as if (\ref{interdit}) were well-defined\footnote{
In later publications of Lukaszyk (and on Wikipedia, etc.), where $D(X,Y)$ refers to $\exy$, we find the notation $D(X,X) > 0$, which proves that the identically distributed variables of ${\cal L}_1(\real)$ are implicitly identified. This leads to the logical contradiction we have put forward.     }.
On this erroneous basis, Lukaszyk claims to have used a new operator which, as such, would deserve a special denomination. This does not make sense and the so-called Lukaszyk-Karmowski metric is none other than the good old $L_1$-norm distance. Note that this clarification does not greatly affect the merit of this author's 2004 article, where otherwise conclusive results in applied physics are presented. Lukaszyk correctly computes the distance $D(X,Y) = \exy$ between independent elements of ${\cal L}_1(\real)$ -- notably when $\xandy$ are Gaussian -- but he should not pretend that the reflexivity condition does not hold.
\section{Expected absolute difference of independent random variables} \label{indep}
\subsection{General considerations} \label{gencons}
Why is the notion of independence so important? This section contains a few remainders and general thoughts about independence. Two random variables are independent if they come from phenomena such that the result observed for one of them has no influence on the other. We should admit that the assumption of independence is often a question of intuition or common sense -- although in this respect caution is needed. When the hypothesis of independence can be made reasonably, great mathematical simplifications follow (resulting in particular from the use of the Fubini-Tonelli theorem).
 
Contrary to a common belief -- at least among researchers having somewhat forgotten the fundamentals of probability theory -- the notion of independence between two random variables $X$ and $Y$ does not imply that they ``have nothing to do with each other''. Indeed, independence only makes sense if these random variables are coupled, i.e. defined on the same probability space. They are coupled to each other, albeit in a particular way, by the fact that they are independent.

In applied sciences, two observations, phenomena or experiments can be perceived as independent. Independence may be imposed from the outside or may be organized in full awareness by the experimenter. In both cases, he or she will be interested in forming a probability model for a joint experiment such that the original two experiments are carried out independently.

Calculating a distance between independent random variables turns out to be often very useful. For example two measurement devices $X$ and $Y$ independently measure unknown quantities with some random error, or multiple researchers independently measure the same object and compare their results. To take an example related to economic inequalities, suppose that $X$ (resp. $ Y $) is the income of a household drawn at random from a statistical population ${\cal P} _1$ (resp. ${\cal P} _2$). Let $\muandnu$ denote the income distribution of $\xandy$, respectively. The random variables $\xandy$ are assumed to be independent, and we use this information to compute $\exy$, which is interpreted as a measure of income disparity between the two populations. In other words, $X=x$ and $Y=y$ are independently observed and $\exy$ is the weighted average of the $|x-y|$'s.
We have already mentioned that a metric such as $GK(X,Y)$, which depends only on the stand-alone distributions of $X$ and $Y$, cannot take account of the independence information.\footnote{We will see in Section~\ref{normalized} that in case of a single population, i.e. if ${\cal P}_1 = {\cal P} _2$ and $\mu = \nu$, and if the independent random variables $X$ and $Y$ are non-negative, then $\exy$ is none other than the Gini mean difference, a measure of income inequality within a population. The normalized Gini mean difference is the celebrated Gini index.} 
\subsection{Expected absolute difference in the context of almost sure equality, equality of distribution and independence} \label{exyas}
Consider $(X,Y) \in {\cal L}_1(\real^2)$, the space of all pairs of integrable real-valued random variables defined on a probability space $\probamodel$. It is well-known that $X\stackrel{a.s.}{=} Y$ ($X = Y$ almost surely) if and only if $\exy =0$. A pair $(X,Y) \in {\cal L}_1(\real^2)$ with $\exy =0$ is such that the probability mass $P_{(X,Y)}$ is concentrated on the diagonal $\Delta:=\{(x,x):x\in\real\}$ of $\real^2$, (that $P_{(X,Y)}(\Delta ) = 1$ is formalized in Subsection~\ref{diagcoupling}, Proposition~\ref{ascase}).

Here are some remarks about the values $\exy$ can take: if the distribution of the  random variables $X$ and $Y$ differ, then $X$ and $Y$ cannot be almost surely equal, and consequently $\exy > 0$. Moreover, the fact that $X$ and $Y$ have the same distribution by no means implies that $\exy =0$: suppose that $X$ and $Y$ are two continuous independent and identically distributed (i.i.d.) random variables. In this case $P({X\neq Y}) = 1$, i.e. $X \neq Y$ a.s., which implies that $\exy  > 0$. For example, take two i.i.d. random variables $\xandy$ having standard normal distribution. Then (\ref{gmdnorm}) -- a consequence of Theorem~\ref{distnorm1} -- implies that $\exy = 2/\sqrt{\pi}$. 

It is interesting to take a closer look at the behavior of $\exy$ when $X$ and $Y$ are independent.
\begin{prop} \label{equivalence} (Proof in the appendix)
Let $X, Y \in {\cal L}_1(\real)$ be  independent. The following statements are equivalent
\begin{description}
\item [{\rm a)}] $X\stackrel{a.s.}{=} Y$.
\item [{\rm b)}] $X$ and $Y$ are a.s. equal to a same constant, i.e. there exists $c\in \real$ such that $X\stackrel{a.s.}{=} Y\stackrel{a.s.}{=}c$.
\end{description}
\end{prop} 	
A random variable is said to be degenerate if it is almost surely constant. So, if $\xandy$ are independent, then $\exy = 0$ occurs if and only if they are (identically distributed and) degenerate, i.e. if and only if their distribution in concentrated on the same constant.

\begin{figure}[t]
  \centering
  \includegraphics[width=6.5cm,height=4cm]{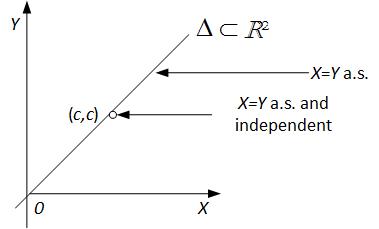}
  \caption{\small Let $(X,Y)$ be a random pair taking values in $\real^2$ (or in a subset of $\real^2$). If $X\stackrel{a.s.}{=} Y$, then the distribution of $(X,Y)$ concentrates on $\Delta$. If in addition we impose independence, then the distribution concentrates on a single point $(c,c)\in \Delta$.}
  \label{fig:D}
\end{figure}

The above considerations can be summarized as follows: \newline
\[
\exy = 0 \Rightarrow \left\{ \begin{array}{l}
                (X,Y) \in \Delta \hbox{ with probability 1 (dependent case)}   \\
                (X,Y) = (c,c) \in \Delta \hbox{ with probability 1 (independent case),}  \\
                       \end{array} 
               \right.
\]
a situation illustrated in Figure~\ref{fig:D}.
\subsection{Partitioning ${\cal L}_1(\real^2)$ into six categories} \label{sixcat} 
\begin{figure}[t]
  \centering
  \includegraphics[width=7cm,height=7cm]{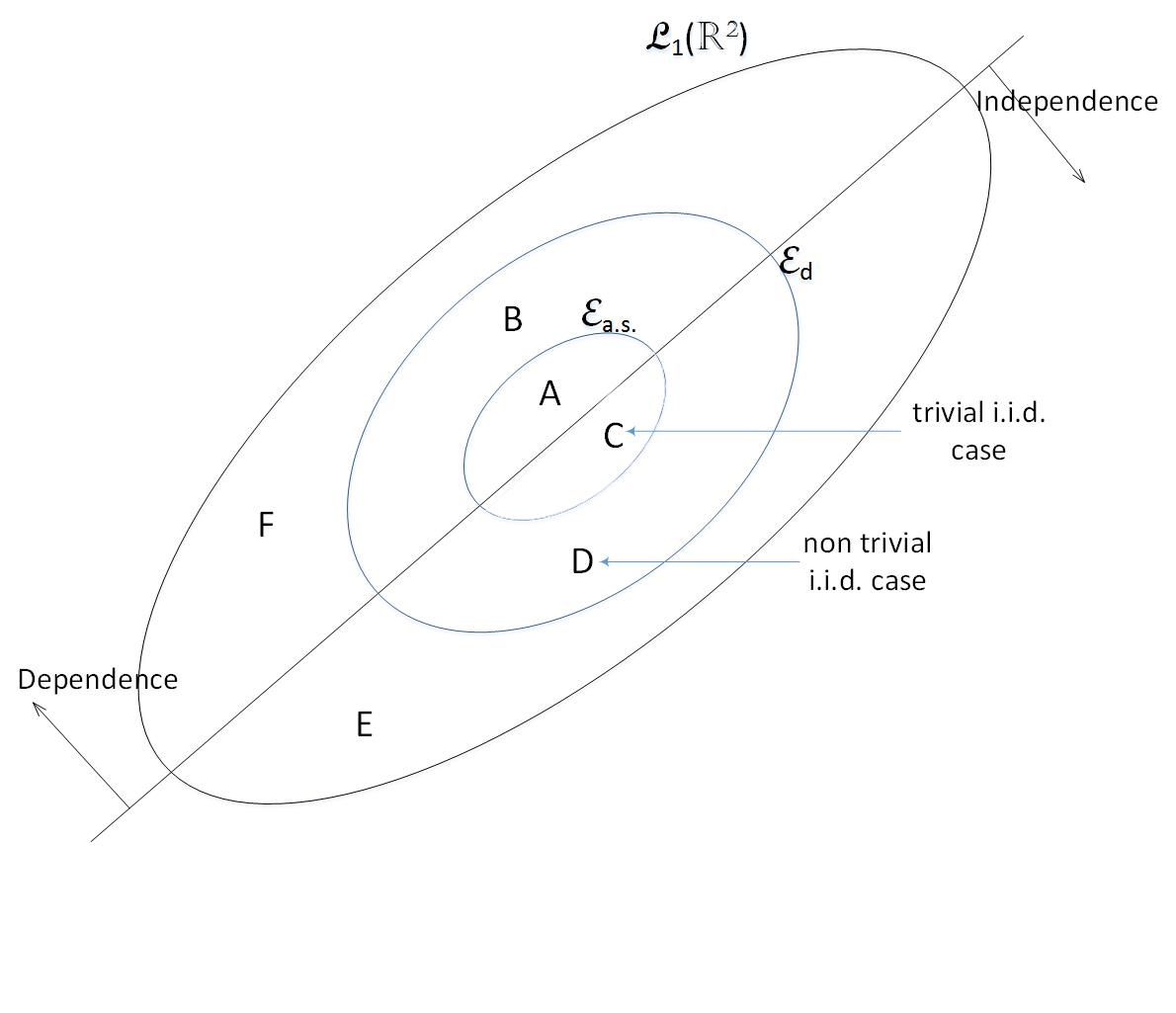}
  \caption{\small The pairs $(X,Y)$ fall into six mutually exclusive categories A, B, C, D, E and F. The Venn diagram represents a partition \{A,B,C,D,E,F\} of ${\cal L}_1(\real^2 )$.}
  \label{fig:abcdf}
\end{figure}

We are interested in random pairs $(X,Y) \in {\cal L}_1(\real^2 )$.
In order to bring together and clarify the concepts encountered in Section~\ref{exyas}, define \newline
${\cal E}_{a.s} = \{ (X,Y) \in {\cal L}_1(\real^2 ) : X \stackrel{a.s.}{=} Y  \}$ and 
${\cal E}_{d} = \{ (X,Y) \in {\cal L}_1(\real^2 ) : X\stackrel{d}{=} Y  \}$. That is, the subsets ${\cal E}_{a.s}$ and ${\cal E}_{d} $ of ${\cal L}_1(\real^2 )$ contain the random pairs whose components are equivalent in the almost sure sense and in the equality of distribution sense, respectively. Taking into account a possible independence between $\xandy$, the  pairs $(X,Y) \in {\cal L}_1(\real^2 )$ fall into six mutually exclusive categories described in Table~\ref{classification}.
\begin{table}[t]
	\centering
		\begin{tabular}{|l|cccc|} \hline
    category $\downarrow$    & independence & a.s. equality   & equality of distribution & $\exy$ \\ \hline
		A                & no             & yes           & yes & $ = 0$ \\ 				
    B                & no             & no            & yes & $ > 0$ \\ 
    C (trivial case) & yes            & yes           & yes & $ = 0$ \\
		D (i.i.d. case)  & yes            & no            & yes & $ > 0$ \\ 				
    E                & yes            & no            & no & $ > 0$ \\ 
    F                & no             & no            & no & $ > 0$ \\ \hline
    \end{tabular}
	\caption{\small Partition of  ${\cal L}_1(\real^2 )$ into six categories according to dependence structure and type of equivalence relation.} 
	\label{classification}
\end{table}
Figure~\ref{fig:abcdf} depicts the six categories of Table~\ref{classification} as a partition of ${\cal L}_1(\real^2 )$.
\begin{table}[h]
	\centering
   \begin{tabular}{ccc}
	   \begin{tabular}{|c|cc|c|} \hline 
                               \textbf{(A)} & 0                 & 1                 & $P_X$  \\ \hline
              0                             &  \emph{0.3}       & \emph{0}          & \textbf{0.3} \\ 
              1                             &  \emph{0}         & \emph{0.7}        & \textbf{0.7} \\ \hline
             $P_Y   $                       & \textbf{0.3}      & \textbf{0.7}      & \\ \hline
   \end{tabular}& 	   \begin{tabular}{|c|cc|c|} \hline
                               \textbf{(B)} & 0                 & 1                 & $P_X$  \\ \hline
              0                             &  \emph{0.1}       & \emph{0.2 }       & \textbf{0.3} \\ 
              1                             &  \emph{0.2}       & \emph{0.5 }       & \textbf{0.7} \\ \hline
               $P_Y$                        & \textbf{0.3}      & \textbf{0.7}      & \\ \hline
   \end{tabular}  & 	   \begin{tabular}{|c|cc|c|} \hline
                               \textbf{(C)} & 0                 & 1                 & $P_X$  \\ \hline
              0                             &  \emph{1}         & \emph{0}          & \textbf{1} \\ 
              1                             &  \emph{0}         & \emph{0}          & \textbf{0} \\ \hline
               $P_Y$                        &  \textbf{1}       & \textbf{0}        & \\ \hline
   \end{tabular}\\ \\
	
	   \begin{tabular}{|c|cc|c|} \hline
                               \textbf{(D)} & 0                 & 1                 & $P_X$  \\ \hline
              0                             &  \emph{0.09}      & \emph{0.21}       & \textbf{0.3} \\ 
              1                             &  \emph{0.21}      & \emph{0.49}       & \textbf{0.7} \\ \hline
              $P_Y $                        & \textbf{0.3}      & \textbf{0.7}      &  \\ \hline
   \end{tabular}& 	   \begin{tabular}{|c|cc|c|} \hline
                               \textbf{(E)} & 0        & 1      & $P_X$  \\ \hline
              0                             &  \emph{0.06}      & \emph{0.14}       & \textbf{0.2} \\ 
              1                             &  \emph{0.24}      & \emph{0.56}       & \textbf{0.8} \\ \hline
               $P_Y$                        & \textbf{0.3}      & \textbf{0.7}      &  \\ \hline
   \end{tabular}  & 	   \begin{tabular}{|c|cc|c|} \hline
                               \textbf{(F)} & 0                 & 1                 & $P_X$  \\ \hline
              0                             &  \emph{0.3}       & \emph{0.1}        & \textbf{0.4}\\ 
              1                             &  \emph{0.4}       & \emph{0.2}        & \textbf{0.6} \\ \hline
               $P_Y$                        &  \textbf{0.7}     & \textbf{0.3}      &  \\ \hline
   \end{tabular}\\
   \end{tabular}
		\caption{\small Example of six bivariate distributions falling into the six categories defined in Table~\ref{classification} and represented in Figure~\ref{fig:abcdf}.} 
	\label{sixdistrib}
	\end{table}
\newpage
\begin{example}  
\vskip .5cm
\emph{Table~\ref{sixdistrib} }displays the distributions of six pairs $(X,Y)$ of (binary) $\{0,1\}$-valued random variables. These bivariate distributions correspond to the categories defined in \emph{Table~\ref{classification}} and represented in \emph{Figure~\ref{fig:abcdf}}.
\end{example}
\subsection{Independence and entropy} \label{indentrop}
It would be incomplete to talk about independence without giving a brief overview of the concept of entropy, whose definition comes from the pioneers of information theory. The entropy of a distribution was established by Shannon for discrete laws and then extended to continuous laws characterized by their density. The notion of entropy has been generalized in many ways and has undergone vast and important developments. 
The real-valued random variables $X$ and $Y$ defined on $\probamodel$, where $\Omega$ is a finite set, are independent when the degree of uncertainty is maximal in their joint distribution. This means that the entropy of the bivariate distribution must be maximal.

Entropy is a measure of uncertainty or randomness. The following equation holds: \newline 
$H(X,Y) = H(X) + H(Y) - I(X,Y) $, where  $H(X,Y)$ is the (joint) entropy of the pair $(X,Y)$, $H(X)$ and $H(Y)$ being the entropy of $X$ and $Y$, respectively. The non-negative $I(X,Y)$ is the so-called \emph{mutual information}, and one has $I(X,Y) = 0$ if and only if $X$ and $Y$ are independent, in which case $H(X,Y)$ is maximal. In the finite case that we are dealing with $H(X) = \sum_i p_i \ln(1/p_i)$, $H(Y) = \sum_j p_j \ln(1/p_j)$ and $H(X,Y) = \sum_i \sum_j p_{ij} \ln(1/p_{ij})$, where $p_i = P(X=x_i)$, $p_j = P(Y=y_j)$ and $p_{ij} = P(X=x_i, Y=y_j)$.

Example~\ref{binary} below illustrates that the distance $\exy$ between dependent $X$ and $Y$  can be both smaller or larger than it is when $X$ and $Y$ are independent.
\newpage
\begin{example} \label{binary} \footnote{Inspired from Prof. Dr. Svetlozar Rachev's online lecture on probability metrics (summer semester 2008), Institute for Statistics and Mathematical Economics, University of Karlsruhe.}
\emph{Table~\ref{minmax}} shows the joint distributions of three pairs $(X,Y)$ of binary $\{0,1\}$-valued random variables.
In the three cases, the marginal distributions are the same. \newline
Case \emph{(a)} refers to a distribution of dependent variables yielding minimal $\exy$. \newline 
Case \emph{(b)} refers to independent random variables. \newline
Case \emph{(c)} displays a distribution of dependent variables yielding maximal $\exy$.
\vskip .1cm
The expected absolute difference, the joint entropy and the Gini-Kantorovich distance for the three distributions are summarized in \emph{Table~\ref{entropy}}. We realize that, unlike entropy, $\exy$ does not culminate with independence.
\end{example}
\begin{table}[t] 
	\centering
   \begin{tabular}{ccc}
   \begin{tabular}{|c|cc|c|} \hline
                     \textbf{(a)}& 0  & 1 & $P_X$\\ \hline
                   0 & 1/4& 1/4& \textbf{1/2} \\
                   1 & 0  & 1/2 & \textbf{1/2}\\ \hline
									 $P_Y$ & \textbf{1/4}  & \textbf{3/4} &  \\ \hline
   \end{tabular}  &  \begin{tabular}{|c|cc|c|} \hline
                    \textbf{(b)} & 0  & 1 & $P_X$\\ \hline
               0 & 1/8& 3/8& \textbf{1/2} \\
                   1 & 1/8  & 3/8 & \textbf{1/2}\\ \hline
									 $P_Y$ & \textbf{1/4}  & \textbf{3/4} &  \\ \hline
                      \end{tabular}
                  & \begin{tabular}{|c|cc|c|} \hline
                     \textbf{(c)}& 0  & 1 & $P_X$\\ \hline
                   0 & 0  & 1/2& \textbf{1/2} \\
                   1 & 1/4  & 1/4 & \textbf{1/2}\\ \hline
									 $P_Y$ & \textbf{1/4}  & \textbf{3/4} &  \\ \hline
                      \end{tabular}
   \end{tabular} 
	\caption{\small Three distributions of random pairs $(X,Y)$ of binary random variables $\xandy$. From left to right, $\xandy$ are: (a) dependent with minimal $\exy$, (b) independent, (c) dependent with maximal $\exy$  .} 
	\label{minmax}
\end{table}
\begin{table}[t]
	\centering
		\begin{tabular}{|l|ccc|} \hline
              & (a)           & (b)          & (c)  \\ \hline
		$\exy$    & \emph{0.250}         & \emph{0.500}        & \emph{0.750} \\				
   $H(X,Y)$   & \emph{1.040}         & \emph{1.255}        & \emph{1.040} \\ 
   $GK(X,Y)$  & \emph{0.250}         & \emph{0.250}        & \emph{0.250} \\ \hline
    \end{tabular}
	\caption{\small Expected absolute difference, joint entropy and Gini-Kantorovich distance for three pairs of variables. (a) $X$ and $Y$ dependent with minimal $\exy$ . (b) $X$ and $Y$ independent. (c) $X$ and $Y$ dependent with maximal $\exy$ . As it ignores the dependence structure, $GK(X,Y)$ is invariant in the three cases.} 
	\label{entropy}
\end{table}
\section{Normalized expected absolute difference} \label{normalized}
This section focuses on the normalized form of the expected absolute difference and its characteristics. In particular, we address the  issue of triangle inequality in relation to this functional. In an interesting paper, Yianilos (2002)\nocite{Yianilos2002} shows that symmetric set difference and Euclidean distance on $\real^d$ have normalized forms that remain metrics. We examine in this section the conditions under which $\exy$ can be $[0,1]$-normalized. The additional difficulty here is that we are not in a deterministic context. 
With our usual notation, ${\cal L}_1(\real)$ is the vector space of all integrable random variables on $\probamodel$ taking values in $(\real, {\cal B}_1)$, and the random variables used below belong to this set. 

It is natural to consider forming relative distance measures. Converting $D(X,Y) := \exy$ to a normalized (or standardized) form may be very useful in the solution of certain problems, especially when relative errors are at stake, as is often the case in numerical analysis. Define a normalized counterpart $D_{norm}(\cdot,\cdot )$ of $D(\cdot,\cdot )$ by
\begin{equation} \label{drelxy}
D_{norm}(X,Y) = \left\{ 
	\begin{array}{ll}
	\frac{\exy}{E|X| + E|Y|} & if \ \  E|X|> 0 \ \  or \ \  E|Y| > 0  \\
	\ \ 0 & \ \ otherwise, \\
	\end{array}   \right.
\end{equation}
so that $0\le D_{norm}(X,Y) \le 1$. The upper bound is reached when, say, $Y\stackrel{a.s.}{=}0$ with $E(|X|) > 0$, while the lower bound is reached when $X$ and $Y$ are almost surely equal and $E(|X|)$ $(= E(|Y|))$ is strictly positive. It is not our intention to comment here on the pros and cons of using a normalized distance measure. We will simply check whether the generally accepted axioms for a distance are verified. What we can say though is that $D_{norm}$ is likely to share the strengths and weaknesses of relative deterministic measures such as the Canberra metric (Lance and Williams (1967))\nocite{Lance1967}. Clearly, $D_{norm}$ is a \emph{distance} since it satisfies nonnegativity, reflexivity ($D_{norm}(X,X) = 0$) and symmetry ($D_{norm}(X,Y) = D_{norm}(Y,X)$). 
Let $(X_i)_{i \in I} \subset {\cal L}_1(\real)$ denote a finite family of independent random variables. We show in Subsection~\ref{ineqtri} that the triangle inequality holds in case of independence, but does not hold in general. More precisely, while $({\cal L}_1(\real),D_{norm})$ is a distance space, we show that $((X_i)_{i \in I},D_{norm})$ is a semimetric space (i.e. $D_{norm}$ is a distance satisfying the triangle inequality).
\subsection{The triangle inequality issue} \label{ineqtri}
\subsubsection{The independent case} \label{trieqind}
A corollary of Theorem~\ref{ineqtri1} below is that $D_{norm}(\cdot  ,\cdot )$ defined in (\ref{drelxy}) satisfies the triangle inequality in the independence case.
\begin{theor} \label{ineqtri1}
(Proof in the appendix) Let $X,Y,Z \in {\cal L}_1(\real)$ be three (mutually) independent random variables and assume that at most one of these variables is almost surely zero. Then 
$E|X| + E|Z| > 0 $, $E|X| + E|Y| > 0 $ and $E|Y| + E|Z| > 0 $,  and the following property is realized
\begin{equation}
\frac{E|X-Z|}{E|X| + E|Z|} \le \frac{\exy}{E|X| + E|Y|} + \frac{E|Y-Z|}{E|Y| + E|Z|} .   \label{ineqtri2}
\end{equation}
\end{theor}
Proving this inequality was a particularly thorny exercise (see the proof in the appendix). Using (\ref{ineqtri2}) and the definition of $D_{norm}$ in (\ref{drelxy}), one can easily check that
\begin{equation}
D_{norm}(X,Z) \le D_{norm}(X,Y) + D_{norm}(Y,Z)   \label{ineqtri3}
\end{equation}
for mutually independent $X,Y,Z$. Note that independence of these variables implies that the three pairs $(X,Z)$, $(X,Y)$ and $(Y,Z)$ intervening in the triangle inequality are the two-dimensional projections of the three-dimensional random vector $(X,Y,Z)$ having the product distribution 
$P_X \otimes P_Y \otimes P_Z$. In other words, the three pairs can be consistently embedded in a three-dimensional vector so that the triangle inequality makes sense (more details on this are given below). 
\subsubsection{The general case} \label{trieqgen}
A question naturally arises: would the triangle inequality (\ref{ineqtri2}) hold in all cases if the assumption of independence were lifted in Theorem~\ref{ineqtri1}~? The answer is negative: with a little of craftmanship, one can find counterexamples such as the one resulting from the three bivariate distributions \textbf{(A)}, \textbf{(B)}, \textbf{(C)} of pairs of dependent random variables shown in Table~\ref{pxpypz1}.
\begin{table}[t]
	\centering
   \begin{tabular}{ccc}
	   \begin{tabular}{|c|cc|c|} \hline 
                               \textbf{(A)} & -1                & 1                 & $P_X$  \\ \hline
              -1                            &  \emph{0.1}       & \emph{0.3}        & \textbf{0.4} \\ 
              1                             &  \emph{0.6}       & \emph{0}        & \textbf{0.6} \\ \hline
             $P_Z   $                       & \textbf{0.7}      & \textbf{0.3}      & \\ \hline
   \end{tabular}& 	   \begin{tabular}{|c|cc|c|} \hline
                               \textbf{(B)} & -1                & 1                 & $P_X$  \\ \hline
              -1                            &  \emph{0.3}       & \emph{0.1 }       & \textbf{0.4} \\ 
              1                             &  \emph{0}         & \emph{0.6 }       & \textbf{0.6} \\ \hline
               $P_Y$                        & \textbf{0.3}      & \textbf{0.7}      & \\ \hline
   \end{tabular}  & 	   \begin{tabular}{|c|cc|c|} \hline
                               \textbf{(C)} & -1                & 1                 & $P_Z$  \\ \hline
              -1                            &  \emph{0.11}       & \emph{0.59}        & \textbf{0.7} \\ 
              1                             &  \emph{0.19}         & \emph{0.11}        & \textbf{0.3} \\ \hline
               $P_Y$                        &  \textbf{0.3}     & \textbf{0.7}      & \\ \hline
   \end{tabular}\\ \\
	\begin{tabular}{|c|cc|c|} \hline 
                               \textbf{(D)} & -1                & 1                 & $P_X$  \\ \hline
              -1                            &  \emph{0.2}       & \emph{0.2}        & \textbf{0.4} \\ 
              1                             &  \emph{0.5}       & \emph{0.1}        & \textbf{0.6} \\ \hline
             $P_Z   $                       & \textbf{0.7}      & \textbf{0.3}      & \\ \hline
   \end{tabular}& 	   \begin{tabular}{|c|cc|c|} \hline
                               \textbf{(E)} & -1                & 1                 & $P_X$  \\ \hline
              -1                            &  \emph{0.3}       & \emph{0.1 }       & \textbf{0.4} \\ 
              1                             &  \emph{0}         & \emph{0.6 }       & \textbf{0.6} \\ \hline
               $P_Y$                        & \textbf{0.3}      & \textbf{0.7}      & \\ \hline
   \end{tabular}  & 	   \begin{tabular}{|c|cc|c|} \hline
                               \textbf{(F)} & -1                & 1                 & $P_Z$  \\ \hline
              -1                            &  \emph{0.3}       & \emph{0.4}        & \textbf{0.7} \\ 
              1                             &  \emph{0}         & \emph{0.3}        & \textbf{0.3} \\ \hline
               $P_Y$                        &  \textbf{0.3}     & \textbf{0.7}      & \\ \hline
   \end{tabular}\\ \\
\end{tabular}
		\caption{\small Two examples of bivariate distributions where the normalized expected absolute difference does not satisfy the triangle inequality. The first example (distributions \textbf{(A)}, \textbf{(B)}, \textbf{(C)}) is valid, for these distributions abide by the consistency rule. The second example (distributions \textbf{(D)}, \textbf{(E)}, \textbf{(F)}) is invalid because these three distributions infringe the consistency rule.    } 
	\label{pxpypz1}
\end{table}
It turns out that $E|X| = E|Y| = E|Z| = 1$ and
\begin{equation}
\frac{\exy}{2} + \frac{E|Y-Z|}{2} - \frac{E|X-Z|}{2} = -0.02,   \label{contrex}
\end{equation}
which contradicts (\ref{ineqtri2}). The statement that the triangle inequality should hold for any $X,Y,Z$ is actually pretty vague. As $\exy$ is a \emph{compound probability metric} (see Definition~\ref{prisico}), the choice of the three pairs $(X,Z)$, $(X,Y)$ and $(Y,Z)$ cannot be totally free. Indeed, suppose we fix the distributions \textbf{(A)} and \textbf{(B)} in Table~\ref{pxpypz1}. Then the choice of distribution \textbf{(C)} cannot be arbitrary, because the (internal) dependence structure of $(Y,Z)$ depends on the dependence structures of $(X,Y)$ and $(X,Z)$. Rachev et al. (2007, p. 93)
\nocite{Rachev2007} give the following consistency rule: ''The three pairs of random variables $(X,Z)$, $(X,Y)$ and $(Y,Z)$ should be chosen in such a way that there exists a consistent three-dimensional random vector $(X,Y,Z)$ and the three pairs are its two-dimensional projections.'' In other words, if this rule is respected, then the three pairs can be safely embedded in a three-dimensional random vector. To validate our counterexample, we must make sure that the distributions \textbf{(A)}, \textbf{(B)}, \textbf{(C)} abide by the consistency rule. This is indeed the case, for the matrix 
\begin{equation} \label{covmat}
V=\left( 
\begin{array}{ccc}        
var(X)&cov(X,Y)&cov(X,Z) \\
cov(X,Y)&var(Y)&cov(Y,Z) \\
cov(X,Z)&cov(Y,Z)&var(Z)
\end{array}
\right)
\end{equation}
is positive definite (with eigenvalues $\lambda_1 = 0.072$, $\lambda_2 = 0.44$ and $\lambda_3 = 2.128$), which means that $V$ is a valid covariance matrix. 
\subsubsection{An illegitimate counterexample} \label{illegitimate}
For completeness, we give what in appearance only is a counterexample. The distributions \textbf{(D)}, \textbf{(E)}, \textbf{(F)} in Table~\ref{pxpypz1} are such that the triangle inequality does not hold because \newline
$(\exy + E|Y-Z| - E|X-Z|)/2 = -0.2$. However, \textbf{(D)}, \textbf{(E)}, \textbf{(F)} do not constitute a valid counterexample, because the matrix $V$ in (\ref{covmat}) resulting from these distributions is indefinite (with eigenvalues  $\lambda_1 = -0.076$, $\lambda_2 = 1.093$ and $\lambda_3 = 1.623$), and thus simply cannot be a covariance matrix of a three-dimensional vector. 

We end this subsection with an example illustrating the importance of the independence assumption in Theorem~\ref{ineqtri1}. In Table~\ref{pxpypz2}, $(X,Z)$, $(X,Y)$ and $(Y,Z)$ are now three pairs of independent variables with the same marginal distributions as those shown in Table~\ref{pxpypz1}. 
\begin{table}[t]
	\centering
   \begin{tabular}{ccc}
	   \begin{tabular}{|c|cc|c|} \hline 
                               \textbf{(G)} & -1                 & 1                 & $P_X$  \\ \hline
              -1                            &  \emph{0.28}       & \emph{0.12}       & \textbf{0.4} \\ 
              1                             &  \emph{0.42}       & \emph{0.18}       & \textbf{0.6} \\ \hline
             $P_Z   $                       & \textbf{0.7}       & \textbf{0.3}      & \\ \hline
   \end{tabular}& 	   \begin{tabular}{|c|cc|c|} \hline
                               \textbf{(H)} & -1                 & 1                 & $P_X$  \\ \hline
              -1                            &  \emph{0.12}       & \emph{0.28 }      & \textbf{0.4} \\ 
              1                             &  \emph{0.18}       & \emph{0.42 }      & \textbf{0.6} \\ \hline
               $P_Y$                        & \textbf{0.3}       & \textbf{0.7}      & \\ \hline
   \end{tabular}  & 	   \begin{tabular}{|c|cc|c|} \hline
                               \textbf{(I)} & -1                 & 1                 & $P_Z$  \\ \hline
              -1                            &  \emph{0.21}       & \emph{0.49}       & \textbf{0.7} \\ 
              1                             &  \emph{0.09}       & \emph{0.21}       & \textbf{0.3} \\ \hline
               $P_Y$                        &  \textbf{0.3}      & \textbf{0.7}      & \\ \hline
   \end{tabular}\\ \\
\end{tabular}
		\caption{\small Example of three bivariate distributions of pairs of independent random variables. They comply with the consistency rule and the triangle inequality holds.} 
	\label{pxpypz2}
	\end{table}
From the distributions \textbf{(G)}, \textbf{(H)}, \textbf{(I)}, we find that 
$(\exy + E|Y-Z| - E|X-Z|)/2 = 0.5$ (instead of -0.02 in (\ref{contrex})), which means that the triangle inequality holds. Note also that $V$ is the three-dimensional diagonal matrix \emph{diag}(0.96,0.84,0.84), which is of course a valid covariance matrix.
\subsection{Link with the Gini index} \label{gindex}
The computation of a distance between identically distributed random variables of ${\cal L}_1(\real)$ is helpful in various domains. Examples are the Gini mean difference (GMD)\footnote{The GMD is twice the $L$-scale (the second $L$-moment). It is sometimes considered as a competitor of the standard deviation. } and the Gini index, used in particular in inequality economics to measure the amount of inequality included in a distribution of income (alternatively consumption or wealth, etc.). Let $\mu$ be such a distribution. The GMD and the Gini index are defined as GMD$(\mu) = \exy$ ($= D(X,Y))$ and $Gini(\mu) = \exy/[2E(X)]$ ($= D_{norm}(X,Y)$), where $X \sim \mu$ and $Y \sim \mu$ are assumed to be independent and (usually) non-negative (see e.g. Yitzhaki (1998)\nocite{Yitzhaki1998}, Yitzhaki and Schechtman (2013)\nocite{YitzhakiSchechtman2013}, or Xu (2003)\nocite{Xu2003})\footnote{When the range of $X$ encroaches on $(-\infty ,0]$, we know from (\ref{drelxy}) that there is no mathematical reason why we should not define $Gini(\mu) = \exy/[2E|X|]$.}. Looking at (\ref{drelxy}), we can say that the Gini index is the distance (semimetric) $D_{norm}$ from $X$ to an i.i.d. ``copy'' of itself. In that sense, it is sometimes called an ``autodistance'' in the literature. However, a copy must be clearly distinguished from the original and there is some confusion on this point. ``Copy'' is to be understood here in the equality of distribution sense (in the almost sure equality sense, the distance is trivially zero). 

Independence of $\xandy$ allows in many cases to use Fubini-Tonelli to represent the GMD and the Gini index in closed form. Independence also implies that, except in the degenerate case where  
$X\stackrel{a.s.}{=} Y\stackrel{a.s.}{=}c$ for some $c \in \real_+$,
$GMD(\mu) > 0$. Although seemingly simple, the Gini index is actually a quite proteiform measure of inequality. It can be expressed in an astonishing number of ways, some of which can be found in Yitzhaky (1998)\nocite{Yitzhaki1998}. 

We end this subsection by probabilistic considerations on the values the Gini index can take. Let $X\sim \mu$ and $Y\sim \mu$ be two i.i.d. non-negative random variables where $\mu$ is (say) the income distribution of a population. Then 
$Gini(\mu) = \exy/[2E(X)]$ is not defined if and only if $X\stackrel{a.s.}{=}0$. Leaving this uninteresting case aside,
$Gini(\mu)= 0 \Leftrightarrow \exy = 0 \stackrel{Prop.\ref{equivalence}}{\Longleftrightarrow} \exists c > 0$ such that $X\stackrel{a.s.}{=} Y\stackrel{a.s.}{=}c$. Moreover, the triangle inequality $\exy \leq |X| + |Y|$ implies that $Gini(\mu) \in [0,1]$.

In economic applications, non-negative real numbers $x_1,x_2,\ldots ,x_n$ are typically incomes earned respectively by individuals $i_1,i_2,\ldots , i_n$ belonging to a population or a statistical sample.
It is known that $0\le g \leq (n-1)/n$, where $g$ is the Gini index. A low value of $g$ corresponds to a more equal income distribution, with 0 indicating perfect equality (all individuals have the same income). The most unequal society is the one where a single individual receives all the income and the remaining individuals receive nothing. In that case, $g = (n-1)/n$. Generalizing a bit, we can say that a tiny proportion $\epsilon > 0$ of the population receives an income $b>0$, while a proportion $1 - \epsilon$ gets nothing. Such an extreme case can be described thanks to i.i.d. random variables $\xandy$ having distribution $\mu = (1 - \epsilon) \delta_0 + \epsilon \delta_b$ where $b$ is a strictly positive income level and where $\epsilon > 0$, $\epsilon$ small, is the probability that $X$ (or $Y$) takes the value $b$. The distribution of the pair $(X,Y)$ is summarized in Table~\ref{epsilon}. 
\begin{table}[t]
	\centering
\begin{tabular}{|c|cc|c|} \hline 
                                & 0                     & $b$                     & $P_X$  \\ \hline
              0                 &  $(1-\epsilon )^2$    & $\epsilon (1-\epsilon )$& $1-\epsilon$ \\ 
              $b$               &$\epsilon (1-\epsilon)$& $\epsilon^2$            & $\epsilon$ \\ \hline
             $P_Y$              & $1-\epsilon$          & $\epsilon$              & \\ \hline
\end{tabular}
\caption{\small Distribution of the pair $(X,Y)$ of independent random variables having marginal distribution $\mu = (1 - \epsilon) \delta_0 + \epsilon \delta_b$. The probability $\epsilon > 0$ is small and $b$ is a strictly positive real number. This distribution describes extreme inequality where the Gini index nears 1.} 
	\label{epsilon}
	\end{table}
The Gini index then becomes $Gini(\mu) = \exy/[2E(X)] = 1 - \epsilon$. The upper bound 1 corresponding to $\epsilon = 0$ cannot be reached, because in such a case $X$ would take the value 0 with probability 1 , and the Gini index would not be defined ($E(X)$ would be zero).
\section{Alternative interpretation of the ${\cal L}_p$-distance} \label{interpretation}
For $p \ge 1$, let ${\cal L}_p (\real)$ denote the class of all real-valued random variables on a probability space $\probamodel$ that have finite $p$-th moment, and consider $X,Y\in {\cal L}_p (\real)$. Suppose that $\xandy$ are identically distributed. In this section, we show in particular that the ${\cal L}_p$-distance $(E|X -Y|^p)^{1/p}$ certainly represents a distance between $\xandy$, but also -- in a sense to be specified -- tells us how far from almost sure equality these variables are. In a symbolic way, we will show that $(E|X -Y|^p)^{1/p}$  can be conceived as a distance 
$Dist(X \stackrel{d}{=} Y,X \stackrel{a.s.}{=} Y)$ between two possible states of the pair $(X,Y)$. In \ref{diagcoupling}, we introduce the so-called \emph{diagonal coupling} of a probability measure with itself, and we clarify its connection to almost sure equality. In \ref{primary}, we define a distance\footnote{in fact a semimetric, i.e. a distance satisfying the triangle inequality.} $\eta_p(\pi_1, \pi_2)$ between bidimensional probability measures $\pi_1$ and $\pi_2$ having finite $p$-th moment. In \ref{etapcouplings}, we recall the notion of coupling and observe what becomes of $\eta_p$ when $\pi_1$ and  $\pi_2$ are couplings of unidimensional probability measures $\mu$ and $\nu$ of finite $p$-th moment. Equating $\mu$ and $\nu$ and taking $\pi_2$ as the diagonal coupling of $\mu$ with itself allows then to interpret $\eta_p$ as a distance indicating how far $X\sim \mu$ and $Y\sim \mu$ are from almost sure equality. The results are illustrated with the bivariate normal distribution. Finally, thanks to $\eta_p$, we propose a probabilistic representation of the Gini mean difference and of the Gini index.
\subsection{Diagonal coupling and almost sure equality} \label{diagcoupling}
Consider a probability space $(\real, {\cal B}_1,\mu)$. We denote by $\mu \triangle\mu$ a probability measure  on the product space $(\real^2, {\cal B}_2)$ defined by
\begin{equation}
(\mu \triangle\mu)(B_1\times B_2) = \mu(B_1\cap B_2) \ \ \ \ \forall B_1, B_2 \in {\cal B}_1. \label{diagmeasure}
\end{equation}
Then $\mu \triangle\mu$ can be extended to the whole ${\cal B}_2$ by using Carath\'eodory's theorem (Kalikow and McCutchean (2010) \nocite{Kalikow2010}).
\begin{defi} \label{diagcoupl}
   $\mu \triangle\mu$, as defined above, is called the \emph{diagonal coupling} of $\mu$ with itself.
\end{defi}
Actually, $(\real, {\cal B}_1,\mu )$ and $(\real^2, {\cal B}_2,\mu \triangle \mu )$ are closely related: the map 
$s:\real \longrightarrow \real^2$ defined by $s(x) = (x,x)$ is a measurable isomorphism from $(\real, {\cal B}_1,\mu )$ to $(\real^2, {\cal B}_2,\mu \triangle\mu )$. An example of diagonal coupling is given in Table~\ref{sixdistrib}~\textbf{(A)}, where $\mu$ is defined by $\mu (0) = 0.3$ and $\mu (1) = 0.7$). Incidentally, 
Proposition~\ref{equivalence} tells us that if $X\stackrel{a.s.}{=} Y$, with $\xandy$ independent, then there exists $c \in \real$ such that $X\stackrel{a.s.}{=} c$ and $Y\stackrel{a.s.}{=} c$, which implies that
$\mu \otimes \mu = \mu \triangle \mu = \delta_{(c,c)}$ (the Dirac delta measure concentrated on $(c,c)$ ), where $\mu \otimes \mu$ is the product measure of $\mu$ with itself. So the probability measures $\mu \otimes \mu$ and $\mu \triangle \mu$ on $(\real^2,{\cal B}_2)$ coincide in this trivial case.

\begin{figure}[t]
  \centering
  \includegraphics{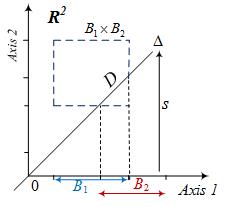}
  \caption{\small A more visual representation of the diagonal coupling of a probability measure with itself, where $D = (B_1 \times B_2)\cap \Delta$. We have $(\mu \triangle \mu)(B_1\times B_2) = \mu(s^{-1}(D)$.    }
  \label{fig:visual}
\end{figure}
Let $\Delta$ denote the diagonal $\{(x,x):x\in\real\}$ of $\real^2$. One can show quite easily that $B_1\cap B_2 = s^{-1}((B_1 \times B_2)\cap \Delta)$ for any subsets $B_1,B_2$ of $\real$. Therefore, the definition of the diagonal coupling in (\ref{diagmeasure}) may be replaced by
$(\mu \triangle\mu)(B_1\times B_2) = \mu(s^{-1}((B_1 \times B_2)\cap \Delta))$, which is visually more telling (see Figure~\ref{fig:visual}).
Let $X,Y:\probamodel \rightarrow (\real, {\cal B}_1)$ (e.g.  $X,Y \in {\cal L}_p (\real)$ if $\xandy$ have finite $p$-th moment) be two identically distributed random variables. Then the diagonal coupling of $P_X$ with itself is the distribution of the pair $(X,Y)$ if and only if $X\stackrel{a.s.}{=} Y$:\footnote{By definition $X\stackrel{a.s.}{=} Y$ makes sense only if $\xandy$ are defined on the same probability space.  } the entire probability mass of $P_{(X,Y)}$ is concentrated on the diagonal $\Delta$. This result is in line with our intuition. It is formally stated in Proposition~\ref{ascase}.
\begin{prop} \label{ascase} (Proof in the appendix)
Let $\xandy$ be two identically distributed random variables defined on $\probamodel$ taking values in 
$(\real, {\cal B}_1)$. Then
$X\stackrel{a.s.}{=} Y$ if and only if $P_{(X,Y)} =  P_X \triangle P_X$.
\end{prop} 
Lemma~\ref{samespace} below -- added for completeness -- is not directly related to what we need in this article. However, we would like to answer the following question: how can we construct a probability space on $\Delta$ consistent with $(\real^2, {\cal B}_2, \mu \triangle \mu)$? Two ways come to mind: (i) take the trace space of $(\real^2, {\cal B}_2, \mu \triangle \mu)$ with respect to $\Delta$, or (ii) take the pushforward space of $(\real, {\cal B}_1, \mu )$ induced by $s$ ($s$ defined above). It turns out that the two methods produce the same space, as evidenced by Lemma~\ref{samespace}.
\begin{lemme} \label{samespace}
(Proof in the appendix) Consider the trace probability space 
$(\Delta, {\cal B}_2 \cap \Delta, (\mu \triangle\mu)\left\lfloor_\Delta \right.)$ of 
$(\real^2, {\cal B}_2, \mu \triangle \mu)$, where $(\mu \triangle\mu)\left\lfloor_\Delta \right.$ denotes the probability measure $\mu \triangle\mu$ restricted to $\Delta$,  and the pushforward probability space 
$(\Delta, s({\cal B}_1), \mu_s)$
of $(\real, {\cal B}_1, \mu )$ induced by $s:\real \longrightarrow \real^2$ defined by $s(x) = (x,x)$. Then 
the trace space and the pushforward space coincide, i.e. \newline
\emph{(a)} ${\cal B}_2 \cap \Delta = s({\cal B}_1)$ (equality of $\sigma$-fields) \newline
\emph{(b)} $(\mu \triangle\mu)\left\lfloor_\Delta  \right. = \mu_s$ (equality of measures)\footnote{Incidentally,   $\mu_s$ is a so-called \emph{deterministic transport plan} in the Monge-Kantorovich transport problem. Denote by $Id$ the identity map on $\real$. Let $\pi_T := \mu_{(Id,T)}$ be the pushforward probability measure of $\mu$ induced by the function $(Id,T):\real \rightarrow \real^2$, where $T$ is a transport map. $\pi_T$ is called a \emph{deterministic transport plan}. In reference to the optimal transport problem, we have here $\mu_s = \mu_{(Id,T)}$ where $T = Id$, i.e. $\mu_s = \mu_{(Id,Id)}$.   }. 
\end{lemme} 
\subsection{A primary metric defined on distributions of pairs of random variables} \label{primary}
Let $(X_1,Y_1)$ (resp. $(X_2,Y_2)$) be two pairs of random variables having joint distribution $\pi_1$ (resp.     $\pi_2$) on $(\real^2, {\cal B}_2)$. What is meant by $\pi_1$ and $\pi_2$ being ``close to each other''? A possible answer -- serving what we wish to show in this subsection -- is to measure their proximity by using a \emph{primary metric}, i.e. to consider that $\pi_1$ and $\pi_2$ coincide when they share a given set of relevant characteristics. Accordingly, we consider here that the distance between $\pi_1$ and $\pi_2$ is zero if (i) the centers (mathematical expectations, means) of $(X_1,Y_1)$ and $(X_2,Y_2)$ are the same and (ii) the deviation between $X_1$ and $Y_1$ is the same as the deviation between $X_2$ and $Y_2$. 

For $(X_1,Y_1)\sim \pi_1$ and $(X_2,Y_2)\sim \pi_2$, assume that the marginals of $\pi_1$ and $\pi_2$ have finite $p$-th moment, $p\in [1,\infty )$. For $i=1,2$, we will use the following notations: \newline
$E|X_i -Y_i|^p = \int_{\real^2} |x_i -y_i|^p d\pi_i(x_i,y_i)$,
$EX_i = \int_{\real^2} x_i d\pi_i(x_i,y_i)$, $EY_i = \int_{\real^2} y_i d\pi_i(x_i,y_i)$ and
$C(\pi_i) = E([X_i, Y_i]) = [EX_i, EY_i]$. We can now define 
\begin{equation} 
\eta_p(\pi_1, \pi_2)  = ||C(\pi_1) - C(\pi_2)|| +
\left|   (E|X_1 -Y_1|^p)^{1/p} - (E|X_2 -Y_2|^p)^{1/p} \right|    \label{etapi1},
\end{equation}
where $||\cdot||$ is the Euclidean norm on $\real^2$. So $\eta_p$, which takes finite values, integrates two sources of deviation: \emph{between pairs} and \emph{within pairs} of random variables. Obviously, the characteristics entering the definition of $\eta_p$ in (\ref{etapi1}) do not fully describe what differentiates $\pi_1$ from $\pi_2$.
\begin{prop} \label{eta}
Let ${\cal P}(\real^2)$ denote the set of probability measures on $(\real^2,{\cal B}_2)$. For $p\in [1,\infty )$, let
${\cal P}_p(\real^2)$ denote the set of probability measures on the Borel subsets of $\real^2$ whose marginals have finite moment of order $p$, i.e. \newline
${\cal P}_p(\real^2) = \{\pi \in {\cal P}(\real^2): \int_{\real^2} |x|^p d\pi (x,y)< \infty \hbox{ and } \int_{\real^2} |y|^p d\pi (x,y)< \infty \}$. Then $\eta_p : {\cal P}_p(\real^2) \times {\cal P}_p(\real^2) \rightarrow \real_+$ defined in {\rm (\ref{etapi1})} is a semimetric, i.e. $\eta_p$ satisfies non-negativity, reflexivity \emph{(}$\eta_p(\pi, \pi) = 0$\emph{)}, symmetry 
\emph{(}$\eta_p(\pi_1, \pi_2) = \eta_p(\pi_2, \pi_1)$\emph{)} and triangle inequality \newline
\emph{(}$\eta_p(\pi_1, \pi_3) \le \eta_p(\pi_1, \pi_2) + \eta_p(\pi_2, \pi_3)$ for all $\pi_1,\pi_2,\pi_3 \in {\cal P}_p(\real^2)$.
\end{prop} 	
The (easy) proof of Proposition~\ref{eta} is omitted. As $\eta_p$ is in particular non-negative, reflexive and symmetric, it can be called a \emph{distance} between  elements of ${\cal P}_p(\real^2)$.\footnote{Formally, as already mentioned, the difference between a semimetric and a distance is the relaxation of the triangle inequality.} 
Moreover, $\eta_p$ is a semimetric without being a metric\footnote{As a semimetric, $\eta_p$ can be transformed into a metric between equivalence classes. Define an equivalence relation between the elements of ${\cal P}_p(\real^2)$ by $\pi_1 \sim \pi_2 :\Leftrightarrow \eta_p(\pi_1, \pi_2) = 0$. Then $\tilde{\eta_p}([\pi_1], [\pi_2]) := \eta_p(\pi_1, \pi_2)$ is a metric on the set 
$\left\{ [\pi] : \pi \in {\cal P}_p(\real^2)   \right\}$ of classes. One can show that equivalent elements are equidistant from any other element of ${\cal P}_p(\real^2)$.     }. Indeed, $\eta_p$ does not satisfy the reverse reflexivity condition 
$\eta_p(\pi_1, \pi_2) = 0 \Rightarrow \pi_1 = \pi_2$. Example~\ref{separa} shows us that there exists probability measures $\pi_1,\pi_2 \in {\cal P}_p(\real^2)$ such that
$\pi_1 \neq \pi_2$ while $\eta_p(\pi_1, \pi_2) =0$.
\begin{example} \label{separa}
Let $(X_1,Y_1)\sim \pi_1$ and $(X_2,Y_2)\sim \pi_2$, where $\pi_1, \pi_2$ are described in \emph{Table~\ref{pi1pi2}}. 
Clearly, $\pi_1, \pi_2 \in {\cal P}_p(\real^2)$ for any $p\ge 1$. Note that $X_1$, $Y_1$, $X_2$ and $Y_2$ all have the same distribution $\mu$ given by $\mu (0) = .20$, $\mu (1) = .35$ and $\mu (2) = .45$. Since $\eta_p(\pi_1, \pi_2) = 0$, $\pi_1$ and $\pi_2$ are in the same equivalence class when classes are defined with respect to the equivalence relation $\pi_1 \sim \pi_2 : \Leftrightarrow \eta_p(\pi_1, \pi_2) = 0$.
\end{example}
\begin{center}
\begin{table}[t]
	\centering
   \begin{tabular}{cc}
   \begin{tabular}{|c|ccc|c|} \hline
          $\pi_1$     & 0     & 1    & 2    & $\mu$ \\ \hline
                   0 & 0.05  & 0.10 & 0.05 &\emph{0.20} \\
                   1 & 0.05  & 0.20 & 0.10 &\emph{0.35}\\ 
									 2 & 0.10  & 0.05 & 0.30 &\emph{0.45}\\ \hline
				 	     $\mu$ & \emph{0.20}  & \emph{0.35} & \emph{0.45} & $\Sigma = 1$\\ \hline
   \end{tabular}  &  \begin{tabular}{|c|ccc|c|} \hline
          $\pi_2$     & 0     & 1    & 2    &  $\mu$\\ \hline
                   0 & 0     & 0.10 & 0.10 &\emph{0.20} \\
                   1 & 0.15  & 0.20 & 0    &\emph{0.35}\\ 
									 2 & 0.05  & 0.05 & 0.35 &\emph{0.45}\\ \hline
						   $\mu$ & \emph{0.20}  & \emph{0.35} & \emph{0.45}&$\Sigma = 1$\\ \hline
   \end{tabular}
   \end{tabular} 
	\caption{\small Counterexample showing that the reverse reflexivity axiom does not hold. Since $C(\pi_1) = C(\pi_2)$ and 
	$E|X_1 -Y_1|^p = E|X_2 -Y_2|^p= 0.3 + 0.15*2^p$, we have $\eta_p(\pi_1, \pi_2) = 0$, while $\pi_1$ and $\pi_2$ are not equal. } \label{pi1pi2}
\end{table}
\end{center} 
\subsection{The special case of couplings} \label{etapcouplings}
\subsubsection{Couplings between distributions and between random variables} \label{coupldef}
We now need the general definition of coupling, of which the diagonal coupling (Definition~\ref{diagcoupl}) is a special case.
\begin{defi} \label{couplprob}
(coupling of probability measures on the real line) A coupling of two given probability measures $\muandnu$ on $(\real, {\cal B}_1)$ is any probability measure $\pi$ on $(\real^2, {\cal B}_2)$ whose marginals are $\muandnu$, that is, $\mu =\pi \circ q_1^{-1}$ and $\nu =\pi \circ q_2^{-1}$, where the $q_i$'s are the projection functions defined by $q_i(x_1,x_2) = x_i$  for all $(x_1,x_2)\in \real^2$, $i=1,2$.
\end{defi}
By definition, couplings are multiple. The class of all \emph{couplings} between $\muandnu$ is denoted by $\pimunu$. For example, the distributions 
$\pi_1$ and $\pi_2$ shown in Table~\ref{pi1pi2} belong to the set $\Pi (\mu,\mu)$, i.e. $\nu = \mu$ in this case. 
\begin{defi} \label{couplvar}
(coupling of real-valued random variables) A coupling of two given random variables $\xandy$ taking values in $(\real, {\cal B}_1)$ is any pair of random variables $(\tilde{X},\tilde{Y})$ taking values in $(\real^2, {\cal B}_2)$ such that 
$\tilde{X}$ and $\tilde{Y}$ are defined on the same probability space $(\tilde{\Omega},\tilde{\cal A},\tilde{P})$, with 
$\tilde{X}\stackrel{d}{=} X$ and $\tilde{Y}\stackrel{d}{=} Y$.
\end{defi}
We observe that the law $\tilde{\pi}$ of $(\tilde{X},\tilde{Y})$ is a coupling of the laws $\mu$ of $X$ and $\nu$ of $Y$. An important point of Definition~\ref{couplvar} is that the coupled random variables are defined on the same probability space, while $\xandy$ may not be defined on a common probability space. If $\xandy$ are defined on the same probability space $\probamodel$, then $(X,Y)$ is also defined on $\probamodel$ and $P_{(X,Y)}$ is a coupling of $P_X$ and $P_Y$.  Two trivial couplings are (i) the \emph{diagonal coupling} $\mu \triangle\mu$  of $\mu$ with itself defined in (\ref{diagmeasure}), and (ii) the \emph{product coupling} $\mu \otimes \nu$. If 
$X \sim \mu$ and $Y \sim \nu$ are independent, then the law of $(X,Y)$ is the product probability measure $\mu \otimes \nu$. Table~\ref{sixdistrib} \textbf{(D)} and \textbf{(E)} are examples of product couplings.
\subsubsection{Distance between couplings, $\muandnu$ arbitrary} \label{distinctmarg}
Let $\Pi_p (\mu,\nu)$ denote the set of couplings of two probability measures $\muandnu$ of finite $p$-th moment, both defined on $(\real, {\cal B}_1)$. For $\pi_1,\pi_2 \in \Pi_p (\mu,\nu)$, suppose that     $(X_1,Y_1)\sim \pi_1$ and $(X_2,Y_2)\sim \pi_2$. As couplings of $\mu$ and $\nu$ have identical centers, $||C(\pi_1) - C(\pi_2)|| $ disappears in (\ref{etapi1}) and we have
\begin{equation}
\eta_p(\pi_1, \pi_2) = \left| (E|X_1 -Y_1|^p)^{1/p} - (E|X_2 -Y_2|^p)^{1/p} \right|. \label{cpleta}
\end{equation}
\begin{example} \label{discretemunu}
(Discrete case) Equation~\emph{(\ref{cpleta})} has a particularly simple form when $p=1$ and when $\mu$ and $\nu$ are discrete. Suppose that $X_1$ and $X_2$ take values in $\{ s_i:i=1,2,\ldots ,n_X \}$, and that $Y_1$ and $Y_2$ take values in $\{ t_j:j=1,2,\ldots ,n_Y \}$. Then \emph{(\ref{cpleta})} becomes 
\begin{equation}
\eta_1(\pi_1, \pi_2) = \left| \sum_{i=1}^{n_X} \sum_{j=1}^{n_Y} |s_i - t_j| (p_{ij} - q_{ij}) \right| , \label{etacpl}
\end{equation}
where 
$p_{ij} = \pi_1(\{ (s_i,t_j) \})$ and $q_{ij} = \pi_2(\{ (s_i,t_j) \})$. Equation \emph{(\ref{etacpl})} indicates that $\pi_1 = \pi_2$ implies $\eta_p(\pi_1, \pi_2) = 0$, but we know that the converse is not true \emph{(}see the counterexample in \emph{Table~\ref{pi1pi2})}.
\end{example}
\subsubsection{Distance between couplings when $\nu = \mu$ and $\pi_2 = \mu \triangle \mu$} \label{iddistras}  
Let us now consider the case where $\nu = \mu$, when $\muandnu$ have finite $p$-th moment. Assume that $(X_1,Y_1)\sim \pi_1$ and $(X_2,Y_2) \sim   \mu \triangle \mu$. Note that $\mu \triangle \mu \in \Pi_p (\mu,\mu)$, and assume that $\pi_1 \in \Pi_p (\mu,\mu)$. For ease of reading, write $(X,Y) \sim \pi$ instead of $(X_1,Y_1)\sim \pi_1$. Clearly, $X,Y,X_2$ and $Y_2$ have all the same distribution $\mu$. We wish to measure the distance $\eta_p$ between $\pi$ and $\mu \triangle \mu$. As $X_2\stackrel{a.s.}{=} Y_2$, we have $E|X_2-Y_2|^p=0$ and (\ref{cpleta}) becomes  
\begin{equation}
\eta_p(\pi, \mu \triangle \mu) = (E|X -Y|^p)^{1/p} = ||X -Y||_p .           \label{psimu}
\end{equation} 
That is, the ${\cal L}_p$-distance $(E|X -Y|^p)^{1/p}$ between identically distributed $\xandy$ also represents a distance between the law $\pi$ of $(X,Y)$ and the diagonal coupling built from the marginals of $\pi$. Note that both the ${\cal L}_p$-distance and $\eta_p$ are semimetrics.
\vskip .2cm
\noindent \underline{Distance between equality of distribution and almost sure equality}
\vskip .2cm
Suppose that $X$, $Y$ (and therefore $(X,Y)$) are defined on some probability space $\probamodel$. Moreover, suppose that $\xandy$ are identically distributed. The following question comes to mind: how far from almost sure equality are $\xandy$?  Writing $\pi = P_{(X,Y)}$ and $\mu = P_X$ in (\ref{psimu}), we obtain
\begin{equation}
\eta_p(P_{(X,Y)}, P_X \triangle P_X) = (E|X -Y|^p)^{1/p}.           \label{PxPy}
\end{equation}
That is, when jointly distributed random variables $X$ and $Y$ are \emph{identically distributed}, $||X-Y||_p$ is actually a distance between the distribution of the pair $(X,Y)$ and the \emph{diagonal coupling} of $P_X$ with itself. It is in this sense that $||X-Y||_p$ may be symbolically interpreted as a distance between $X\stackrel{d}{=} Y$ and $X\stackrel{a.s.}{=} Y$.
\vskip .2cm
\noindent \underline{Illustration of the above developments: the case of bivariate normal distributions}
\vskip .2cm
Consider the bivariate normal density function
\begin{equation}
f_{XY}(x,y) = \frac{1}{2\pi \sigma_X \sigma_Y (1-\rho^2)^{1/2}} \times     \nonumber
\end{equation}
\begin{equation}
\exp \left\{ - \frac{1}{2(1-\rho^2)} \left[ \left( \frac{x-m_X}{\sigma_X} \right)^2 -2\rho 
             \frac{(x-m_X)(y-m_Y)}{\sigma_X \sigma_Y } + \left(\frac{y-m_Y}{\sigma_Y}\right)^2   \right] \right\}   \label{bivdens}			
\end{equation}
with parameters $m_X$, $m_Y$ (marginal means), $\sigma_X > 0$, $\sigma_Y > 0$ (marginal standard deviations) and    $|\rho| < 1$ (correlation coefficient). Let $(X,Y) \sim \pi$, where $\pi$ is characterized by the density function 
(\ref{bivdens}). The marginals $\muandnu$ of $\pi$ are $\mu = N(m_X,\sigma^2_X)$ and $\nu = N(m_Y,\sigma^2_Y)$. Importantly, they do not depend on $\rho$, which means that each bivariate normal distribution $\pi$ is a coupling of $\muandnu$. An infinite number of couplings can be created by just changing the value of $\rho$.	

When $m_X = m_Y$ and $\sigma_X = \sigma_Y$, i.e. when $\xandy$ are identically distributed, then $\mu \triangle \mu$ is the diagonal coupling of $N(m_X,\sigma^2_X)$	with itself, which is supported on the diagonal $\Delta$ of $\real^2$. Setting for example $p=1$ in (\ref{PxPy}), we obtain
\begin{equation}
\eta_1\left(\pi, N(m_X,\sigma^2_X) \triangle N(m_X,\sigma^2_X) \right) = E|X -Y|,    \nonumber
\end{equation}
from which we draw the conclusion that the smaller the value of $\exy$, the more the graph of $f_{XY}(x,y)$ is concentrated along the diagonal $\Delta$.
\begin{figure}[t]
  \centering
  \includegraphics[width=17cm,height=7cm]{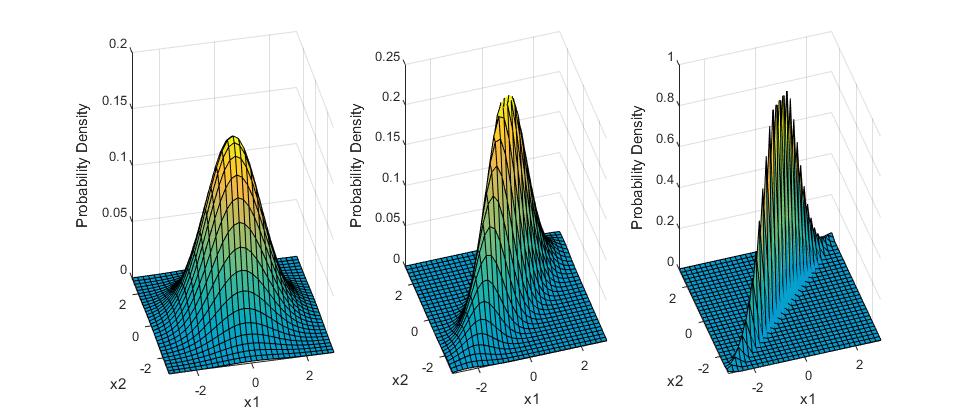}
  \caption{\small Graph of the pdf's of three bivariate normal distributions, all three having $N(0,1)$ as marginals. From left to right: $\rho = 0$, $\rho = 0.8$ and $\rho = 0.99$, respectively. Clearly, the graph in the center is closer to almost sure equality of $\xandy$ than the graph on the left, which corresponds to the distribution of a pair of i.i.d. random variables having a standard normal distribution. It can be shown that the distance $\eta_1$ between the distribution on the left and the distribution on the right is very close to the Gini mean difference $\exy = 2/\sqrt{\pi}$, where $\xandy$ are i.i.d. standard normal random variables.  }
  \label{fig:troisnormales}
\end{figure} 
Figure~\ref{fig:troisnormales} graphs the densities of three bivariate normal distributions, all three having parameters $m_X = m_Y =0$ and $\sigma_X = \sigma_Y =1$. Consequently, the two marginal laws of each of these distributions are the standard normal $N(0,1)$. The only difference between the three bivariate distributions of Figure~\ref{fig:troisnormales} is the value of $\rho$. The distribution on the left, where $\rho = 0$, corresponds to i.i.d. standard normal random variables $\xandy$. Looking at the distribution on the right, where $\rho = 0.99$, we can guess the bell-shaped form the distribution will have when $\rho = 1$. In this case
$\mu \triangle \mu = N(0,1) \triangle N(0,1)$, defined on $\real^2$, concentrates its probability mass on  $\Delta$.

It is interesting to visualize towards which distribution the densities of Figure~\ref{fig:troisnormales} converge when $\rho\nearrow 1$ and $\Delta$ is seen as a standalone line of numbers. In this case the manifold $\Delta$ differs from $\real$ only by its coordinate system: $\real$ is ``stretched'' to obtain $\Delta$ and $\Delta$ can be seen as $\real$ with a new coordinate system given by $y=y(x)=\sqrt{2}x$. Let $f$ (resp. $g$) be the pdf representing $\mu = N(0,1)$ in the old (resp. new) coordinate system. Then, for any $B\in {\cal B}_1$, $\int_B f(x)dx = \int_B g(y)dy$, where, using the Jacobian rule for a change of coordinates, 
$g(y) = |\frac{dx}{dy}| f(x) = \frac{1}{\sqrt{2}} f (\frac{y}{\sqrt{2}} )$, which is the density of the distribution $N(0,2)$. That is, we have shown that the bivariate normal density represented on the right of Figure~\ref{fig:troisnormales} is close to the density $N(0,2)$ defined on $\Delta$ when $\Delta$ is considered as a simple line of numbers. 
\subsubsection{Effect of independence} \label{effindep} 
If, in addition to being identically distributed, two jointly distributed random variables $\xandy$ are assumed to be independent (i.e. if $\xandy$ are i.i.d.), then $P_{(X,Y)} = P_X \otimes P_Y = P_X \otimes P_X$. Moreover, Proposition~\ref{equivalence} implies that, for $c:=E(X)$, $P_X \triangle P_X = \delta_{(c,c)}$, the Dirac delta measure concentrated on $(c,c)\in \Delta$.
In that case, (\ref{etapi1}) becomes
\begin{equation}
\eta_p(P_X \otimes P_X, \delta_{(E(X),E(X))}) = (E|X -Y|^p)^{1/p},           \label{trivcpl}
\end{equation}
i.e. $(E|X -Y|^p)^{1/p}$, whose value depends only on $P_X$, is a measure of the distance between the product coupling of $P_X$ with itself and the measure concentrated on a point of the diagonal of $\real^2$. Note that 
$\delta_{(E(X),E(X))}$ is not in $\Pi (P_X,P_X)$. However, since $\pi_1 := P_X \otimes P_X$ and 
$\pi_2 := \delta_{(E(X),E(X))}$ have the same center, $||C(\pi_1) - C(\pi_2)|| = 0$ in (\ref{etapi1}).
\subsubsection{The Gini mean difference as a distance between measures} \label{probGMD}
Consider two random variables $X,Y \in {\cal L}_1 (\real)$.
As we have seen in Subsection~\ref{gindex}, the Gini mean difference (GMD) of an income distribution $\mu$ is defined as $\exy$, where $X\sim \mu$ and $Y\sim \mu$ are non-negative and independent. We are now able to give a probabilistic definition of the GMD, perhaps the most general definition that can be given to this index. Looking at (\ref{trivcpl}), and setting $p=1$, we obtain
\begin{equation}
GMD(\mu) = \exy = \eta_1(\mu \otimes \mu,\delta_{(E(X),E(X))}) .    \label{mumu}
\end{equation} 
In other words, the GMD of an income (or a wealth, etc.) distribution $\mu$ represents a distance between the product coupling $\mu \otimes \mu$ of $\mu$ with itself and the Dirac delta measure supported on the single point 
$(E(X),E(X))$. Consequently, GMD$(\mu) = \exy$ measures how far the i.i.d. $X\sim \mu$ and $Y\sim \mu$ are from almost sure equality. Note that $\mu \otimes \mu$ distributes its mass symmetrically with respect to $\Delta$. The GMD thus measures a distance between the distribution $\mu \otimes \mu$ on $\real^2$ (or on a set of $A\times A$, $A\subset \real$) and the measure concentrated on the center of this same distribution.
\section{Expected absolute difference for independent random variables: applications to physics and to economics} \label{appl}
Many examples can be given of the usefulness of $\exy$ for independent variables, and they range from physics to economics. In physics, for instance, Lukaszyk (2004)\nocite{Lukaszyk2004} presents a modified Shephard-Liszka\nocite{Liszka1984,Shephard1967} approximation, where $\exy$ proves to be more reliable than a plain Euclidean metric, suggesting that an analogous improvement can be achieved in various numerical methods, and in particular in approximation algorithms. In the same paper, Lukaszyk suggests further applications in fringe pattern analysis, or in quantum mechanics, to estimate the distance of two quantum particles described by their wave functions.
As pointed out in paragraph~\ref{gindex}, $\exy$ has also important applications in inequality economics. Other applications include clustering systems, pattern recognition or finance.

In Subsection~\ref{formulas}, we indicate the forms in which $\exy$ is usually represented in scientific publications, especially of physics, engineering, finance and economics. In \ref{analytic}, we express $\exy$ in analytic form when $\xandy$ are two independent normally distributed random variables. This new result generalizes specific formulas already present in the literature, in particular that of physics. In \ref{pointsrect}, we give closed-form formulas for the average distance and the normalized average distance between coordinates of points falling at random into a rectangle of the plane. In \ref{distintervals}, we take advantage of results of \ref{pointsrect} to express in closed form the mean value of the set $\{|a-b|: a \in A, b \in B \}$, where $A$ and $B$ are bounded intervals of real numbers.
\subsection{Formulations in use in applied fields} \label{formulas}
For independent $X,Y \in {\cal L}_1(\real )$, let $P_{(X,Y)} = \mu \otimes \nu$ be the product distribution of two probability distributions $\muandnu$ defined on $(\real,{\cal B}_1)$. The Fubini-Tonelli theorem applies, and we can interchange the order of integration or summation so that
\begin{eqnarray} 
\exy &=& \int_{\real^2}|x-y|dP_{(X,Y)} (x,y)) = \int_{\real^2} |x-y| d(\mu \otimes \nu)(x,y)  \nonumber   \\ \nonumber
     &=& \int_{\real} \{ \int_{\real} |x-y|d\nu(y) \} \ d\mu(x)
		  = \int_{\real} \{ \int_{\real} |x-y|d\mu(x) \} \ d\nu(y).   \\  \label{deltaw}
\end{eqnarray}
Thereafter, $F$ (resp. $G$) will refer to the cumulative distribution function (cdf) of $X$ (resp. $Y$), and $f$ (resp. $g$) will refer to the probability density function (pdf) of $X$ (resp. $Y$). Particularly interesting are the cases where independent $X$ and $Y$ are (i) both discrete, (ii) both absolutely continuous, or (iii) one of them is discrete and the other absolutely continuous.

\noindent (i) The (independent) \emph{discrete} random variables $X$ (resp. $Y$), take values in the countable sets   $\Omega_X =\{ x_1,x_2,\ldots,x_i,\ldots \} \subset \real$ (resp. $\Omega_Y =\{ y_1,y_2,\ldots,y_j,\ldots \} \subset \real$). In that case, $f$ (resp. $g$) are \emph{probability mass functions}\footnote{i.e. have densities with respect to the counting measure on $\Omega_X$ (resp. $\Omega_Y$). }. 
In this context, (\ref{deltaw}) can be rewritten as $\exy=\sum_i\sum_j\left|x_{i}-y_{j}\right|p_{i}q_{j}$,
where $p_i = P(X = x_i)$ and $q_j = P(Y = y_j)$\footnote{This double sum is just an integral with respect to the counting measure on $\Omega_X \times \Omega_Y$.}. In the finite equiprobable case, when independent random variables $X$ and $Y$ both take the non-negative values $x_1,x_2,\ldots , x_N$ with respective probabilities $p_1=p_2=\cdots = p_N=1/N$, then
$$
\exy= \frac{1}{N^2}\sum_{i=1}^N\sum_{j=1}^N\left|x_{i}-x_{j}\right|
$$
is the Gini mean difference in its \emph{discrete} form, see for example Gini (1912)\nocite{Gini1912}, Kendall and Stuart (1958)\nocite{Kendall1958}, or Xu (2003)\nocite{Xu2003}.
	
\noindent (ii) The independent $X \sim \mu$ and $Y \sim \nu$ are both \emph{absolutely continuous}\footnote{i.e. absolutely continuous with respect to the Lebesgue measure.}. Equation (\ref{deltaw}) usually appears in the following form in the literature (notably in physics and economics)
$$
\exy= \int^{\infty}_{x=-\infty} \int^{\infty}_{y=-\infty}|x-y|f(x)g(y)dxdy.
$$
When $X,Y \in {\cal L}_1(\real )$ are i.i.d. non-negative absolutely continuous random variables ($f=g$),  
$$
\exy= \int^{\infty}_{x=-\infty} \int^{\infty}_{y=-\infty}|x-y|f(x)f(y)dxdy.
$$
is known in the economic literature as the \emph{continuous} Gini mean difference (see, for example, Yitzhaki (1998)\nocite{Yitzhaki1998}, or Yitzhaki and Schechtman (2013)\nocite{YitzhakiSchechtman2013}).

\noindent (iii) $X$ is \emph{discrete}, whereas $Y$, independent of $X$, is \emph{absolutely continuous}. Then
$$
\exy= \sum_i p_i \int^{\infty}_{y=-\infty}|x_i-y|g(y)dy.
$$
A special case of this formula was used by Lukaszyk (2004)\nocite{Lukaszyk2004}, when he proposed a modified Liszka method to handle an experimental mechanics issue.
\subsection{Analytic form of the expected absolute difference between two independent normally distributed random variables} \label{analytic}
The normal case plays a crucial part in a great many of the techniques used in applied statistics. The Central-limit theorem alone ensures that this will be the case, but there are other important reasons extensively discussed in the literature. 

We begin this section by writing $\exy$ in a form facilitating the calculation of its analytic expression when independent $\xandy$ are both absolutely continuous.
\begin{prop} \label{covXcovY1}
(Proof in the appendix) Let $X,Y \in {\cal L}_1(\real )$ be two independent absolutely continuous random variables with means $\mu_X = E(X)$ (resp. $\mu_Y = E(Y)$) and cdf's $F$ (resp. $G$). Then
\begin{equation}
\exy = 2 \left\{E[XG(X)] + E[YF(Y)] \right\} - \mu_X - \mu_Y  \label{covXcovY2}.
\end{equation}
\end{prop}
Consider the special case where $\xandy$ in Proposition~\ref{covXcovY1} are i.i.d. Then $F=G$, $\mu_X = \mu_Y$, and (\ref{covXcovY2}) becomes
\begin{equation}
\exy = 4 E[XF(X)] - 2\mu_X \label{XFX}.
\end{equation}
Moreover, since $X$ absolutely continuous $\Rightarrow$ $F$ continuous $\Rightarrow$ $F(X) \sim U(0,1)$ $\Rightarrow$
$E[F(X)] = 1/2$, we have: $\cov[X,F(X)] = E[XF(X)] - \mu_X/2 \stackrel{(\ref{XFX})}{=} \exy/4$, i.e.
\begin{equation}
\exy = 4 \cov[X,F(X)]   \label{covar}.
\end{equation}
This result -- of which (\ref{covXcovY2}) is a generalization -- can be found in Lerman and Yitzhaki (1984)\nocite{Lerman1984}.

In the following (new) theorem (Theorem~\ref{distnorm1}), we give the analytic form of $\exy$ for normally distributed independent random variables. 
\begin{theor} \label{distnorm1}
(Two alternative proofs can be found in the appendix) Assume that $X\sim N(\mu_X,\sigma^2_X)$ and $Y\sim N(\mu_Y,\sigma^2_Y)$ are independent normally distributed random variables. Let $\phi$ (resp. $\Phi$) be the pdf (resp. the cdf) of the standard normal distribution. Then the expected absolute difference between $X$ and $Y$ is given by
\begin{eqnarray} 
\exy &=& \frac{2\sxx}{\sqrt{\B}} \phi \left( \frac{|\A|}{\sigma_Y} \right) \exp \left( \frac{\sxx (\A)^2 }{2\syy (\B)} \right)    \nonumber \\
      &+& \frac{2\syy}{\sqrt{\B}} \phi \left( \frac{|\A|}{\sigma_X} \right) \exp \left( \frac{\syy (\A)^2 }{2\sxx (\B)} \right) \nonumber \\
		  &+& 2|\A| \Phi \left( \frac{|\A|}{\sqrt{\B}} \right) - |\A| .       \label{distnorm2} \\  \nonumber
\end{eqnarray}
\end{theor}
Equation (\ref{distnorm2}) can also be written
\begin{equation}
\exy = \mxy \left[ 2\Phi\left(\frac{\mxy}{\sqrt{\B}}\right) -1 \right] +2\cdot\sqrt{\B}\cdot \phi \left(\frac{\mxy}{\sqrt{\B}} \right)  \label{distnorm3}.
\end{equation}
Note that (\ref{distnorm2}) is the formula we end up with if we use Proposition~\ref{covXcovY1}. To obtain (\ref{distnorm3}), we used the fact that the convolution of two Gaussian distributions is a Gaussian distribution. The proof of (\ref{distnorm3}) is shorter than that of (\ref{distnorm2}) resulting from Proposition~\ref{covXcovY1}. However, Proposition~\ref{covXcovY1} applies to absolutely continuous variables and we wanted to test it in the particular case where the variables are Gaussian. It is left to the reader to show that (\ref{distnorm2}) and (\ref{distnorm3}) are equivalent.

The rest of Subsection~\ref{analytic} is devoted to corollaries of Theorem~\ref{distnorm1}.
Equation (\ref{distnorm2}) generalizes formulas that have already proven their usefulness in the applied sciences, especially in physics, as illustrated by the following examples. First, consider the case of a degenerate normal random variable $Y$ with mean $\mu_Y$ and variance $\sigma^2_Y \searrow 0$. Symbolically, we may write 
$Y\sim N(\mu_Y,0) = \delta_{\mu_Y}$, where $\delta_{\mu_Y}$ is the Dirac measure supported on the singleton $\{\mu_Y\}$. Let us write $\sigma$ instead of $\sigma_X$ and $\mu_{XY}$ instead of $|\mu_X-\mu_Y|$. Taking the limit $\sigma^2_Y \searrow 0$ in (\ref{distnorm2}) and (\ref{distnorm3}), we obtain the respective formulations (\ref{degenerate}) and (\ref{direct1}) below
\begin{eqnarray} 
\exy &=& \frac{\sqrt{2}\sigma}{\sqrt{\pi}}\exp \left(- \frac{\mu_{XY}^2 }{2\sigma^2} \right) 
                + 2\mu_{XY} \Phi \left( \frac{\mu_{XY}}{\sigma} \right) - \mu_{XY}  \label{degenerate}  \\ 
       &=& \mu_{XY} \left[ 2\Phi\left(\frac{\mu_{XY}}{\sigma}\right) -1 \right] +2\sigma\cdot \phi \left(\frac{\mu_{XY}}{\sigma} \right) . \label{direct1}
\end{eqnarray}
Using the equality  ${\displaystyle \Phi(z)=1-\erfc\left(\frac{z}{\sqrt{2}}     \right)/2  }$, (\ref{degenerate}) becomes
\begin{equation}
\exy = \mu_{XY} + \frac{\sqrt{2}\sigma}{\sqrt{\pi}}\exp \left(- \frac{\mu_{XY}^2 }{2\sigma^2} \right) 
                - \mu_{XY} \ \erfc\left( \frac{\mu_{XY}}{\sqrt{2}\sigma}  \right) .  \label{degenerate2}
\end{equation}
Lukaszyk (2004)\nocite{Lukaszyk2004} used (\ref{degenerate2}) to successfully implement a modified Liszka approximation method to experimental mechanics. Note that (\ref{degenerate}), unlike (\ref{direct1}), leads directly to Lukaszyk's formula (\ref{degenerate2}).

If $X\sim N(\mu,\sigma^2)$, a direct consequence of (\ref{degenerate}) is that $E|X - \mu| = \sqrt{2}\sigma/\sqrt{\pi} \approx 0.7979 \sigma$. Indeed, setting $Y\sim \delta_{\mu}$ implies that $\mu_X = \mu_Y = \mu$ and $\mu_{XY} = 0$. 

In the same paper, Lukaszyk studied the case of two normal distributions having the same variance\footnote{The assumption of homoscedasticity greatly facilitated the integral calculations undertaken by Lukaszyk, as can be seen in his PH.D. thesis (2001)\nocite{Lukaszyk2001} .}, i.e. $X\sim N(\mu_X,\sigma^2)$ and 
$Y\sim N(\mu_Y,\sigma^2)$. Setting $\sigma_X=\sigma_Y=:\sigma$ in (\ref{distnorm2}) and (\ref{distnorm3}), we obtain successively
\begin{eqnarray} 
\exy &=& 2\sqrt{2}\sigma \phi \left( \frac{\mu_{XY}}{\sigma} \right) \exp \left(\frac{\mu_{XY}^2 }{4\sigma^2} \right) + 2\mu_{XY} \Phi \left( \frac{\mu_{XY}}{\sqrt{2}\sigma} \right) - \mu_{XY} \label{equalvar} \\ 
       &=& \mu_{XY} \left[ 2\Phi\left(\frac{\mu_{XY}}{\sqrt{2}\sigma}\right) -1 \right] +2\sqrt{2}\sigma\cdot \phi \left(\frac{\mu_{XY}}{\sqrt{2}\sigma} \right) . \label{direct2}
\end{eqnarray} 
In Lukaszyk's paper $\exy$ appears in the form
\begin{equation}
\exy = \mu_{XY} + \frac{2\sigma}{\sqrt{\pi}} \exp \left(-\frac{\mu_{XY}^2 }{4\sigma^2} \right) 
     - \mu_{XY} \ \erfc\left( \frac{\mu_{XY}}{2\sigma}  \right) ,  \label{equalvarluk}
\end{equation}
which follows directly from (\ref{equalvar}).

So Lukaszyk (2001, 2004) found (and used) the formulas when $\sigma_X = \sigma_Y =: \sigma$. When $\sigma_X$ and $\sigma_Y$ are arbitrary, but $\mu_X = \mu_Y$, (\ref{distnorm2}) or (\ref{distnorm3}) directly imply
\begin{equation}
\exy = \sqrt{\frac{2}{\pi}}\sqrt{\B}  .           \label{muxeqmuy}
\end{equation}
In other words,  ${\displaystyle ||X-Y||_1 = \sqrt{\frac{2}{\pi}} ||X-Y||_2 \approx 0.7979 ||X-Y||_2   }$: the ${\cal L}_1$-distance between $\xandy$ is approximately one fifth smaller than the ${\cal L}_2$-distance in this case. We used the fact that $\var (X-Y) = \B$, since $\xandy$ are independent and that $E(X) = E(Y)$ implies that $\var (X-Y) = E(X-Y)^2$.

Next, consider the case where $\mu_X = \mu_Y =: \mu$ and $\sigma_X = \sigma_Y =: \sigma$, i.e. $\xandy$ are i.i.d. and both follow $N(\mu,\sigma^2)$. Setting $\sigma_X = \sigma_Y = \sigma$ in (\ref{muxeqmuy}), we obtain
\begin{equation}
\exy = \frac{2\sigma}{\sqrt{\pi}},                  \label{gmdnorm}
\end{equation}
which is the Gini mean difference when the underlying income distribution is Gaussian, see e.g. Yitzhaki and Schechtman (2013)\nocite{YitzhakiSchechtman2013}. Note that $\exy$ in (\ref{gmdnorm}), which does not depend on $\mu$, is a measure of dispersion of the same nature as the standard deviation $\sigma$.

Next, if $\sigma^2_X \searrow 0$ and $\sigma^2_Y \searrow 0$, i.e. if $X\sim N(\mu_X,0) = \delta_{\mu_X}$ and $Y\sim N(\mu_Y,0) = \delta_{\mu_Y}$, equations (\ref{distnorm2}) or (\ref{distnorm3}) become 
\begin{equation}
\exy = |\A |.                  \label{distreal}
\end{equation}
Of course, instead of taking the limits 
$\sigma^2_X \searrow 0$ and $\sigma^2_Y \searrow 0$ in (\ref{distnorm2}) or (\ref{distnorm3}), one can calculate directly 
$\exy = \int_{\real} \ \{ \int_{\real} |x-y| \delta(y-\mu_Y) dy\}\ \delta(x-\mu_X) dx = |\A |$. When $X\stackrel{a.s.}{=} a$ and $Y\stackrel{a.s.}{=} b$, $a,b \in \real$, $\exy$ simply transforms into $|a-b|$, i.e. $E|\cdot - \cdot|$ becomes the metric on $\real$ induced by the norm $|\cdot|$.

We end this subsection by observing that equation (\ref{direct1}) provides an analytic formula for $E|X|$, which enables to express $D_{norm}(X,Y)$ in (\ref{drelxy}) in analytic form when $X$ and $Y$ are independent and normally distributed. To get $E|X|$, assume that $Y\stackrel{a.s.}{=} 0$, i.e. that $Y \sim \delta_0$. Then $\mu_{XY} = |\mu|$ and (\ref{direct1}) becomes 
\begin{equation}
E|X| = |\mu| \left[ 2\Phi\left(\frac{|\mu|}{\sigma}\right) - 1\right] + 2\sigma \phi \left( \frac{|\mu|}{\sigma} \right).
\label{mubarre}
\end{equation}
\subsection{Average distance between coordinates of points falling at random into a proper rectangle of $\real^2$} \label{pointsrect}
In this section, uppercase letters $A,B$ refer to bounded proper intervals of real numbers and $L_A,L_B$ refer to their respective length. A \emph{proper interval} is an interval that is neither empty (an example of empty interval is $[a,a[ = \emptyset$,   for some $a \in \real$)\footnote{To avoid any confusion, we use the notation $]a,b[$ instead of $(a,b)$ in Subsections~\ref{pointsrect} and \ref{distintervals}. } nor degenerate (i.e. of the form $[a,a] = \{a\}$). A proper bounded rectangle in $\real^2$ is the cartesian product $A \times B$ of two proper bounded intervals. We are interested in univariate or bivariate continuous uniform distributions such as $U(A)$ or $U(A \times B)$. Moreover, for $a \in \real$, we identify $U(\{a\})$ to the Dirac delta measure $\delta_{a}$ supported on $\{a\}$.

For a given set $C$, let $I_C$ denote the indicator function (defined by $I_C(z) = 1$ if $z\in C$, $I_C(z) = 0$ if $z\notin C$). Consider the random pair $(X,Y) \sim U(A \times B)$. Noting that $L_A^{-1}L_B^{-1} I_{A\times B}(x,y) = L_A^{-1}I_A(x) \cdot L_B^{-1}I_B(y)$ for all $(x,y) \in \real^2$, one can easily show the following (intuitive) equivalence: 
[$(X,Y) \sim U(A \times B)$] if and only if [$X \sim U(A)$, $Y \sim U(B)$, $\xandy$ independent].
We are now ready to ask the question: a point $(x,y)$, which is a realization of $(X,Y)$, falls at random into $A\times B$. What is the average distance between its coordinates? Answer: $\exy$. Put slightly differently, $\exy$ reflects the expected absolute difference of two independent variables $\xandy$ following continuous uniform distributions $X\sim U(A)$ and $Y\sim U(B)$. Theorem~\ref{distunif} below expresses $\exy$ in closed form. Without loss of generality, the intervals $A$ and $B$ are assumed to be open in the theorem.
\begin{theor} \label{distunif}
(Proof in the appendix) Let $A=]a_1,a_2[$ and $B=]b_1,b_2[$ be two open bounded intervals of real numbers, let $L_A = a_2 - a_1$, $L_B = b_2 - b_1$ be their lengths, and $m_A=(a_1+a_2)/2$, $m_B=(b_1+b_2)/2$ be their midpoints. Assume that $X\sim U(A)$ and $Y\sim U(B)$ are independent (or, equivalently, that $(X,Y) \sim U(A \times B)$) and consider the following three possible cases
\[
\begin{array}{lll} 
\textbf{Case 1} & a_1 \leq b_1 <a_2 < b_2 & \hbox{ (overlap without inclusion)}\\
\textbf{Case 2} & a_1 \leq b_1 < b_2 \leq a_2 & B \subset A \hbox{ (inclusion, with $B=A$ possible)} \\
\textbf{Case 3} & a_1 < a_2 < b_1 < b_2 & A \cap B = \emptyset \hbox{ (separation).}
\end{array}
\]
Then the expected absolute difference of $\xandy$ is given in closed form by \newline \noindent
$\exy =$
\[
\begin{array}{lr} 
L_A^{-1}L_B^{-1}[(b_2-b_1)(b_1-a_1)(b_2-a_1)/2 + (b_2-a_2)(a_2-b_1)(b_2-b_1)/2 + \frac{(a_2-b_1)^3}{3}] & (\textbf{Case 1} )\\
L_A^{-1}L_B^{-1}[(b_2-b_1)(b_1-a_1)(b_2-a_1)/2 - (b_2-a_2)(a_2-b_1)(b_2-b_1)/2 + \frac{(b_2-b_1)^3}{3}] & (\textbf{Case 2} )\\
|m_A-m_B|  .                                               & (\textbf{Case 3} )
\end{array}
\]
Moreover,
$$
   \begin{array}{lc}
 	E|X| = \left\{ 
	\begin{array}{ll}
	L_A^{-1} (a_1^2 + a_2^2)/2 & if \hskip .5cm 0\in A \\
	|m_A|& if \hskip .5cm 0\notin A 
	\end{array}   \right.
    & \hbox{ and } \ \ E|Y| = \left\{ 
	\begin{array}{ll}
	L_B^{-1} (b_1^2 + b_2^2)/2 & if \hskip .5cm 0\in B \\
	|m_B|& if \hskip .5cm 0\notin B. 
	\end{array}   \right.
   \end{array} 
$$
\end{theor}
Theorem~\ref{distunif} can also be used to give closed formulas for $\exy$ when one of the two intervals $A$ or $B$ is degenerate. Suppose for example that $X\sim U(A)$ and $Y\sim \delta_b$. One way of expressing $\exy$ is then to take the cases 2 and 3 of Theorem~\ref{distunif} and to calculate the limit $b_2\searrow b_1$. We obtain
\begin{equation}
	 \label{distptint}
\exy = \left\{ \begin{array}{l}
            \frac{(b-a_1)^2+(a_2-b)^2}{2L_A} \hbox{ if } b \in A \hbox{ (using $b_2 \searrow b_1$ in case 2) }  \\
            |b-m_A| \ \ \ \ \ \ \, \hbox{ if } b\notin A \hbox{ (using $b_2 \searrow b_1$ in case 3). }\\
               \end{array} 
               \right.	
\end{equation}
A more direct way to proceed is to use the Lebesgue integral with the coupling $\pi = \mu \otimes \nu$ as the measure used for integration, where $\mu = U(A)$ and $\nu = \delta_b$. Indeed, let $h:\real^2 \rightarrow \real$ be given by $h(x,y) = |x-y|$. Then 
$\exy = \int h d\pi = \int_A \{ \int_{\real} |x-y| \delta(y-b)dy\} L_A^{-1}I_A(x) dx = L_A^{-1} \int_A |x-b| dx$. For $A=]a_1,a_2[$, distinguishing the respective situations where (i) $b\in A$ and (ii) $b\notin A$ with $b \le a_1$ and $b \ge a_2$, we find the formulas of (\ref{distptint}). If both intervals are degenerate, i.e. of type $\{ a \}$ and $\{ b \}$, then $\mu = \delta_a$, $\nu = \delta_b$, $\pi = \delta_a \otimes \delta_b = \delta_{(a,b)}$, and $\int h d\pi = |a-b|$.

Theorem~\ref{distunif} allows to easily calculate the normalized average distance $D_{norm} (X,Y) = \exy/(E|X| + E|Y|) \in [0,1]$ defined in (\ref{drelxy}). As evidenced by Theorem~\ref{ineqtri1}, the fact that $\xandy$ (and an additional $Z$) are independent implies that $D_{norm}(\cdot , \cdot)$ satisfies the triangle inequality.
\subsection{The average of the distances $|a-b|$, $ a \in A, b \in B$, is not a distance } \label{distintervals}
Let $\mathcal I$ denote the set of bounded proper intervals of real numbers. As $A\in \mathcal I$ and the probability measure $U(A)$ are in bijection, the closed formulas for $\exy$ in Theorem~\ref{distunif} can also be taken in a purely deterministic sense to compute the ``mean value'' $D(A,B)$ of the set $\{|a-b|: a \in A, b \in B \}$, where $A,B \in \mathcal I$. Formally, we just have to adopt new notations: replace, say, $\exy$ by $D(A,B)$, $E|X|$ by $S_A$, $E|Y|$ by $S_B$ and $\exy/(E|X| + E|Y|)$ by $D_{norm}(A,B) = D(A,B)/(S_A + S_B)$ in Theorem~\ref{distunif}. More precisely, we have the weighted means $D(A,B) = \int_A \int_B |x-y| w(x,y) dxdy / \int_A \int_B w(x,y) dxdy$,
$S_A = \int_A |x|u(x)dx / \int_A u(x)dx$ and $S_B = \int_B |y|v(y)dy / \int_B v(y)dy$, where $w$, $u$ and $v$ are the respective weight functions $w=I_{A \times B}$, $u = I_A$ and $v = I_B$. As $D(A,B)$ stems from $\exy$ and $D_{norm}(A,B)$ stems from $\exy/(E|X| + E|Y|)$, $D(A,B)$ is invariant under location transformations, while 
$D_{norm}(A,B)$ is invariant under scale transformations.

It turns out that the functionals $D$ and $D_{norm}$, both ${\mathcal I} \times {\mathcal I} \longrightarrow \real_+$, satisfy the axioms of symmetry and triangle inequality (the fact that $D_{norm}(\cdot,\cdot)$ satisfies the triangle inequality, far from obvious, is a consequence of Theorem~\ref{ineqtri1}).  However, reflexivity does not hold: indeed $D(A,A) > 0$ and $D_{norm}(A,A) > 0$ for any $A \in {\mathcal I}$, which means that $D$ and $D_{norm}$
are neither distances, nor -- a fortiori -- metrics on $\mathcal I$.\footnote{Actually, $D$ and $D_{norm}$ are \emph{metametrics}. The term ``metametric'' is specified in Deza and Deza (2014)\nocite{Deza2014}. Metametrics appear in the study of Gromov hyperbolic metric spaces. They were first defined by V\"ais\"al\"a (2005)\nocite{Vaisala2005}.}

Finally, let $b$ be any fixed real number. To obtain the closed-form formula for the ``mean value'' $D(A,\{ b\})$ of the set $\{ |a-b|$, $ a \in A \}$, replace $\exy$ by $D(A,\{ b\})$ in (\ref{distptint}).
\section{Optimal transport problem for probability measures on the real line} \label{formulation}
The following brief introduction to the problem of optimal transport is intended for the many non-specialists in the field. We will stay at a rather heuristic level, focusing on the founding ideas of transport theory. For a detailed account of the theory, the reader is referred to Villani (2003, 2008)\nocite{Villani2008} for example. 
The problem of optimal transport can be presented in two related ways. The formulation of Monge is ancient and dates back to the 18th century. Kantorovich's work is much more recent and was published during World War II. It can be interpreted as a generalization or a relaxation of Monge's approach. In practice, the latter seems to be more direct and easier to interpret, but its resolution is mathematically more complicated. 

This text is designed for a broad readership and focuses on the main ideas of the optimal transport problem. In this perspective of relative simplicity, our discussion here is limited to probability measures $\muandnu$ defined on the real line rather than on more general spaces, so as not to lose sight of the main issues. The originality of this short presentation consists in exploiting known results (if possible with a slightly shifted look) while using notations familiar to practitioners of applied statistics, physics or econometrics. This does not prevent some new results from emerging.

The optimal mass transport problem tries to find the most efficient way to transport a \emph{source measure} $\mu$ over a \emph{target measure} $\nu$ taking account of a given cost function. A transport cost determines in some way the difference or distance between these measures. For the sake of simplicity, the cost function $c:\real^2 \rightarrow \real_+$ that will be used in this article is mainly of type $c(x,y) = |x-y|$, with sometimes a slight generalization: $c(x,y) = h(x-y)$, where $h$ is convex and continuous. We will see later how the choice of a cost function of type $c(x,y) = |x-y|$ leads to the expected absolute difference $\exy$ between $\xandy$.

It should be borne in mind that the optimal transport problem is much more general than the particular cases treated here. This concerns notably -- as we have pointed out -- the type of spaces on which $\muandnu$ are defined, but no less significantly the characteristics of the cost function. The restriction to the one-dimensional case and the use of a simple cost function make it possible to define the problem without technicalities -- sometimes severe -- related to more general cases. When $\muandnu$ are defined on $\real$, or on a subset of $\real$, the problems of Monge and Kantorovich have easily interpretable closed form solutions in some important cases. This is a significant property as it alleviates the need for optimization.
 
When, as in this article, the source measure $\mu$ and the target measure $\nu$ are defined on the same space, the transport of measures has applications in many fields. For instance, if objects are initially distributed according to $\mu$, then they are arranged after transport according to $\nu$. In inequality economics, $\mu$ represents a distribution of income and the problem is to find a planning carrying $\mu$ over a less unequal target distribution $\nu$. In finance, $\mu$ can be the return distribution of a portfolio of stocks and $\nu$ the return of another portfolio or a benchmark.

In the next two sections, we present in detail how the Monge and Kantorovich approaches of the optimal transport unfold when $\muandnu$ are probability measures on the real line (or have a support on the real line).
\subsection{The Monge formulation} \label{mongeform}
In his \emph{M\'emoire sur la Th\'eorie des D\'eblais et Remblais}, Monge (1781),\nocite{Monge1781} was interested in minimizing the cost of transporting sand from a dune to fill a ditch, or transporting stones from an excavation to build a fortification. Monge's historical modeling was in $\real^3$ and the cost function was the Euclidean distance. In the generalizations that followed, $\real^3$ became for example a Polish metric space and the cost function took various forms that were quite different from the original Euclidean distance. Above all, the piles of sand or pebbles and the cavities to be filled became over time distributions of probability, of income, of wealth, configurations of physical particles, return distributions of financial assets, and so on. 

In order to state Monges's problem in the one-dimensional real case, we need the following definition.
\begin{defi}\label{appltransf} (transport map)
Consider $\mu, \nu \in {\cal P}(\real)$, where ${\cal P}(\real)$ is the set of probability measures on              $(\real ,{\cal B}_1)$. We say that a measurable map $T:(\real ,{\cal B}_1) \rightarrow (\real ,{\cal B}_1)$ transports $\mu$ to $\nu$, and we call $T$ a \emph{transport map}, if $\nu(B) = \mu(T^{-1}(B))$ for all Borel subsets $B$ of $\real$.
\end{defi}
When $T$ transports $\mu$ to $\nu$, we use the notation $\mu_T = \nu$ (rather than $T_{\#}\mu = \nu$). We denote by ${\cal T}(\mu,\nu) = \{ T: (\real, {\cal B}_1,\mu) \rightarrow (\real, {\cal B}_1)| T \hbox{ measurable with } \mu_T = \nu\}$ the set of transport maps pushing forward $\mu$ to $\nu$.
\begin{defi}\label{minmonge} (Monge's formulation of the transport problem for probability measures on the real line -- or with support on the real line)
Let ${\cal P}_1(\real)$ be the set of probability measures on $(\real, {\cal B}_1)$ that have finite first moment.
Given $\mu, \nu \in {\cal P}_1(\real)$ and the cost function $c(x,y) = |x-y|$, find a transport map that realizes the infimum
\begin{equation}
\inf\{\int_{\real}|x-T(x)| \ d\mu(x): T \in {\cal T}(\mu,\nu) \}. \label{infmonge}
\end{equation}
\end{defi}
Unfortunately, we may not find any measurable map such that $\mu_T = \nu$. In other words, ${\cal T}(\mu,\nu)$ may be empty. We do not have to look very far: let us take $\mu = \delta_{x_1}$ (the Dirac delta measure supported on 
$\{x_1\}$) and $\nu = \frac{1}{2}\delta_{y_1} + \frac{1}{2}\delta_{y_2}$, with $y_1\neq y_2$. In this case, no map $T$ with $\mu_T = \nu$ can be found. Important cases where ${\cal T}(\mu,\nu) \ne \emptyset$ are (Thorpe 2018)\nocite{Thorpe2018}: (i) the discrete case when 
$\mu = \frac{1}{n}\sumin \delta_{x_i}$ and $\nu = \frac{1}{n}\sumjn \delta_{y_j}$, i.e. when $\muandnu$ are supported on the same number of points with equal mass. And (ii) the absolutely continuous case, when $d\mu (x) = f(x)dx$ and $d\nu (y) = g(y)dy$. Moreover, even if ${\cal T}(\mu,\nu) \ne \emptyset$, the constraint in Monge's problem is usually very non-linear and difficult to handle with the classical tools of the calculus of variations. Kantorovich's approach alleviates these problems by seeking an optimal transport plan rather than an optimal transport map.
\subsection{The Kantorovich formulation (often called Monge-Kantorovich formulation)} \label{kantoform}
\begin{defi}\label{plan} (Transport plan)
Consider $\mu, \nu \in {\cal P}(\real)$, where ${\cal P}(\real)$ is the set of probability measures on              $(\real ,{\cal B}_1)$. Let ${\cal P}(\real^2)$ denote the set of probability measures on $(\real^2, {\cal B}_2)$. We say that a probability measure $\pi \in {\cal P}(\real^2)$ whose marginals are $\muandnu$, transports  $\mu$ to $\nu$. The measure $\pi$ is called a \emph{transport plan}. We say that $\pi$ has first marginal $\mu$ and second marginal $\nu$ if  $\pi (A\times \real) = \mu(A)$ and $\pi (\real\times B) = \nu(B)$ for all $A,B \in {\cal B}_1$. Equivalently, if $q_1(x,y)=x$ and $q_2(x,y)=y$ are the first and second projection functions, respectively, then $\mu = \pi_{q_1}$, $\nu = \pi_{q_2}$, and $\mu(A) = \pi_{q_1}(A) = \pi(q_1^{-1}A)$, $\nu(B) = \pi_{q_2}(B) = \pi(q_2^{-1}B)$ for all $A,B\in {\cal B}_1$. The class of transport plans is denoted by $\pimunu$; it is also called the class of all \emph{couplings} between $\muandnu$.
\end{defi}
Note that the set $\pimunu$ of transport plans is never empty since it contains the trivial plan $\mu \otimes \nu$.
For any $A,B\in {\cal B}_1$, the quantity $\pi (A\times B)$ tells us how much mass in set $A$ is being moved to set $B$. The total amount of mass removed from $A$ has to be equal to $\mu(A)$ and the total amount of mass moved to $B$ must be $\nu (B)$. Hence the constraints: $\pi (A\times \real) = \mu(A)$ and $\pi (\real\times B) = \nu(B)$ for all $A,B \in {\cal B}_1$.

Kantorovich (1942)\nocite{Kantorovich1942} proposed a general formulation of the problem by considering optimal transport plans \emph{which allow mass to be split}. This is a very important difference between the two approaches; Monge's problem, unlike Kantorovich's, requires that each mass in $x$ is sent to a single position $y$: there is no possible separation of a unit of mass of $\mu$ into several pieces during the transport. Still restricting ourselves to probabilities $\muandnu$ defined on the real line, or having a support on the real line, we can formulate the following definition.
\begin{defi}\label{minkanto} (Kantorovich's form of the transport problem)
Given $\mu, \nu \in {\cal P}_1(\real)$ and the cost function $c(x,y) = |x-y|$, find a transport plan that realizes the infimum
\begin{equation}
\inf\{\int_{\real^2}|x-y| \ d\pi(x,y): \pi \in \Pi (\mu,\nu) \}. \label{infkanto}
\end{equation}
\end{defi}
The term $\int_{\real^2}|x-y| \ d\pi(x,y)$ represents the transport cost from $\mu$ to $\nu$ under $\pi$. We can think of $d\pi (x,y)$ as the amount of mass transferred from $x$ to $y$. 
Actually, as $c(x,y)=|x - y|$ is continuous\footnote{or even lower semi-continuous, a weaker condition implying the existence of a minimizer.}, a minimizer $\pi^*$ always exists (Gangbo (2004), th. 2.4)\nocite{Gangbo2004} and we can replace ``inf'' by ``min'' in (\ref{infkanto}). 
\subsubsection{Probabilistic point of view: some helpful clarifications} \label{probapoint}
\noindent 
Before resuming the substantive discussion on the problem of transport, we turn for a moment to considerations of a purely formal or didactic nature. The optimal transport domain is mainly conceived in terms of (probability) measures. Our experience is that many newcomers to this field feel more comfortable with concepts such as random variables or mathematical expectation, while the language of measure theory seems less telling to them, at least initially. We think that a brief development will clarify some aspects that specialists may consider as futile or obvious. 

(i) Classically, if we have a workspace $(\real^2,{\cal B}_2, \pi)$, but wish to reason in terms of ramdom variables, we formally introduce a general probability space $\probamodel$ and a pair of random variables $(X,Y)$ so as to obtain the scheme $\probamodel \stackrel{(X,Y)}{\longrightarrow} (\real^2,{\cal B}_2, \pi)$, where $\Omega :=\real^2$, ${\cal A} := {\cal B}_2$ and $(X,Y) := Id_2$, the identity map on $\real^2$. As a consequence, 
$\pi = P_{(X,Y)} = P_{(Id_2)} = P$, noting the obvious fact that $Id_2 = (q_1,q_2)$, (as in Definition~\ref{plan}, we use the notations $q_1$ and $q_2$ for the projection functions). By doing so, the representation $\probamodel \stackrel{(X,Y)}{\longrightarrow} (\real^2,{\cal B}_2, P_{(X,Y)})$ becomes
$(\real^2,{\cal B}_2, \pi) \stackrel{Id_2}\longrightarrow (\real^2,{\cal B}_2, \pi)$, i.e., in particular, the variables $\xandy$ transform themselves into canonical projections\footnote{This reminds us that the term ``random variable'' is rather unfortunate since a random variable is neither random nor a variable.   }. So $\xandy$ can be interpreted indifferently as random variables or as projections. For instance, a notation such as $E_\pi |X-Y|$ instead of $\int_{\real^2}|x-y| d\pi(x,y))$  makes sense  as a mean absolute deviation between two random variables. This convention will be applied below, notably in Figure~\ref{fig:Figscheme}, which will help us articulate the Kantorovich relaxation of the Monge problem.

(ii) A transport plan $\pi \in \Pi (\mu,\nu)$ is a coupling between $\muandnu$, i.e. a joint distribution $(X,Y) \sim \pi$ such as marginally $X \sim \mu$ and $Y \sim \nu$. By abuse of language the expression ``coupling between $\xandy$'' is used to mean ``coupling between the distribution of $X$ and the distribution of $Y$''. Considered in probabilistic form, the problem stated in (\ref{infkanto}) amounts to finding the distribution of a pair of random variables $(X^*,Y^*)$ minimizing the mean absolute deviation $\exy$ among all jointly distributed pairs $(X,Y)$ such that $P_X = \mu$ and $P_Y = \nu$. 
\subsubsection{Using Kantorovich's relaxation to solve Monge's problem} \label{determi}
\noindent 
In some special cases, Kantorovich's approach can be used to solve the (difficult) Monge optimisation problem. In this context, \emph{deterministic} plans play an essential role.
\begin{defi}\label{detplan}
(Deterministic transport plan). Let $\mu, \nu \in {\cal P}(\real)$, and denote by $Id$ the identity map on $\real$. Let $\pi_T := \mu_{(Id,T)}$ be the pushforward probability measure of $\mu$ induced by the function
$(Id,T):\real \rightarrow \real^2$, where $T$ is a transport map that pushes forward $\mu$ to $\nu$. Then  $\pi_T \in \Pi (\mu,\nu)$, and $\pi_T$ is called a \emph{deterministic} transport plan.
\end{defi}
Note that $\pi_T = \mu_{(Id,T)}$ is supported on the graph of $T$. Let us check that the marginals of $\mu_{(Id,T)}$ are $\mu$ and $\nu$, respectively. In this regard, consider $A,B \in {\cal B}_1$. Then, using in particular Lemma~\ref{properties} (ii) in Subsection~\ref{onedim}: \newline
$\pi_T (A\times \real) = \mu_{(Id,T)} (A\times \real) = \mu [(Id,T)^{-1}(A\times \real)] = 
\mu [Id^{-1}(A) \cap T^{-1}(\real)) = \mu (A)$.\newline
$\pi_T (\real \times B) = \mu_{(Id,T)} (\real \times B) = \mu [(Id,T)^{-1}(\real \times B)] = 
\mu (T^{-1}(B)) = \mu_T (B) = \nu (B)$, since $T$ is a transport map.

\begin{figure}[t]
  \centering
  \includegraphics[width=14cm,height=4cm]{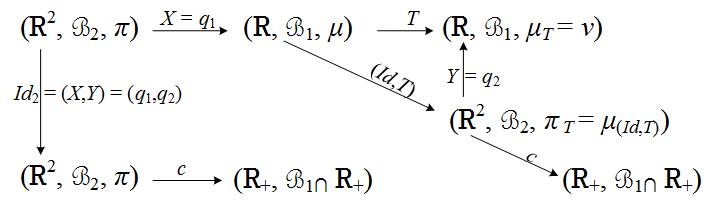}
  \caption{\small Diagram putting plans, deterministic plans, transport maps and (non-negative) cost functions $c$ into perspective. ${\cal B}_1 \cap \real_+$ is the class of Borel subsets of $\real_+$.}
  \label{fig:Figscheme}
\end{figure}

Figure~\ref{fig:Figscheme} provides a convenient overview of the situation. 
In particular, and in relation to Subsection~\ref{probapoint}, it clearly shows that we can express the probability measures in terms of law (${\cal L}$) of random variables: there exists random variables $\xandy$ such that $\pi = {\cal L}((X,Y))$, $\mu ={\cal L}(X)$, $\nu = {\cal L}(Y)$, $\mu_T = {\cal L}(T(X))$ and $\pi_T = {\cal L}((X,T(X)))$. 
This may seem like a detail, but the way to represent the transport cost associated with a transport plan $\pi$ or a transport map $T$ varies according to the research field. Three common types of notations are shown in Table~\ref{totalcost} for the cost function $c(x,y) = |x-y|)$, where $c_{\pi}$ and $c_T$ denote the cost resulting from the transformation of $\mu$ into $\nu$ by means of $\pi$ and $ T$, respectively.
\vskip .4cm
\begin{table}[t] 
	\centering
		\begin{tabular}{|l|l|l|} \hline
  Economics,            & Optimal transport & Measure theory \\
	physics, etc.         & theory            & (Lebesgue integral)\\ \hline
	(a) $c_{\pi} = E_\pi |X-Y|$     & {$\displaystyle c_{\pi} = \int_{\real^2}|x-y|d\pi (x,y)$}   & $c_{\pi} = \int c d\pi$      \\
					& & \\
  (b) $c_T = E_{\pi_T} |X-Y|$ & {$\displaystyle c_T = \int_{\real^2}|x-y|d\pi_T (x,y)$} & $c_T = \int c d\pi_T = \int c\circ (Id,T)\circ X d\pi$ \\ 
	& & \\
  (c) $c_T = E_\pi |X-T(X)|$  & {$\displaystyle c_T = \int_{\real^2}|x-T(x)|d\pi (x,y)$}& $c_T = \int c\circ (Id,T)\circ X d\pi$      \\ 
	& & \\
	(d) $c_T = E_\mu |X-T(X)|$  & {$\displaystyle c_T = \int_{\real}|x-T(x)|d\mu (x)$}    & $c_T = \int c\circ (Id,T) d\mu = \int c\circ (Id,T)                                                                                                 \circ X d\pi$  \\ 
	\hline
    \end{tabular}
	\caption{\small Given the cost function $c(x,y) = |x-y|$), a same object is represented from three points of view: (i) physics, economics, etc., (ii) optimal transport theory and (iii) measure theory. The quantities in (b), (c) and (d) coincide. This table is related with Figure~\ref{fig:Figscheme}.  } 
\label{totalcost}
\end{table}
It turns out that the formulas in (b), (c) and (d) of Table~\ref{totalcost} are different expressions of the same quantity $c_T = \int c\circ (Id,T)\circ X d\pi$. 
This results immediately from a repeated use of the change of variable formula in the context of Figure~\ref{fig:Figscheme}, recalling that $X = q_1$ under the convention adopted in Subsection~\ref{probapoint}. For example, let us show that (c) = (b). Remembering that $c(x,y) = |x-y|$, we have $E_\pi |X-T(X)|= \int c \circ (Id,T) \circ X d\pi$ (see Figure~\ref{fig:Figscheme}). On the other hand, $E_{\pi_T} |X-Y| = \int c d\pi_T = \int c d\mu_{(Id,T)} = \int c\circ (Id,T) d\mu = \int c\circ (Id,T) d\pi_{q_1} = \int c \circ (Id,T) d\pi_X = \int c\circ (Id,T)\circ X d\pi$. 

A heuristic interpretation of (a) is as follows: a pair of random variables $(X,Y)$ is defined on a probability space $\probamodel = (\real^2,{\cal B}_2, \pi)$. We observe independently an infinity of realizations $(x,y)$ of $(X,Y)$ falling on $\real^2$ according to $\pi$, we calculate the distance $|x-y|$ between the coordinates and average these distances to obtain $E_\pi |X-Y|$. The three equivalent expressions (b), (c) and (d) are interpreted in an almost identical way. Take $E_{\pi_T} |X-Y|$: we observe independently an infinity of realizations $(x,y)$ of $(X,Y) \sim \pi$. But this time, we forget $y$ to replace it by $T(x)$. In other words, an infinite number of pairs $(x,y)$ fall (almost surely) on the graph of $T$. We calculate the distances between the coordinates of these pairs and average them to find $E_{\pi_T} |X-Y|$.

The most important equality in Table~\ref{totalcost} is 
\begin{equation}
E_\mu |X-T(X)|  = E_{\pi_T} |X-Y|, \label{EpiTmu} 
\end{equation}
which shows that any transport map $T$ induces a transport plan of the same cost, i.e. can be canonically embedded into the set of transport plans. Adopting the convention that $\inf \emptyset = \infty$ if ${\cal T}(\mu,\nu) = \emptyset$, this means that
\begin{equation}
\min_{\pi \in \Pi (\mu,\nu)} E_\pi |X-Y| \leq \inf_{S \in {\cal T} (\mu,\nu)} E_\mu |X-S(X)| . \label{infinf}
\end{equation}
Equality in (\ref{infinf}) holds under fairly general assumptions when any plan can be approximated by transport maps (see e.g. Ambrosio and Pratelli (2003)\nocite{Ambrosio2003})\footnote{The presence of atoms can seriously impede the existence of transport maps. Under fairly general assumptions, it can be shown that if $\mu$ is atomless (in our setting, this means that $\mu (\{ x\}) = 0 \ \forall x\in \real$), then the set $\{ \pi_T : \mu_T = \nu \}$ is weak-*dense in $\Pi (\mu,\nu)$, which implies equality in (\ref{infinf}) (see Carlier (2010)\nocite{Carlier2010} or Ambrosio et al. (2004)\nocite{Ambrosio2004}). Ambrosio (2002)\nocite{Ambrosio2002} notes that the infimum of the Kantorovich problem ``is attained on an extremal element of $\Pi (\mu,\nu)$''. However, all extremal points are not induced by transport maps ``otherwise one would get existence of transport maps directly from the Kantorovich formulation''. It can be shown that deterministic transport plans are extremal in $\Pi (\mu,\nu)$. ``Unfortunately, the extremal points of $\Pi (\mu,\nu)$ are not all transport plans, except in very particular cases. It turns out that the existence of optimal transport maps depends not only on the geometry of $\Pi (\mu,\nu)$, but also (in a quite sensible way) of the choice of the cost function $c$''.}. 

We are now ready to show the following result: (i) if the Kantorovich problem admits an optimal plan (minimizer) $\pi$, and (ii) this plan turns out to be deterministic i.e. of the form $\pi_T$, then the transport map $T$ is Monge-optimal. To see that, let us assume that $\pi_T$ is Kantorovich-optimal. Then \newline
$\inf_{S \in {\cal T} (\mu,\nu)} E_\mu |X-S(X)| \leq E_\mu |X-T(X)| = E_{\pi_T} |X-Y| =^{(\hbox{hyp.})}
\min_{\pi \in \Pi (\mu,\nu)} E_{\pi} |X-Y|$ \newline
$ \leq^{(\ref{infinf})}  \inf_{S \in {\cal T} (\mu,\nu)} E_\mu |X-S(X)| $
and therefore
\[
E_\mu |X-T(X)| = \inf_{S \in {\cal T} (\mu,\nu)} E_\mu |X-S(X)| = \min_{S \in {\cal T} (\mu,\nu)} E_\mu |X-S(X)| .
\]
This relaxation of Monge by Kantorovich occurs in a few important cases. We will see two examples below (in \ref{Fcontinuous} and \ref{discrmonge}).
\subsection{Closed-form solution of the optimal transport problem in dimension one} \label{onedim}
The following theorem and its proof can be found in Thorpe (2018), \nocite{Thorpe2018} see also Villani (2003) \nocite{Villani2003} and Santambrogio (2015). \nocite{Santambrogio2015} It is a powerful tool for the treatment of the two cases mentioned above.
\begin{theor} \label{hxy}
Let $\mu , \nu \in {\cal P}(\real)$, with cumulative distribution functions $F$ and $G$, respectively. Assume that  $c(x,y) = h(x - y)$ where $h$ is convex and continuous. Let  $\bar{\pi}$ be the probability measure on $\real^2$ with cdf $H(x,y) = \min \{F(x), G(y) \}$. Then $\bar{\pi} \in \Pi (\mu,\nu)$ and, furthermore, $\bar{\pi}$  is optimal for Kantorovich's optimal transport problem with cost function $c$.
\end{theor}
In this subsection (Subsection \ref{onedim}), we are mainly interested in a cost function of type $c(x,y) = |x-y|$. Nevertheless, all the results obtained remain valid for a cost function such as the one specified in Theorem~\ref{hxy}. 

Still limited to one-dimensional probability measures $\mu , \nu \in {\cal P}(\real)$, we give two examples where the Kantorovich relaxation approach leads to a solution of the Monge problem. We base ourselves on the criterion stated in Theorem~\ref{hxy}.\newline 
(i) When the continuous cost function satisfies a certain convexity condition, an optimal plan is supported on a ``curve'' in $\real^2$ depending on the quantile functions associated with respective cdf's $F$ of $\mu$ and $G$ of $\nu$ (quantile functions are defined in Definition~\ref{geninv}) below. Moreover, if $\mu$ is atomless, that is if $F$ is continuous, then this optimal plan is deterministic, i.e. associated with a Monge-optimal transport map. \newline
(ii) Theorem~\ref{hxy}, applied this time to the special discrete case where $\mu = \frac{1}{n}\sumin \delta_{x_i}$ and $\nu = \frac{1}{n}\sumjn \delta_{y_j}$, also leads to a deterministic optimal plan. The optimization process will provide an interesting by-product in that case (Proposition~\ref{permut}). 

\begin{figure}[t]
  \centering
  \includegraphics[width=12cm,height=5cm]{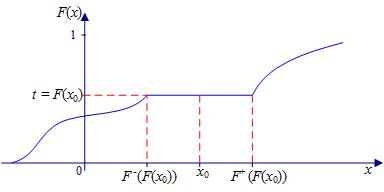}
  \caption{\small Location of the values $F^- \circ F(x_0)$ and $F^+ \circ F(x_0)$ of the two generalized inverses when $x_0$ corresponds to the ``flat part'' of a cumulative distribution function $F$.}
  \label{fig:geninverses}
\end{figure}

We now need the following definition to give an analytic representation of the optimal plan $\bar{\pi}$ referred to in Theorem~\ref{hxy}.
\begin{defi}\label{geninv}
(Quantile function, Karr (1993)\nocite{Karr1993} p. 63). Consider a measure $\mu \in {\cal P}(\real)$ with cumulative distribution function $F$, i.e. $F(x) = \mu((-\infty , x])$. The generalized inverse $F^-$ of $F$, or quantile function associated with $F$, is defined by
\begin{equation}
F^-(t) = \inf \{ x \in \real:F(x) \geq t \} \ \ \ t \in [0,1]. \label{defgeninv}
\end{equation}
Another generalized inverse can be defined:
$$F^+(t) = \sup \{ x \in \real:F(x) \leq t \}.$$
\end{defi}
The function $F^-$ always exists, even when $F$ is not continuous or not strictly increasing. 
As both $F$ and $F^-$ are monotonically increasing, they are also measurable, an important property that we will use later.
The notation $F^{-1}$ instead of $F^-$ is used by many authors. However, it can sometimes be confusing (it may be mistaken for the preimage operator of a set, see below).  If $F$ is continuous and strictly increasing, then the two generalized inverses are equal to the ordinary inverse (on the range of $F$). One can often work with $F^-$ as if it were an ordinary inverse.  Note that (\ref{defgeninv}) implies $F^-(0) = -\infty$, and we adopt the convention that $\inf \emptyset = \infty$. Figure~\ref{fig:geninverses} illustrates the inequalities $F^- \circ F(x_0) \leq x_0 \leq F^+ \circ F(x_0)$ when $x_0$ corresponds to a flat part of $F$, but these two inequalities are in fact valid for any $x_0 \in \real$.
An important property of the quantile function $F^-$ is the following: for each $t$ and $x$
\begin{equation}
F^-(t) \leq x \Leftrightarrow t \leq F(x) , \label{propquant}
\end{equation}
(noting that to prove the implication ``$\Rightarrow$'', one uses the right continuity of $F$).

We are now ready to represent in a more useful way a plan which -- like the one of Theorem~\ref{hxy} -- has a cdf of type $H(x,y) = \min \{F(x), G(y) \}$, where $F$ and $G$ are the cdf's characterizing the probability measures $\mu , \nu \in {\cal P}(\real)$, respectively. To that aim, consider the ``curve'' $K: [0,1]\rightarrow \real^2$ given by $K(t) = (F^-(t),G^-(t))$.\footnote{We use the term ``curve'' by abuse of language, even if $K$ is not continuous. $K$ is a parametric curve in the usual sense when $F^-$ and $G^-$ are continuous, which is true if and only if $F$, resp. $G$, are strictly increasing.     } Note that $K$ is measurable, since $F^-$ and $G^-$ are measurable. Examples of such ``curves'' are given in Figure~\ref{fig:F-G-}. We need the following lemma:
\begin{figure}[t]
  \centering
  \includegraphics[width=10cm,height=9cm]{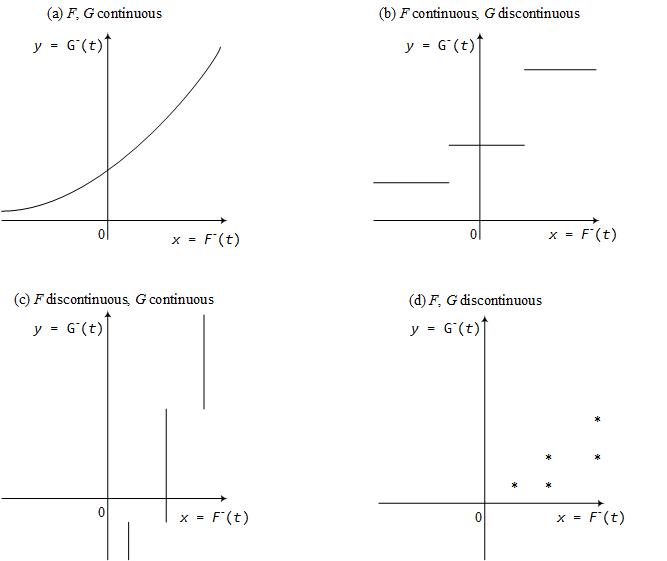}
  \caption{\small Various shapes of $\lambda_K ((0,1]) \subset \real^2$, with $F^-(t)$ on the x-axis and $G^-(t)$ on the y-axis.}
  \label{fig:F-G-}
\end{figure} 
\begin{lemme} \label{properties}
\emph{(i)} For all $b \in [0,1]$, one has $b=\lambda([0,b])$.\newline
\emph{(ii)} If $f = (f_1,f_2)$ is a function $E \rightarrow E_1 \times E_2$, then $f^{-1} (A \times B) = (f^{-1}_1 A)\cap (f^{-1}_2 B)$ for all $A \subset E_1$, $B \subset E_2$. \newline
\emph{(iii)} Define $A_x = (-\infty,x]$ and let $F^-$ be the quantile function associated with the cdf $F$.       Then $(F^-)^{-1} (A_x) = [0,F(x)]$.
\end{lemme}
(i) is trivial, (ii) is well-known. Property (iii) is a consequence of (\ref{propquant}). Indeed, \newline
$(F^-)^{-1} (A_x) = \{ t \in [0,1]: F^-(t) \leq x \} = \{ t \in [0,1]:  t \leq F(x) \} = [0,F(x)]$.

Let us designate by $\lambda$ the Lebesgue measure restricted to $[0,1]$. Then $\lambda_K$ denotes the pushforward probability measure of the Lebesgue measure on $[0,1]$ induced by $K$ (on $(\real^2,{\cal B}_2)$).\footnote{Another way of looking at $\lambda_K$: let $X \sim F$ (resp. $Y \sim G$) be the cdf characterizing $\mu$ (resp. $\nu$), and let $U\sim U(0,1)$ be a random variable uniformly distributed on $[0,1]$. Then $\hat{X} := F^-(U) \sim \mu$, $\hat{Y} := G^-(U) \sim \nu$ and $\lambda_K$ is the law of the pair $(\hat{X},\hat{Y})$. That is, the transport plan $\lambda_K$ is a coupling of $\muandnu$ or, in other words, $(\hat{X},\hat{Y})$ is a coupling of $\xandy$.   } 
Adopting, as in Theorem~\ref{hxy}, the notation $\bar{\pi}$ for a plan with cdf $H(x,y) = \min \{F(x), G(y) \}$, we
show that: (A) $\lambda_K \in \Pi (\mu,\nu)$, and (B) $\lambda_K = \bar{\pi}$. 

To prove (A), i.e. to prove that $\muandnu$ are the marginals of $\lambda_K$, we use the rule governing composition of maps: the first marginal of 
$\lambda_{(F^-,G^-)}$ is $(\lambda_{(F^-,G^-)})_{q_1} = \lambda_{q_1 \circ (F^-,G^-)}) = \lambda_{F^-} = \mu$. 
To see that $\lambda_{F^-} = \mu$, we can use the points (i) and (iii) of Lemma~\ref{properties}: define     $A_x = (-\infty,x]$. We only have to show that $\lambda_{F^-}(A_x) = \mu (A_x)$ for any $x\in \real$. But 
$\lambda_{F^-}(A_x) = \lambda ((F^-)^{-1} (A_x)) \stackrel{(iii)}{=} \lambda([0,F(x)]) \stackrel{(i)}{=} F(x) = \mu (A_x)$. Using the same rule and $q_2$ instead of $q_1$, we see immediately that the second marginal of $\lambda_{(F^-,G^-)}$ is $\lambda_{G^-}  = \nu$.

Next, using (i), (ii) and (iii) of Lemma~\ref{properties}, we are now ready to prove (B), i.e. to prove that $\bar{\pi} = \lambda_K$. For $A_x = (-\infty,x]$ and  $B_y = (-\infty,y]$, all we need to show is that $\bar{\pi}(A_x \times B_y) = 
\lambda_K (A_x \times B_y)$. Then
$\bar{\pi} (A_x \times B_y)
= H(x,y) = \min \{F(x), G(y) \} 
=^{(i)} \lambda([0,\min \{F(x), G(y) \}]) \newline
= \lambda([0,F(x)] \cap [0,G(y)]) 
=^{(iii)}  \lambda([(F^-)^{-1} A_x] \cap [(G^-)^{-1} B_y]) 
=^{(ii)} \lambda( (F^-,G^-)^{-1}(A_x \times B_y)) \newline
= \lambda( K^{-1} (A_x \times B_y)) = \lambda_K( A_x \times B_y )$.

\begin{figure}[t]
  \centering
  \includegraphics[width=12cm,height=7cm]{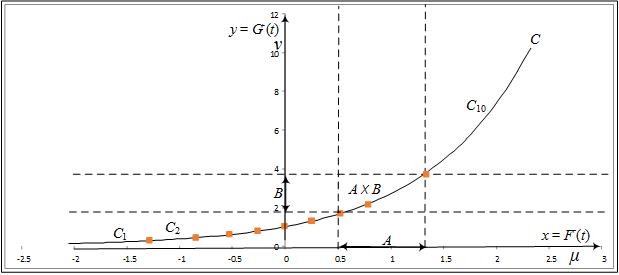}
  \caption{\small Optimal plan supported on a curve $C$ when both $F$ (the cdf of $\mu$) and $G$ (the cdf of $\nu$) are continuous and strictly increasing, with $F^-(t)$ on the x-axis and $G^-(t)$ on the y-axis. Here $\mu$ is the standard normal distribution and $\nu$ is the lognormal distribution with parameters 0 and 1.}
  \label{fig:planoptFGcont2}
\end{figure}

Taking into account the results we have just stated, the great contribution of Theorem~\ref{hxy} is to establish that $\lambda_K$ is Kantorovich-optimal as long as the cost function $c(x,y)$ satisfies the required convexity condition.
\begin{example}  	\label{curve}
In this example, $F$ is the cdf of a random variable $X \sim N(0,1) = \mu$ and $G$ is the cdf of $Y \sim \exp (X)$, i.e. $Y$ has lognormal distribution $logN(0,1) = \nu$. Figure~\ref{fig:planoptFGcont2} shows the curve 
$C = K((0,1]) = (F^-,G^-)((0,1])$. The interval $(0,1]$ has been divided into ten parts of equal length: 
$I_1 = (0,0.1]$, $I_2 = (0.1,0.2], \ldots$, $I_{10} = (0.9,1]$. Consequently, $C$ is divided into ten portions of curve $C_1 = K(I_1)$, $C_2 = K(I_2), \ldots$, $C_{10} = K(I_{10})$. We observe that the mass is stronger on the central portions of $C$ rather than on its extremities. 

Next, let $A\subset \real$ and $B\subset \real_+$ be the intervals shown in Figure~\ref{fig:planoptFGcont2}. Intuitively, $\lambda_K(A \times B)$ can be interpreted as the amount of mass contained in $A$ that is moved to $B$ by $\lambda_K$.
Since $C$ is the support of $\lambda_K$, $0.2=\lambda_K(A\times B) = \lambda_K(A\times \real_+) = \lambda_K(\real\times B) = \mu (A) = \nu (B)$. As expected, $\mu (A)$ -- the total amount of mass removed from $A$ -- and $\nu (B)$ -- the total amount of mass transferred to $B$ -- are equal.
\end{example}
Now, suppose that the cost function $c(x,y) = h(x-y)$ satisfies the convexity condition stated in Theorem~\ref{hxy}. 
Since $\bar{\pi} = \lambda_K$, the optimal cost is given by
\begin{equation}
\bar{c} = \int_0^1 h(F^-(t) - G^-(t))dt . \label{FrecGrec}
\end{equation}
We simply used the change of variables formula: 
$\bar{c} = \int c \ d\lambda_{(F^-,G^-)} = \int c \circ (F^-,G^-) d\lambda = \int_0^1 h(F^-(t) - G^-(t))dt$.
Taking $c(x,y) = |x-y|$ as a special case, (\ref{FrecGrec}) becomes
\begin{equation}
\bar{c} = \int_0^1 |F^-(t) - G^-(t)|dt . \label{varabs}
\end{equation}
Note that $\bar{c}$ in (\ref{varabs}) is not only the $L_1$-distance between quantile functions, it is also the $L_1$-distance between the corresponding cumulative distribution functions, i.e.
\begin{equation}
\bar{c} = \int_{\real} |F(x) - G(x)|dx. \label{FG}
\end{equation}
A proof of this remarkable coincidence is given in Thorpe (2018)\nocite{Thorpe2018}, see also Rachev and Rueschendorf (1998)\nocite{Rachev1998}. It should also be noted that Theorem~\ref{hxy} does not make any particular assumption on $F$ or $G$ and that the optimal plan $\bar{\pi} = \lambda_K$ has not necessarily the deterministic form $\pi = \pi_T =\mu_{(Id,T)}$ for a certain transport map $T$; therefore does not, as it stands, help to solve Monge's problem. The question now is to give additional assumptions about $\muandnu$ (or $F$ and $G$) ensuring that $K((0,1])$ is the graph of a transport map $T$: the very fact that this curve is the graph of a transport map $T$ means that $T$ is Monge-optimal. We give below two special instances where this happens: 
$K((0,1])$ is the graph of a transport map $T$ (i) when $F$ is continuous (Subsection~\ref{Fcontinuous}) and (ii) in the discrete case, when $\muandnu$ are supported on the same number of points of identical mass (Subsection~\ref{discrmonge}). Indeed, these two subsections yield two situations where, under the assumptions of Theorem~\ref{hxy}, $\lambda_{(F^-,G^-)} = \mu_{(Id,T)}$. 
\subsubsection{Deterministic optimal plan when $F$ is continuous } \label{Fcontinuous}
\begin{figure}[t]
  \centering
  \includegraphics[width=7cm,height=10cm]{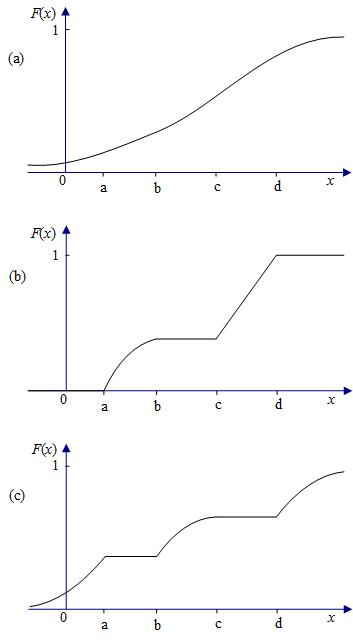}
  \caption{\small Examples of continuous cdf's. $F$ can be strictly increasing everywhere as in (a), but can also alternate intervals over which it is in turn strictly increasing or constant as in (b) and (c). The intervals over which $F$ is constant (resp. strictly increasing) are closed (resp. open). The collection of these intervals is a partition of $\real$. For example $\{ (-\infty, a], (a,b),[b,c],(c,d),[d,\infty) \}$ in case (b) and
	                   $\{ (-\infty, a), [a,b],(b,c),[c,d],(d,\infty) \}$ in case (c). }
  \label{fig:continuous_cdf}
\end{figure}

\noindent 
Examples of continuous cdf's appear in Figure~\ref{fig:continuous_cdf}. We now show that if $F$ is continuous, then the deterministic plan $\pi_{G^-\circ F} = \mu_{(Id,G^-\circ F)}$  has distribution function $H(x,y) = \min \{F(x), G(y)\}$. Theorem~\ref{hxy} then implies that $\mu_{(Id,G^-\circ F)}$  is Kantorovich-optimal for any cost function having the convexity property specified in this theorem. This means that $T = G^-\circ F$ is Monge-optimal for the same cost function. We need the following lemma:
\begin{lemme} \label{Fcont} (Proof in the appendix)
For $\mu \in {\cal P} (\real )$ and $x \in \real$, define $A_x = (-\infty ,x]$ and assume that $F(x) = \mu (A_x)$ is continuous. Then \newline
{\rm (a)} $\mu_F = \lambda$, where $\lambda$ denotes the Lebesgue measure restricted to $[0,1]$.\newline
{\rm (b)} For any Borel subset $C$ of $\real$ and any $x_0 \in \real$,
$\mu (A_{x_0} \cap C) = \mu (\ F^{-1} ([0, F(x_0)]) \cap C \ )$.
\end{lemme} 	
We will use the properties (i), (ii) and (iii) of Lemma~\ref{properties}, as well as the properties (a) and (b) of Lemma~\ref{Fcont}. Let $(x,y) \in \real^2$. According to Theorem~\ref{hxy}, all we have to do is show that 
$\pi_{(G^-\circ F)} (A_x \times B_y) = \min \{ F(x),G(y)\}$, where $A_x=(-\infty ,x]$ and $B_y=(-\infty ,y]$. 

\noindent We have \newline
$\pi_{(G^-\circ F)} (A_x \times B_y) = \mu_{(Id,G^-\circ F)}(A_x \times B_y) 
=^{(ii)} \mu [\ A_x \cap F^{-1} ((G^-)^{-1}(B_y))\ ]$ \newline
$=^{(iii)} \mu (\ A_x \cap F^{-1} ([0,G(y)]\ ) =^{(b)}  \mu (\ F^{-1} ([0, F(x)]) \cap F^{-1} ([0,G(y)] \ )
$ \newline
$= \mu (\ F^{-1} ([0, F(x)] \cap [0,G(y)])\ ) = \mu_F (\ [0, F(x)] \cap [0,G(y)]\ ) =^{(a)} 
                                            \lambda (\ [0,\min \{ F(x),G(y)\}]\ )$ \newline
$=^{(i)}\min \{ F(x),G(y)\} = H(x,y)$.

In Example~\ref{curve}, $F$ (standard normal) and $G$ (lognormal) are both continuous and correspond to case (a) in Figure~\ref{fig:F-G-}. If $c(x,y) = h(x-y)$ is a cost function with $h$ convex and continuous, the curve $C$ in Figure~\ref{fig:planoptFGcont2} is the graph of a (strictly increasing) optimal transport map $T(x) = G^- \circ F (x)$.
\subsubsection{Deterministic optimal plan in a special discrete case} \label{discrmonge}
\noindent 
We now propose a new case where the solution of Monge's problem involves Kantorovich' relaxation. Let the cost function $c(x,y) = h(x-y)$ be as in Theorem~\ref{hxy} (i.e. $h$ is convex and continuous) and assume that $\mu = \frac{1}{n}\sumin \delta_{x_i}$ and $\nu = \frac{1}{n}\sumjn \delta_{y_j}$. As we are dealing here with dimension one, the $x_i$'s and $y_j$'s are real numbers, and we assume that they are ordered: $x_1 \leq x_2 \leq \cdots \leq x_n$ and $y_1 \leq y_2 \leq \cdots \leq y_n$. Define $t_i = F(x_i)$, and $s_j = G(y_j)$,  $i,j \in \{ 1, \ldots , n \}$. As all points $\{x_i\}$ and $\{y_j\}$ have the same mass ($1/n$), $t_i = s_i$, $i = 1, \ldots , n$.
To prove that $\lambda_{(F^-,G^-)} = \mu_{(Id,G^-\circ F)}$, all we have to show is that $\lambda_{(F^-,G^-)}(\{(x_i,y_j)\}) = \mu_{(Id,G^-\circ F)} (\{(x_i,y_j)\})$ for all $i,j \in \{ 1, \ldots , n \}$. Consider the partition $\{(0,t_1],(t_1,t_2], \ldots , (t_{n-1},t_n =1]\}$ of the interval $(0,1]$, each element of the partition being of length $1/n$. Since $(F^-,G^-)((0,t_1]) = (x_1,y_1), (F^-,G^-)((t_1,t_2]) = (x_2,y_2), \ldots , (F^-,G^-)((t_{n-1},t_n]) = (x_n,y_n)$, we have \newline
$\lambda_{(F^-,G^-)}(\{(x_1,y_1)\}) = \lambda_{(F^-,G^-)}(\{(x_2,y_2)\}) = \cdots = \lambda_{(F^-,G^-)}(\{(x_n,y_n)\}) =1/n$. And since \newline  
$\sumin \lambda_{(F^-,G^-)}(\{(x_i,y_i)\}) = 1$, we have 
$\lambda_{(F^-,G^-)}(\{(x_i,y_j)\}) = \frac{1}{n}\delta_{ij}$ where $\delta_{ij}$ is the Kronecker delta 
($\delta_{ij}$ equals one if $i=j$, zero otherwise). On the other hand, note that $x_i \stackrel{F}{\longmapsto} t_i = s_i \stackrel{G^-}{\longmapsto} y_i$ (because $G^-$ is left-continuous), and therefore 
\begin{equation}
G^- \circ F (x_i) = y_i,\ \ \  i = 1, \ldots , n. \label{optiT1}
\end{equation}
Then 
$\mu_{(Id,G^-\circ F)} (\ \{(x_i,y_j)\}\ ) = \mu [\ (Id,G^-\circ F)^{-1} \{(x_i,y_j)\}\ ] = \mu (\ \{x_i\}\cap (G^-\circ F)^{-1} \{y_j\}\ ) =^{(\ref{optiT1})} \mu (\{ x_i\} \cap \{ x_j\}) = \frac{1}{n}\delta_{ij}$, and we have thus proved that $\lambda_{(F^-,G^-)} = \mu_{(Id,G^-\circ F)}$. Denoting by $T^* = G^-\circ F$ the Monge-optimal transport map, one gets from (\ref{optiT1})
\begin{equation}
T^*(x_i) = y_i,\ \ \  i = 1, \ldots , n. \label{optiT2}
\end{equation}
Proposition~\ref{permut} is a consequence of the above.
\begin{prop} \label{permut}
(Proof in the appendix) Consider two ordered sets of real numbers $x_1 \leq x_2 \leq \cdots \leq x_n$ and $y_1 \leq y_2 \leq \cdots \leq y_n$. Let $S$ be the set of permutations $\sigma: \{ \idotn \} \rightarrow \{\idotn \}$ and let $h$ be a convex continuous  function. Then
\begin{equation}
\min_{\sigma \in S} \sumin h(x_i - y_{\sigma(i)}) = \sumin h(x_i - y_i) \label{minsigma}. \hbox{ In particular }
\end{equation}
\begin{equation}
\min_{\sigma \in S} \sumin |x_i - y_{\sigma(i)}| = \sumin |x_i - y_i| = ||x - y||_1 ,\label{Manhattan}
\end{equation}
the Manhattan distance between $x=(x_1, \ldots ,x_n)$ and $y=(y_1, \ldots ,y_n)$.
\end{prop}
Note that the results shown in Subsection~\ref{discrmonge} have a multidimensional extension when the cost function $c$ is general and the optimization problem restricts to discrete measures \newline
$\mu = \frac{1}{n}\sumin \delta_{x_i}$ and $\nu = \frac{1}{n}\sumjn \delta_{y_j}$, where
$x_i , y_j \in \real^d$, $d \ge 2$. However, the $x_i$'s and $y_j$'s cannot be totally ordered in that case and the Monge-optimal transport map 
$T_{\sigma}$ defined by $T_\sigma (x_i) = y_{\sigma(i)}$, $i = \idotn$, is not necessarily the one corresponding to the trivial permutation $\bar{\sigma}(i) := i$. Using the Minkowski-Carath\'eodory Theorem and the Birkhoff Theorem, Thorpe (2018, Th. 2.5 and 2.6)\nocite{Thorpe2018} shows that any solution $\pi^*$ to Kantorovich's optimal transport problem is a permutation matrix, i.e. there exists a permutation $\sigma^* \in S$ such that
$\pi_{ij}^* = \frac{1}{n}\delta_{j=\sigma^*(i)}$, which implies that $T^*: \real^d \rightarrow \real^d$ defined by
$T^*(x_i) = y_{\sigma^*(i)}$ is a Monge-optimal transport map, see also  Villani (2003) \nocite{Villani2003}.

The results of Section~\ref{onedim} can be summarized as follows: let probability measures $\mu ,\nu$ on $(\real, {\cal B}_1)$ have the respective cdf's $F$ and $G$. For a cost function $c(x,y) = |x-y|$, or more generally $c(x,y) = h(x-y)$ with $h$ convex and continuous, the transport plan $\lambda_{(F^-,G^-)}$ supported on the ``curve'' 
$C = (F^-,G^-)(\ (0,1]\ )$ is Kantorovich-optimal. In this context, we gave two examples (special cases) where the Kantorovich-optimal transport plan $\lambda_{(F^-,G^-)}$ becomes $\mu_{(Id,G^-\circ F)}$ and the curve $C$ is the graph of a Monge-optimal transport map $T = G^- \circ F$. In the first example, $F$ is assumed to be continuous, while in the second example, $\muandnu$ are assumed to be discrete and supported on the same number of ordered real numbers having the same mass. Proposition~\ref{permut} is a consequence of the latter case.
\subsection{Optimal transport cost as a metric} \label{wasser}
We add this section for completeness, noting that the optimal transport problem can be put forward to define a distance between $\muandnu$, namely the so-called \emph{Wasserstein metric} (or \emph{Wasserstein distance}). It is also known, in computer sciences, as the \emph{earth mover's distance}.
Like the optimal transport problem discussed in this paper, the definition of the Wasserstein distance is set in a much broader context than the one presented below. This concerns the shape or the properties of the cost function, the characteristics of the spaces on which $\mu$ and $\nu$ are defined (usually $\real^d$ or subsets of $\real^d$), as well as the value of $p$ in the definitions that follow.

The set of probability measures on $(\real ,{\cal B}_1)$ (or on $(E,E \cap {\cal B}_1)$, $E \subset \real$) with finite $p$-th moment is defined as
\[
{\cal P}_p (\real ) = \left\{ \mu \in {\cal P} (\real ) :\int_{\real}|x|^p d\mu(x) < \infty   \right\},
\]
noting that if $E \subset \real$ is bounded, then ${\cal P}_p (E ) = {\cal P} (E )$.
\begin{defi}\label{Wasser}
For $\mu, \nu \in {\cal P}_p (\real )$ and $p\in [1,\infty)$, the Wasserstein distance between $\muandnu$ is defined as
\begin{equation}
W_p(\mu ,\nu) = \left( \inf_{\pi \in \Pi (\mu,\nu)} \int_{\real^2}|x-y|^p d\pi(x,y) \right)^{1/p}, \label{Wasserp}
\end{equation}
that is, the Wasserstein distance is the $p^{th}$ root of the minimum of the Kantorovich optimal transport problem for cost function $c(x,y) = |x-y|^p$.
\end{defi} It can be shown that the distance $ W_p: {\cal P}_p(\real) \times {\cal P}_p(\real) \rightarrow [0,\infty)$ is a metric on ${\cal P}_p(\real)$. 
Under the probabilistic notation adopted above, where the random variables are projections (i.e. $X = q_1$ and $Y = q_2$), (\ref{Wasserp}) becomes
\[
W_p(\mu ,\nu) = \left( \inf_{\pi \in \Pi (\mu,\nu)}  E_\pi |X-Y|^p \right)^{1/p}.
\]
In particular, for $p = $1
\begin{equation}
W_1(\mu ,\nu) = \inf_{\pi \in \Pi (\mu,\nu)} E_{\pi} |X - Y| , \label{Wassere1}
\end{equation}
that is, since $c(x,y) = |x-y|$ is convex and continuous 
\begin{equation}
W_1(\mu ,\nu) = \int_0^1 |F^-(t) - G^-(t)|dt = \int_{\real} |F(x) - G(x)|dx ,
\end{equation}
as we have seen in (\ref{varabs}) and (\ref{FG}).
\section{Concluding remarks} \label{conclusion}
In this article, we focused on the $L_p$ distances and in priority on $L_1$. Its content is intended to be a compromise between theoretical considerations and applications. We did not dwell on the already well-known properties of these distances, but we went back to basics of probability theory to clarify various aspects that applied science papers tend to neglect. We have examined the relationship between the $L_1$ distance and the Gini-Kantorovich distance (an $L_1$ distance between cfd's), as well as certain uses of the former, such as the Gini mean difference or the Lukaszyk-Karmowski metric. Unlike simple metrics such as the Gini-Kantorovich distance, $\exy$ integrates the dependency structure between $\xandy$ and makes it possible in particular to take account of the assumption of independence. We then studied the axiomatic in which $\exy$ is inscribed; this allowed us to uncover a interpretive error that crept into the literature on the subject. The properties of $\exy$ have been clarified in the cases of independence, equality of distribution and almost sure equality. The problem of the normalization of $\exy$ has been solved, especially the question of triangle inequality. The Gini index is a special case of this [0,1]-normalized form. We have also shown that, for identically distributed variables $\xandy$, $\exy$ can be interpreted as a distance to almost sure equality. In a section reserved more specifically for applications, $\exy$ is expressed in analytic form when $X\sim N(\mu_X,\sigma^2_X)$ and $Y\sim N(\mu_Y,\sigma^2_Y)$ are independent. The resulting formula generalizes tools used in applied physics: it allows in particular to lift the assumption of homoscedasticity ($\sigma^2_X = \sigma^2_Y$), and thus allows more flexibility in the use of the Lukaszyk-Karmowski metric. Closed-form formulas are also determined when independently distributed $\xandy$ have continuous uniform distributions. These formulas can be used to compute the average distance between intervals. Moreover, leads are opened for obtaining analytical forms when $\xandy$ follow non-normal distributions, noting that such an attempt can be very complicated or even hopeless in some cases. 
Finally, for two probability measures $\mu$ and $\nu$ defined on the real line, $\exy$ is a key ingredient in the optimal transport problem when the issue is how to transport $\mu$ to $\nu$, whilst minimizing a cost function of the form $c(x,y) = |x-y|$. The question of optimal transport is developed in a relatively complete way within this restricted framework. The way the problem is presented has been thought of as a first approach to a particularly demanding domain. \newpage
\section{Appendix: the proofs} \label{appendix}
\vskip 0.1cm 
\noindent \textbf{{\em Proof of Proposition~\ref{equivalence}} }   
\vskip .1cm
\noindent {\rm a)} $\Rightarrow$ {\rm b)}:  First, let us show that $X\stackrel{a.s.}{=} Y$ implies $X^2\stackrel{a.s.}{=}XY$ . But $X\stackrel{a.s.}{=}Y$ means that $P(X-Y=0)=1$. Define the events $A = \{ X - Y = 0\}$ and $B = \{ X^2 - XY = 0\}$. As $X - Y = 0$ implies $X^2 - XY = 0$, we have $A\subset B$ and $P(A) \leq P(B)$. Since $P(A) = 1$, we conclude that $P(B) = 1$, i.e. $X^2 \stackrel{a.s.}{=} XY$, which in turn implies that 
$E(X^2) = E(XY)$. As $X$ and $Y$ are independent, we have $E(X^2) = E(X) E(Y)$ (which means that $X^2$ as and $XY$ are both integrable). Noting that $E(X) = E(Y)$ (since $X\stackrel{a.s.}{=}Y$), we obtain $[E(X)]^2 = E(X^2)$, which implies that we have equality in the Jensen's inequality $\psi(E(X)) \leq E[\psi(X)]$, $\psi(\cdot)$ being here the square function. Equality in the Jensen's inequality for a strictly convex $\psi$ implies that the concerned random variable is almost surely a constant (Niculescu and Persson (2004)\nocite{Niculescu2004}). Consequently, there exists a real number $c$ such that $X\stackrel{a.s.}{=}c$. As $X\stackrel{a.s.}{=} Y$, we also have $Y\stackrel{a.s.}{=} c$. \newline
\noindent {\rm b)} $\Rightarrow$ {\rm a)}: trivial.
\vskip 0.1cm 
\noindent \textbf{{\em Proof of Theorem~\ref{ineqtri1}} }   
\vskip .1cm
Let $a$, $b$, $c$, $\alpha$, $\beta$, and $\gamma$ be non-negative real numbers such that $(\alpha - \beta  + \gamma) \geq 0$  and $(\alpha a - \beta b + \gamma c) \geq 0$. Then
\begin{equation}
\alpha a^2 - \beta b^2 + \gamma c^2 + (\alpha - \beta  + \gamma) (ab + bc + ac) \geq 0. \label{abc}
\end{equation}  
Indeed, without loss of generality, we set $a\leq c$, an assumption enabling us to examine three cases (instead of six). \newline
\emph{Case 1}: $a\leq c \leq b$ \newline
We let the reader check that 
\[
\alpha a^2 - \beta b^2 + \gamma c^2 + (\alpha - \beta  + \gamma) (ab + bc + ac)
\]
\[
= \underbrace{(\alpha a - \beta b + \gamma c)}_{\geq 0} (a+b+c) + \underbrace{(\alpha - \beta  + \gamma)}_{\geq 0} ac + \gamma a \underbrace{(b-c)}_{\geq 0} + \alpha c\underbrace{(b-a)}_{\geq 0},
\] 
from which (\ref{abc}) follows.

\noindent \emph{Case 2}: $a\leq b \leq c$ \newline
We let the reader check that 
\[
\alpha a^2 - \beta b^2 + \gamma c^2 + (\alpha - \beta  + \gamma) (ab + bc + ac)
\]
\[
= \underbrace{(\alpha a - \beta b + \gamma c)}_{\geq 0} (a+c) + \underbrace{(\alpha - \beta  + \gamma)}_{\geq 0} (ab+bc) +                   \beta \underbrace{(b-a)(c-b)}_{\geq 0},
\] 
from which (\ref{abc}) follows.

\noindent \emph{Case 3}: $b\leq a \leq c$ \newline
Since $\alpha a^2 - \beta b^2 + \gamma c^2 \geq0$, (\ref{abc}) follows immediately, because we have

\[
\alpha a^2 \underbrace{- \beta b^2}_{\geq - \beta a^2} + \underbrace{\gamma c^2}_{\geq \gamma a^2} \geq 
\alpha a^2 - \beta a^2 + \gamma a^2 = \underbrace{(\alpha - \beta  + \gamma)}_{\geq 0} a^2 \geq 0,
\] 
which implies (\ref{abc}).

Next, let $x$, $y$, $z$ be real numbers. Then
\begin{equation}
|y-z||x| - |x-z||y| + |x-y||z| \geq 0. \label{canberra}
\end{equation}  
To prove (\ref{canberra}), note that if (at least) one of the three values $x$, $y$, $z$ is zero, then (\ref{canberra}) is trivial. Now, suppose that the three values $x$, $y$, $z$ are nonzero. Then \newline \noindent
${ \displaystyle 
\left( \frac{1}{y} - \frac{1}{z}\right)^2 = \frac{1}{y^2} + \frac{1}{z^2} - \frac{2}{yz} = \frac{y^2+z^2 -2yz}{y^2z^2} 
= \frac{(z-y)^2}{y^2z^2}    }$  which implies     \newline \noindent
${ \displaystyle \left| \frac{1}{y} - \frac{1}{z}\right| = \frac{|y-z|}{|y||z|}    }$. In the same way, we get
${ \displaystyle \left| \frac{1}{x} - \frac{1}{y}\right| = \frac{|x-y|}{|x||y|}    }$ and
${ \displaystyle \left| \frac{1}{x} - \frac{1}{z}\right| = \frac{|x-z|}{|x||z|}    }$. Using the triangle inequality:     \newline \noindent
${ \displaystyle \left| \frac{1}{x} - \frac{1}{z}\right| \leq \left| \frac{1}{x} - \frac{1}{y}\right| + \left| \frac{1}{y} - \frac{1}{z}\right|    }$, i.e.
${ \displaystyle \frac{|x-z|}{|x||z|} \leq \frac{|x-y|}{|x||y|} + \frac{|y-z|}{|y||z|}     }$. Multiplying by $|x||y||z|$, we have 
$|x-z||y| \leq |x-y||z| + |y-z||x| $, which proves (\ref{canberra}).

For the rest of the proof, let us denote $D(X,Y) = \exy$ and \newline
$D_{norm}(X,Y) = \exy /(E|X| + E|Y|)$ with $E|X| + E|Y| > 0$.
Next, let us show that
\begin{equation}
\theta(X,Y,Z) := D(Y,Z)\overline{\mu}_X - D(X,Z)\overline{\mu}_Y + D(X,Y)\overline{\mu}_Z \geq 0, \label{ptol}
\end{equation}
with the notations $\bar{\mu}_X := E|X|$, $\bar{\mu}_Y := E|Y|$ and $\bar{\mu}_Z := E|Z|$. 
Using Fubini-Tonelli several times: \newline
\noindent
${ \displaystyle \theta(X,Y,Z) = \int_{y=-\infty}^\infty \int_{z=-\infty}^\infty |y-z| dP_Z(z) dP_Y(y)  \cdot          
                             \int_{x=-\infty}^\infty|x| dP_X(x)   }$ \newline \noindent
${ \displaystyle - \int_{x=-\infty}^\infty \int_{z=-\infty}^\infty |x-z| dP_Z(z) dP_X(x)  \cdot  
                              \int_{y=-\infty}^\infty|y| dP_Y(y)                 
}$ \newline \noindent 
${ \displaystyle + \int_{x=-\infty}^\infty \int_{y=-\infty}^\infty |x-y| dP_Y(y) dP_X(x)  \cdot  
                              \int_{z=-\infty}^\infty|z| dP_Z(z)
}$ \newline \noindent
${ \displaystyle = \int_{y=-\infty}^\infty \int_{z=-\infty}^\infty \int_{x=-\infty}^\infty |y-z||x| dP_X(x) dP_Z(z)dP_Y(y)                       
}$ \newline \noindent
${ \displaystyle - \int_{x=-\infty}^\infty \int_{z=-\infty}^\infty \int_{y=-\infty}^\infty |x-z||y| dP_Y(y) dP_Z(z) dP_X(x)
}$ \newline \noindent 
${ \displaystyle + \int_{x=-\infty}^\infty \int_{y=-\infty}^\infty \int_{z=-\infty}^\infty|x-y||z| dP_Z(z) dP_Y(y) dP_X(x) 
}$ \newline \noindent
${ \displaystyle = \int_{x=-\infty}^\infty \int_{y=-\infty}^\infty \int_{z=-\infty}^\infty 
                   \underbrace{|y-z||x| - |x-z||y| + |x-y||z|}_{\geq 0, \hbox{ see (\ref{canberra})}} dP_Z(z) dP_Y(y) dP_X(x)\geq 0   }$. 

\begin{figure}[t]
	\centering
		\includegraphics{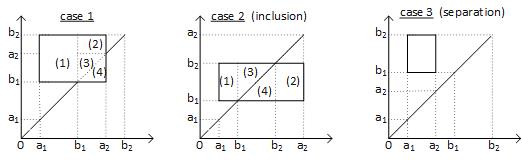}
	\caption{\small Computation of the double integral for the three cases to be taken into account in the proof of Theorem~\ref{distunif}.
	\textbf{Case 1}: $a_1 \leq b_1 <a_2 < b_2$ ($A \cap B\neq \phi$ without inclusion).  \textbf{Case 2}: $a_1 < b_1 < b_2 < a_2$ (inclusion: $B \subset A$ ). \textbf{Case 3}: $a_1 < a_2 < b_1 < b_2$ ($A \cap B = \emptyset$, separation). }
	\label{fig:IAB}
\end{figure}

We are now able to prove that $D_{norm}(X,Y) = D(X,Y)/(\overline{\mu}_X + \overline{\mu}_Y)$ satisfies the triangle inequality. In order to use (\ref{abc}), define  $\gamma = D(X,Y)$, $\alpha = D(Y,Z)$, $\beta = D(X,Z)$, $a=\overline{\mu}_X$, $b=\overline{\mu}_Y$, and $c=\overline{\mu}_Z$. The assumptions of Theorem~\ref{ineqtri1} insure that $a+c>0$, $a+b>0$ and $b+c>0$. We have to show the triangle inequality \newline
$D_{norm}(X,Z) \leq D_{norm}(X,Y) + D_{norm}(Y,Z)$, i.e. with our notation
\[
\frac{\beta}{a+c} \leq \frac{\gamma}{a+b}  + \frac{\alpha}{b+c} \hskip 1cm {\rm or, equivalently,} 
\]
\begin{equation}
\alpha a^2 - \beta b^2 + \gamma c^2 + (\alpha - \beta  + \gamma) (ab + bc + ac) \geq 0. \label{verif}
\end{equation}
Noting that $(\alpha - \beta  + \gamma) \geq 0$ since $D(\cdot,\cdot)$ satisfies the triangle inequality and that, using (\ref{ptol}), $(\alpha a - \beta b + \gamma c) \geq 0$, inequation (\ref{verif}) is valid and the triangle inequality of $D_{norm}(\cdot,\cdot)$ follows from (\ref{abc}). Theorem~\ref{ineqtri1} is now proved.
\vskip 0.1cm 
\noindent \textbf{{\em Proof of Lemma~\ref{inequadri}} }   
\vskip .1cm
\noindent
By triangle inequality, one has $\psi (X,Y) \leq \psi (X,X_1) + \psi (X_1,Y)$ and 
$\psi (X_1,Y) \leq \psi (X_1,Y_1) + \psi (Y_1,Y)$. Taken together, these two inequalities imply
\begin{equation}
\psi (X,Y) - \psi (X_1,Y_1) \leq \psi (X,X_1) + \psi (Y_1,Y). \label{intermed1}
\end{equation}
In the same manner, by inverting $X \leftrightarrow X_1$ and $Y \leftrightarrow Y_1$, we get
\begin{equation}
\psi (X_1,Y_1) - \psi (X,Y) \leq \psi (X_1,X) + \psi (Y,Y_1). \label{intermed2}
\end{equation}
Joining (\ref{intermed1}) and (\ref{intermed2}), and using symmetry and non-negativity, (\ref{quadrineq}) follows immediately.
\vskip 0.1cm 
\noindent \textbf{{\em Proof of Proposition~\ref{ascase}} }   
\vskip .1cm
\noindent
$\xandy$ are defined on $\probamodel$, take their values in $(\real, {\cal B}_1)$ and are such that $P_X = P_Y$. Let us write $P_X = P_Y = \mu$ for simplicity. \newline \noindent
``$\Rightarrow$'': Consider $B_1$, $B_2 \in {\cal B}_1$. Then
$P_{(X,Y)} (B_1\times B_2) = P(X\in B_1, Y\in B_2)$\newline [[using $P(\{ X=Y \}) = 1$]] $=P(X\in B_1, Y\in B_2, \{ X=Y \})$
$= P(X\in B_1, X\in B_2) = P(X^{-1}(B_1\cap B_2)) = P_X(B_1\cap B_2) = \mu (B_1\cap B_2)$
$= (\mu \triangle\mu ) (B_1\times B_2) $. Since $P_{(X,Y)}$ and $\mu \triangle\mu $ coincide on the $\pi$-system ${\cal B}_1\times {\cal B}_1$, and by virtue of the Dynkin's $\pi$-$\lambda$ theorem, they actually coincide on the whole ${\cal B}_1\otimes {\cal B}_1 = {\cal B}_2$. \newline \noindent
``$\Leftarrow$'': 
Denote by $Id$ the identity map on $\real$ and consider the injective measurable function 
$(Id,Id) : \real \longrightarrow \real^2$ defined by $(Id,Id)(x) = (x,x)$, which is such that 
$(Id,Id)(\real) = \Delta$, the main diagonal of $\real^2$, and $(Id,Id)^{-1}(\Delta) = \real$. First, note that 
$\mu \triangle\mu = \mu_{(Id,Id)}$, where $\mu_{(Id,Id)}$ is the pushforward distribution of $\mu$ on 
$(\real^2,{\cal B}_2)$ induced by $(Id,Id)$. Indeed, for all $B_1,B_2 \in {\cal B}_1$,
$\mu_{(Id,Id)} (B_1 \times B_2) = \mu ((Id,Id)^{-1} (B_1 \times B_2)) =  \mu (B_1 \cap B_2) = 
(\mu \triangle\mu )(B_1 \times B_2)$. 
Then $P(\{ X=Y \}) = P((X,Y) \in \Delta) = P_{(X,Y)}(\Delta )$
[[ hypothesis]] $= (\mu \triangle \mu) (\Delta ) = \mu_{(Id,Id)} (\Delta ) = \mu ((Id,Id)^{-1}(\Delta)) 
= \mu (\real ) =1$, i.e. $X\stackrel{a.s.}{=} Y$.
\vskip 0.1cm 
\noindent \textbf{{\em Proof of Lemma~\ref{samespace}} }   
\vskip .1cm
\noindent
The $\sigma$-field ${\cal B}_2 \cap \Delta$ is generated by the set 
${\cal C}_\Delta := \{ (B_1 \times B_2)\cap \Delta : B_1, B_2 \in {\cal B}_1 \}$, i.e. 
${\cal B}_2 \cap \Delta = \sigma ({\cal C}_\Delta )$. On the other hand, $s({\cal B}_1)$ is a $\sigma$-field on $\Delta$, because $s:\real \longrightarrow \Delta$ is bijective, and $s({\cal B}_1)$ is generated by the set
${\cal C}:= \{ s(B_1): B_1 \in {\cal B}_1 \}$. All we have to show is that ${\cal C}_\Delta = {\cal C}$.\newline
(a) `$\subset$': Let $D \in {\cal C}_\Delta$. There exists $B_1, B_2 \in {\cal B}_1$ such that $D = (B_1\times B_2)\cap \Delta = s(B_1 \cap B_2)$. As $B_1 \cap B_2\in {\cal B}_1$, $D = s(B_1 \cap B_2) \in {\cal C}$.  \newline
`$\supset$': Let $D \in {\cal C}$, i.e. there exists $B_1 \in {\cal B}_1$ such that $D = s(B_1)$. There also exists $B_2 \in {\cal B}_1$ such that $D = (B_1 \times B_2)\cap \Delta$, which means that $D \in {\cal C}_\Delta$.\newline
(b) We only have to show that the two probability measures agree on the $\pi$-system 
${\cal C}_\Delta$ generating the $\sigma$-field ${\cal B}_2 \cap \Delta$. Take any $D \in {\cal C}_\Delta$, i.e. there exists $B_1, B_2 \in {\cal B}_1$ such that $D = (B_1 \times B_2)\cap \Delta$. Then
$(\mu \triangle\mu)\left\lfloor_\Delta \right.(D) = (\mu \triangle\mu)(D \cap \Delta) / (\mu \triangle \mu)(\Delta) = (\mu \triangle\mu)(D) = $ [[using $(\mu \triangle \mu)((B_1 \times B_2)\cap \Delta) = (\mu \triangle \mu)(B_1 \times B_2)$]] $(\mu \triangle \mu)(B_1 \times B_2) =$
 [[definition of $\mu \triangle \mu$]] $\mu (B_1 \cap B_2) = $ [[using $B_1 \cap B_2 = s^{-1}((B_1 \times B_2) \cap \Delta) $]]
$\mu (s^{-1}((B_1 \times B_2)\cap \Delta) = \mu_s((B_1 \times B_2)\cap \Delta) = \mu_s (D)$.
\vskip 0.1cm 
\noindent \textbf{{\em Proof of Proposition~\ref{covXcovY1}} }   
\vskip .1cm
Let $f$ (resp. $g$) be the probability density function (pdf) of $X$ (resp. $Y$).
Using Fubini-Tonelli extensively, and defining $\psi(x,y) = f(y)g(x) + f(x)g(y)$,
\begin{eqnarray*} 
\exy &=& \int_{\real^2}|x-y|f(x)g(y)dxdy \\
       &=& \int_{x=-\infty}^{\infty} \int_{y=-\infty}^{x}|x-y|f(x)g(y)dydx  + \int_{x=-\infty}^{\infty} \int_{y=-\infty}^{x}|y-x|f(y)g(x)dydx \\
       &=& \int_{x=-\infty}^{\infty} \int_{y=-\infty}^{x}|x-y|\psi(x,y)dydx = \int_{x=-\infty}^{\infty} \int_{y=-\infty}^{x}(x-y)           \psi(x,y)dydx .
\end{eqnarray*} 
Let us write $\exy = I_1 + I_2$, where $I_1$ and $I_2$ are defined hereunder:
\begin{eqnarray*} 
I_1 &:=& \int_{x=-\infty}^{\infty} \int_{y=-\infty}^{x} xf(y)g(x)dydx  
        - \int_{x=-\infty}^{\infty} \int_{y=-\infty}^{x} yf(y)g(x)dydx\\
    &=& \int_{x=-\infty}^{\infty} xg(x) \underbrace{\int_{y=-\infty}^{x} f(y)dy}_{F(x)} dx  
        - \int_{x=-\infty}^{\infty} g(x)H(x)dx ,
\end{eqnarray*} 				
 where $H(x) = \int_{y=-\infty}^{x} yf(y)dy$, i.e. $H'(x) = xf(x)$. Applying integration by parts to the second integral, we finally get			
\begin{eqnarray} 
I_1 &=& \int_{x=-\infty}^{\infty} xF(x)g(x)dx - \underbrace{G(x)H(x) \left|^{\infty}_{-\infty}\right. }_{\mu_Y} 
        + \int_{x=-\infty}^{\infty} xG(x)f(x)dx  \nonumber  \\
    &=&	E[YF(Y)]  -	\mu_Y + E[XG(X)]. \label{esp1}
\end{eqnarray} 							
Next, let us calculate
\begin{eqnarray*} 
I_2 &:=& \int_{x=-\infty}^{\infty} \int_{y=-\infty}^{x} xf(x)g(y)dydx  
        - \int_{x=-\infty}^{\infty} \int_{y=-\infty}^{x} yf(x)g(y)dydx\\
    &=& \int_{x=-\infty}^{\infty} xf(x) \underbrace{\int_{y=-\infty}^{x} g(y)dy}_{G(x)} dx  
        - \int_{x=-\infty}^{\infty} \int_{y=-\infty}^{x} f(x)K(x)dxdy ,
\end{eqnarray*} 				
 where $K(x) = \int_{y=-\infty}^{x} yg(y)dy$, i.e. $K'(x) = xg(x)$. Applying integration by parts to the second integral,			
\begin{eqnarray} 
I_2 &=& \int_{x=-\infty}^{\infty} xG(x)f(x)dx - \underbrace{F(x)K(x) \left|^{\infty}_{-\infty}\right. }_{\mu_X} 
        + \int_{x=-\infty}^{\infty} xF(x)g(x)dx  \nonumber  \\
    &=&	E[XG(X)] -	\mu_X + E[YF(Y)] . \label{esp2}
\end{eqnarray} 
Adding (\ref{esp1}) and \ref{esp2} one gets
$$
\exy = I_1+I_2 = 2E[XG(X)]  + 2E[YF(Y)] -	\mu_X -	\mu_Y.
$$
\vskip 0.1cm 
\noindent \textbf{{\em Proof of Theorem~\ref{distnorm1}} }   
\vskip .1cm
Let $F$ (resp. $G$) be the cdf of $X$ (resp. $Y$). We first calculate $E[XG(X)]$.
\[
E[XG(X)] = \frac{1}{2\pi \sigma_X \sigma_Y}\int_{x=-\infty}^{\infty} \int_{y=-\infty}^{x} x
             \exp \left\{-\frac{1}{2} \left[ \left(\frac{x-\mu_X}{\sigma_X}\right)^2 + \left(\frac{y-\mu_Y}{\sigma_Y}\right)^2 \right] \right\}dydx. \\  
\]				
Set $s=(x-\mu_X)/\sigma_X$ and $t=(y-\mu_Y)/\sigma_Y$. The news bounds are $-\infty < s < \infty$ and 
$-\infty < t < \frac{\sigma_X}{\sigma_Y}s + \frac{\mu_X - \mu_Y}{\sigma_Y}$ and the Jacobian of the linear transformation is     $\sigma_X\sigma_Y$. For ease of reading, write $\theta = \sigma_X/\sigma_Y$ and $\tau = (\mu_X-\mu_Y)/\sigma_Y$. Then
\begin{eqnarray*} 
E[XG(X)] &=& \frac{1}{2\pi}\int_{s=-\infty}^{\infty} \int_{t=-\infty}^{\theta s + \tau} (\sigma_X s+\mu_X)
              \exp \left\{-\frac{1}{2} \left[ s^2 + t^2\right] \right\} dtds       \\
    &=& \frac{\sigma_X}{2\pi}\int_{s=-\infty}^{\infty} \int_{t=-\infty}^{\theta s + \tau} s
              \exp \left\{-\frac{1}{2} \left[ s^2 + t^2\right] \right\} dtds \\
		    &&  + \  \frac{\mu_X}{2\pi}\int_{s=-\infty}^{\infty} \int_{t=-\infty}^{\theta s + \tau}
              \exp \left\{-\frac{1}{2} \left[ s^2 + t^2\right] \right\} dtds \\
				&=:& 	A_1 + A_2.		
\end{eqnarray*}
\begin{eqnarray*} 
A_1 &=& \frac{\sigma_X}{2\pi}\int_{s=-\infty}^{\infty} \int_{t=-\infty}^{\theta s + \tau} s
              \exp \left\{-\frac{1}{2} \left[ s^2 + t^2\right] \right\} dtds \\
		    &=&  \sigma_X \int_{-\infty}^{\infty} s \phi(s) \Phi(\theta s + \tau) ds      \\
				&& 	[\hbox{using integration by parts and noting that }  s \phi(s) = -\phi'(s)]   \\
        &=& - \sigma_X \int_{-\infty}^{\infty} \phi'(s) \Phi(\theta s + \tau) ds      \\
        &=& - \sigma_X \underbrace{\phi(s) \Phi(\theta s + \tau)\left|^{\infty}_{-\infty}\right.}_{0}     
				    + \theta \sigma_X \int_{-\infty}^{\infty} \phi(s) \phi(\theta s + \tau) ds            \\				
				&=&  \frac{\theta \sigma_X}{2\pi} \int_{-\infty}^{\infty} \exp \left\{-\frac{1}{2} \left[ s^2 +                           (\theta s +\tau)^2\right] \right\}ds     \\
				&=&  \frac{\theta \sigma_X}{2\pi} \int_{-\infty}^{\infty} \exp \left\{-\left[ \frac{1+\theta^2}{2}s^2 +                           \theta \tau s + \frac{\tau^2}{2}\right] \right\}ds     \\ 
				&=&  \frac{\theta \sigma_X}{\sqrt{2\pi}} \frac{1}{\sqrt{1+\theta^2}} 
				     \exp \left\{ \frac{\theta^2\tau^2}{2(1+\theta^2)} - \frac{\tau^2}{2} \right\}       \\ 				
        &=&  \frac{\theta \sigma_X}{\sqrt{1+\theta^2}} \phi (\tau)
				     \exp \left\{ \frac{\theta^2\tau^2}{2(1+\theta^2)}  \right\}.
\end{eqnarray*} 
On the other hand
\begin{eqnarray} 
A_2 &=& \frac{\mu_X}{2\pi}\int_{s=-\infty}^{\infty} \int_{t=-\infty}^{\theta s + \tau} 
              \exp \left\{-\frac{1}{2} \left[ s^2 + t^2\right] \right\} dtds  \nonumber  \\  
		    &=&  \mu_X \int_{-\infty}^{\infty} \phi(s) \Phi(\theta s + \tau) ds \label{phiPhi}     \\ 
				&=& 	\mu_XE[\Phi(\theta S + \tau) ], \hbox{ where } S\sim N(0,1),       \nonumber  
\end{eqnarray}
$\Phi(\theta S + \tau)$ being a random variable of which we seek the mean. We could directly calculate the integral appearing in (\ref{phiPhi}) by using sheer mathematical analysis, but relying on the probabilistic meaning of $\phi$ and $\Phi$ will help us to avoid tedious computations. Let $Y$ be a standard normal random variable, independent of $S$. Since $\Phi(\theta s + \tau) = P(Y \le \theta s + \tau)$, we have that $E[\Phi(\theta S + \tau)] = P(Y - \theta S \le \tau)$. Define     $U =Y - \theta S $, which implies that $ U\sim N(0,1+\theta^2)$. We have    \newline \noindent 
${\displaystyle E[\Phi(\theta S + \tau)] = P(U \le \tau) = P(\frac{U}{\sqrt{1+\tau^2}} \le \frac{\tau}{\sqrt{1+\tau^2}}) 
		    = \Phi(\frac{\tau}{\sqrt{1+\tau^2}})  }$, and finally \newline \noindent
${\displaystyle A_2 = \mu_X \Phi(\frac{\tau}{\sqrt{1+\tau^2}}) }$.  \newline \noindent
Then ${\displaystyle E[XG(X)] = A_1 + A_2 = \frac{\theta \sigma_X}{\sqrt{1+\theta^2}} \phi (\tau)
			\exp \left\{ \frac{\theta^2\tau^2}{2(1+\theta^2)}  \right\} + \mu_X \Phi(\frac{\tau}{\sqrt{1+\tau^2}}) }$.\newline \noindent 
			Since $\theta=\sigma_X / \sigma_Y$ and $\tau = (\mu_X-\mu_Y)/\sigma_Y$, one gets \newline \noindent
${ \displaystyle E[XG(X)] =
  \frac{\sxx}{\sqrt{\B}} \phi \left( \frac{\A}{\sigma_Y} \right) \exp \left\{ \frac{\sxx (\A)^2 }{2\syy (\B)} \right\} + \mu_X \Phi \left(\frac{\A}{\sqrt{\B}} \right)   }$. Now, we can simply interchange $X$ and $Y$ to obtain  
	\newline \noindent 
	${ \displaystyle E[YF(Y)] =      
  \frac{\syy}{\sqrt{\B}} \phi \left( \frac{\AAA}{\sigma_X} \right) \exp \left\{ \frac{\syy (\A)^2 }{2\sxx (\B)} \right\} + \mu_Y \Phi \left(\frac{\AAA}{\sqrt{\B}} \right)   }$.\newline
From (\ref{covXcovY2}), writing $a:=\mu_X-\mu_Y$ and $b:=\sigma_X^2+\sigma_Y^2$, and noting that $\phi (-z) = \phi (z)$ and 
$\Phi (-z) = 1-\Phi (z)$, we get  \newline \noindent
${ \displaystyle \exy = \frac{2\sxx}{\sqrt{b}} \phi \left( -\frac{a}{\sigma_Y} \right) \exp \left\{ \frac{\sxx a^2 }{2b\syy} \right\}    
        + \frac{2\syy}{\sqrt{b}} \phi \left( \frac{a}{\sigma_X} \right) \exp \left\{ \frac{\syy a^2 }{2b\sxx} \right\}   
				+ 2\mu_X \Phi \left(\frac{a}{\sqrt{b}} \right)  }$
\newline \noindent
${ \displaystyle + \ 2\mu_Y \Phi \left(-\frac{a}{\sqrt{b}} \right)  -\mu_X - \mu_Y  }$    \newline \noindent
${ \displaystyle = \frac{2\sxx}{\sqrt{b}} \phi \left( \frac{a}{\sigma_Y} \right) \exp \left\{ \frac{\sxx a^2 }{2b\syy} \right\}
+ \frac{2\syy}{\sqrt{b}} \phi \left( \frac{a}{\sigma_X} \right) \exp \left\{ \frac{\syy a^2 }{2b\sxx} \right\} 
+ 2a \Phi \left( \frac{a}{\sqrt{b}} \right) -a     }$.  \newline \noindent
Finally, as $\exy$ does not change if we replace $a$ by $-a$, we can substitute $|a|$ for $a$ in the last line above.
\vskip .1cm 
\underline{Alternative proof of Theorem~\ref{distnorm1}}
\vskip .1cm
First, let us show the following result:
Let $X\sim N(\mu,\sigma^2)$, and let $\phi$ (resp. $\Phi$) be the pdf (resp. the cdf) of the standard normal distribution. Then
\begin{equation}
E|X| = |\mu| \left[2\Phi\left(\frac{|\mu|}{\sigma}\right) - 1 \right] + 2\sigma \phi \left(\frac{|\mu|}{\sigma}\right) .	 \label{mubarre2}
\end{equation}
Let us write $\overline{\mu}$ instead of $E|X|$. To prove (\ref{mubarre2}), we must consider the cases a) and b) below:  \newline \noindent
 a) $x\geq 0$. Define 
 ${\displaystyle 
 \bar{\mu}_{1} = \frac{1}{\sigma \sqrt{2\pi}} \int^{\infty}_{0}
 x e^{-\frac{1}{2}\left(\frac{x-\mu}{\sigma} \right)^2 }dx  
 = \frac{1}{\sqrt{2\pi}} \int^{\infty}_{-\mu/\sigma} (\mu + \sigma y) e^{-\frac{y^2}{2} }dy     }$	\newline									
 \centerline{[we have set $y=(x-\mu)/\sigma$]}		\newline
 ${\displaystyle   
 = \frac{\mu}{\sqrt{2\pi}} \int^{\infty}_{\frac{-\mu}{\sigma}} e^{-\frac{y^2}{2} }dy     
 + \frac{\sigma}{\sqrt{2\pi}} \int^{\infty}_{\frac{-\mu}{\sigma}} y e^{-\frac{y^2}{2} }dy
 = \mu[1-\Phi(-\mu/\sigma)] - \frac{\sigma}{\sqrt{2\pi}} e^{-\frac{y^2}{2}}\left|^{\infty}_{-\mu/\sigma} \right.
 }$	 \newline										
 ${ \displaystyle = \mu[1-\Phi(-\mu/\sigma)] + \frac{\sigma}{\sqrt{2\pi}} e^{-\frac{\mu^2}{2\sigma^2}} 
 = \mu[1-\Phi(-\mu/\sigma)] + \sigma\phi(-\mu/\sigma).
 }$ 		\newline
 b) $x<0$. Define 
 ${\displaystyle 
 \bar{\mu}_{2} = \frac{-1}{\sigma\sqrt{2\pi}} \int^{0}_{-\infty}
 x e^{-\frac{1}{2}\left(\frac{x-\mu}{\sigma} \right)^2 }dx  
 = \frac{-1}{\sqrt{2\pi}} \int^{\frac{-\mu}{\sigma}}_{-\infty} (\mu + \sigma y) e^{-\frac{y^2}{2} }dy     
 }$	\newline									
 ${\displaystyle   
 = - \frac{\mu}{\sqrt{2\pi}} \int^{\frac{-\mu}{\sigma}}_{-\infty} e^{-\frac{y^2}{2} }dy     
   - \frac{\sigma}{\sqrt{2\pi}} \int^{\frac{-\mu}{\sigma}}_{-\infty} y e^{-\frac{y^2}{2} }dy
 = - \mu \Phi(-\mu/\sigma) + \frac{\sigma}{\sqrt{2\pi}} e^{-\frac{y^2}{2}}\left|^{-\mu/\sigma}_{-\infty} \right.
 }$	 \newline										
 ${ = - \mu \Phi(-\mu/\sigma) + \frac{\sigma}{\sqrt{2\pi}} e^{-\frac{\mu^2}{2\sigma^2}} 
 = \sigma\phi(-\mu/\sigma) - \mu \Phi(-\mu/\sigma).
 }$ Hence \newline $\bar{\mu} = \bar{\mu}_{1} + \bar{\mu}_{2}$
 $= \mu[1-\Phi(-\mu/\sigma)] + \sigma\phi(-\mu/\sigma)
      + \sigma\phi(-\mu/\sigma) - \mu \Phi(-\mu/\sigma)$ \newline 
 $ = \mu[1-2\Phi(-\mu/\sigma)] + 2\sigma\phi(-\mu/\sigma) = \mu[2\Phi(\mu/\sigma) - 1] + 2\sigma\phi(\mu/\sigma) $. 
Noting that $\bar{\mu}(-\mu) = \bar{\mu}(\mu)$, we can replace $\mu$ by $|\mu|$. The proof of (\ref{mubarre2}) is now complete. \newline
Next, $X\sim N(\mu_X,\sigma^2_X)$ and $Y\sim N(\mu_Y,\sigma^2_Y)$ are independent. Define $Z=X-Y$. A well-known consequence of the convolution theorem is that $Z\sim N(\mu_X - \mu_Y,\sigma^2_X + \sigma^2_Y)$. Replacing $\mu$ by $\mu_X - \mu_Y$ and $\sigma^2$ by $\sigma^2_X + \sigma^2_Y$ in (\ref{mubarre2}), we get (\ref{distnorm3}).
\vskip 0.1cm 
\noindent \textbf{{\em Proof of Theorem~\ref{distunif}} }   
\vskip .1cm
An alternative proof would use the fact that the convolution of two uniform laws is a triangular law. We give here a direct proof. Without loss of generality, the bounded proper intervals $A$ and $B$ are open. Define $I(A,B) = L_AL_B \exy = \int_A \int_B |x-y|dxdy$, which we wish to express explicitly. The three possible cases are given in figure~\ref{fig:IAB}. We give the proof only for the first case, the proof for the two other cases being analogous. 
Let us consider Figure~\ref{fig:IAB}, case 1, and the partition $\{ (1),(2),(3),(4) \}$ of $A\times B$. Then
\begin{eqnarray*} 
(1)&=&\int_{a_1}^{b_1} \int_{b_1}^{b_2} (y-x)dxdy = \int_{a_1}^{b_1} \{\int_{b_1}^{b_2}ydy\} dx - \int_{a_1}^{b_1} x \{\int_{b_1}^{b_2}dy\} dx  \\
&=& (b_2^2-b_1^2)(b_1-a_1)/2 - (b_2-b_1)(b_1^2-a_1^2)/2 = (b_1-a_1)(b_2-b_1)(b_2-a_1)/2.  \\
(2)&=&\int_{b_1}^{a_2}\int_{a_2}^{b_2} (y-x)dxdy = \int_{b_1}^{a_2}\{\int_{a_2}^{b_2} ydy\} dx - \int_{b_1}^{a_2}x \{\int_{a_2}^{b_2} dy\} dx  \\
&=& (b_2^2-a_2^2)(a_2-b_1)/2 - (b_2-a_2)(a_2^2-b_1^2)/2 = (a_2-b_1)(b_2-a_2)(b_2-b_1)/2. \\
(3)&=&\int_{b_1}^{a_2}\int_x^{a_2} (y-x)dxdy = \int_{b_1}^{a_2}\{\int_x^{a_2} ydy\} dx - \int_{b_1}^{a_2}x \{\int_x^{a_2} dy\} dx  \\
&=&\frac{1}{2}\int_{b_1}^{a_2}(a_2^2-x^2) dx - \int_{b_1}^{a_2}x(a_2-x)dx  \\
&=&\frac{a_2^2}{2}(a_2-b_1)- \frac{1}{6} (a_2^3-b_1^3) - a_2\int_{b_1}^{a_2}x dx + \int_{b_1}^{a_2}x^2 dx\\
&=&\frac{a_2^2}{2}(a_2-b_1)- \frac{1}{6} (a_2^3-b_1^3) - \frac{a_2}{2}(a_2^2-b_1^2) + \frac{1}{3}(a_2^3-b_1^3)\\
&=&\frac{a_2^2}{2}(a_2-b_1-a_2^2+b_1^2) + \frac{1}{6} (a_2^3-b_1^3). \\
(4)&=&\int_{b_1}^{a_2}\int_{b_1}^x (x-y)dxdy = \int_{b_1}^{a_2}x \{\int_{b_1}^x dy\} dx - \int_{b_1}^{a_2}\{\int_{b_1}^x y dy\} dx  \\
&=&\int_{b_1}^{a_2}x(x-b_1)dx -\frac{1}{2}\int_{b_1}^{a_2}x^2 dx + \frac{b_1^2}{2}(a_2-b_1)\\
&=&\int_{b_1}^{a_2}x^2 dx - b_1\int_{b_1}^{a_2}x dx - \frac{1}{6} (a_2^3-b_1^3) + \frac{b_1^2}{2}(a_2-b_1)\\
&=&\frac{1}{3} (a_2^3-b_1^3) - \frac{b_1}{2} (a_2^2-b_1^2) - \frac{1}{6} (a_2^3-b_1^3) + \frac{b_1^2}{2}(a_2-b_1)\\
&=&\frac{1}{6} (a_2^3-b_1^3) + \frac{b_1}{2}(a_2b_1-a_2^2).
\end{eqnarray*}
Then $I(A,B) = (1)+(2)+(3)+(4) = (b_2-b_1)(b_1-a_1)(b_2-a_1)/2
+ (b_2-a_2)(a_2-b_1)(b_2-b_1)/2 - b_1a_2^2 +a_2b_1^2 +\frac{1}{3} (a_2^3-b_1^3) = (b_2-b_1)(b_1-a_1)(b_2-a_1)/2 + (b_2-a_2)(a_2-b_1)(b_2-b_1)/2 + \frac{(a_2-b_1)^3}{3}$.\newline
We still have to express $E|X|$ and $E|Y|$ explicitly. We give the proof for $E|X|$, the proof for $E|Y|$ being identical. To compute $E|X|$, we must consider three cases:\newline
(i) $a_1 \geq 0$. Then $E|X| = L_A^{-1}\int_A |x|dx = L_A^{-1}\int_A x dx = \frac{L_A^{-1}}{2} (a_2^2-a_1^2) =L_A^{-1} \frac{a_1+a_2}{2}L_A= m_A$.
\newline
(ii) $a_2 \leq 0$. Then $E|X| = -L_A^{-1} \int_A x dx = -L_A^{-1}\frac{a_1+a_2}{2}L_A = -m_A$. \newline
Considering (i) and (ii) together, i.e. if $0\notin A$, we have $E|X| = |m_A|$. \newline
(iii) If $0\in A$, we have $E|X| = - L_A^{-1}\int_{a_1}^0 x dx + L_A^{-1}\int_0^{a_2} x dx = (a_1^2 + a_2^2)/(2L_A)$. \newline
Following the same lines, we can show that $E|Y| = |m_B|$ if  $0\notin B$ and $E|Y| = (b_1^2 + b_2^2)/(2L_B)$ if $0\in B$.
\vskip 0.1cm 
\noindent \textbf{{\em Proof of Lemma~\ref{Fcont}} }   
\vskip .1cm
\noindent
(a) Let $F$ be the cdf of a random variable $X$, i.e. $F$ is part of the configuration 
$(\real^2,{\cal B}_2, \pi )\stackrel{X=q_1}{\rightarrow} (\real,{\cal B}_1, \mu = \pi_{q_1}) \stackrel{F}{\rightarrow} 
([0,1], {\cal B}_1 \cap [0,1] )$. It is well-known that if $F$ is continuous, then $F(X)\sim U(0,1) = \lambda$ (see e.g. Karr (1993) \nocite{Karr1993} or Embrechts and Hofer (2014)\nocite{Embrechts2014}). Hence \newline ${\displaystyle\lambda = {\cal L}(F(X)) = \pi_{(F\circ X)} = \pi_{(F\circ q_1)} = (\pi_{q_1})_F = \mu_F           }$.
\newline
(b) $F$ can be strictly increasing everywhere, but can also alternate intervals over which it is in turn strictly increasing or constant (see Figure~\ref{fig:continuous_cdf}). The intervals over which $F$ is constant (resp. strictly increasing) are closed (resp. open). The collection of these intervals is a partition of $\real$. 
Let $x_0$ be a real number. Two cases are possible:
(i) $x_0$ is either in an open interval over which $F$ is strictly increasing, or
(ii) $x_0$ is in a closed interval over which $F$ is constant.
The very definitions of $F^-$ and $F^+$ imply that $F^- \circ F(x_0) \leq x_0 \leq F^+ \circ F(x_0)$. In case (i), the three values in question are equal and, since $(-\infty ,F^+ \circ F(x_0)] = F^{-1}([0, F(x_0)])$, we have $A_{x_0} = F^{-1}([0, F(x_0)])$ and  $\mu (A_{x_0}) = \mu (\ F^{-1}([0, F(x_0)])\ )$.
In case (ii), at least two of the three values $F^- \circ F(x_0)$, $x_0$ and $F^+ \circ F(x_0)$ will differ. However, since $\mu$ is atomless, $\mu$ cancels out on the interval $[F^- \circ F(x_0),F^+ \circ F(x_0)]$  and, in particular, on the interval $(x_0,F^+ \circ F(x_0)]$, resulting in $\mu (A_{x_0}) = \mu (\ (-\infty, F^+ \circ F(x_0)]\ ) = \mu (\ F^{-1} ([0, F(x_0)])\ )$.
So we always have $\mu (A_{x_0}) = \mu (\ F^{-1} ([0, F(x_0)])\ )$. Let $C$ be a Borel subset of $\real$. Noticing that $A_{x_0} \subset F^{-1} ([0, F(x_0)])$, we have $\mu (A_{x_0} \cap C) = \mu (\ F^{-1} ([0, F(x_0)]) \cap C\ )$.
\vskip 0.1cm 
\noindent \textbf{{\em Proof of Proposition~\ref{permut}} }   
\vskip .1cm
It was shown in Subsection~\ref{discrmonge} that the deterministic transport plan $\pi_{T^*}$ with $T^* = G^-\circ F$ is Kantorovich-optimal (see (\ref{optiT2})). However, $\pi_{T^*}$ is not the only deterministic plan in $\Pi (\mu,\nu)$. By definition, there are the same number of deterministic plans in $\Pi (\mu,\nu)$ as there are transport maps in     ${\cal T} (\mu,\nu)$, and this number is $n!$. Indeed, let  $T: \{x_1, \ldots , x_n\}\rightarrow \{ y_1, \ldots , y_n \}$ be a transport map. Then $T$ is necessarily bijective. To see this, suppose  the opposite, i.e. that the function $T$ is not injective or not surjective. If $T$ is not injective, then there exists $y_j$ and $x_i\ne x_k$ such that 
$T^{-1}(\{ y_j \}) \supset \{ x_i,x_k \}$ and therefore $\mu (T^{-1}(\{ y_j \})) \ge \mu (\{ x_i,x_k \}) = 2/n > 1/n = \nu(\{ y_j \})$, 
which means that $T\notin {\cal T} (\mu,\nu)$: contradiction. Suppose now that $T$ is not surjective, i.e. there exists $y_j$ such that 
$T^{-1}(\{ y_j \}) = \emptyset$. Then $\nu(\{ y_j \}) = 1/n \ne \mu (\emptyset) = 0$: contradiction. The number of bijections $T: \{x_1, \ldots , x_n\}\rightarrow \{ y_1, \ldots , y_n \}$ being $n!$, there are exactly $n!$ transport maps in ${\cal T} (\mu,\nu)$, and they are of the form $T_\sigma (x_i) = y_{\sigma(i)}$, $i = \idotn$, with $\sigma \in S$. Consider the trivial permutation $\bar{\sigma}(i) := i$. Using this notation, the optimal transport map of (\ref{optiT2}) becomes $T_{\bar{\sigma}} (x_i) = y_i$, $i = 1, \ldots ,n$, and corresponds to the optimal deterministic plan $\pi_{T_{\bar{\sigma}}}$. It follows that for any $\sigma \in S$ we have $E_{\pi_{T_{\bar{\sigma}}}} h(X-Y)\leq E_{\pi_{T_\sigma}} h(X-Y)$, i.e. 
$E_{\mu} [h(X-T_{\bar{\sigma}} (X)] \leq E_\mu [h(X-T_{\sigma} (X)]$, which implies 
$\sumin h(x_i - y_i) \leq \sumin h(x_i - y_{\sigma(i)})$, and Proposition~\ref{permut} is now proved. 

\singlespacing

\tableofcontents

\end{document}